\documentclass[a4paper,english,british,conference, 10pt, final, onecolumn]{IEEEtran}
\usepackage[T1]{fontenc}
\usepackage[latin9]{luainputenc}
\usepackage{amsmath}
\usepackage{amsthm}
\usepackage{amssymb}
\usepackage{graphicx}
\usepackage{setspace}
\makeatletter

\@ifundefined{pageheight}{\let\pageheight\pdfpageheight}{}
\@ifundefined{pagewidth}{\let\pagewidth\pdfpagewidth}{}
\pageheight\paperheight
\pagewidth\paperwidth

\theoremstyle{plain}
\newtheorem{thm}{\protect\theoremname}
  \theoremstyle{definition}
  \newtheorem{defn}[thm]{\protect\definitionname}

\usepackage{color}
\usepackage{soul}

\makeatother

\usepackage{babel}
  \addto\captionsbritish{\renewcommand{\definitionname}{Definition}}
  \addto\captionsbritish{\renewcommand{\theoremname}{Theorem}}
  \addto\captionsenglish{\renewcommand{\definitionname}{Definition}}
  \addto\captionsenglish{\renewcommand{\theoremname}{Theorem}}
  \providecommand{\definitionname}{Definition}
\providecommand{\theoremname}{Theorem}

\begin{document}

\title{Linear System Identification in a Nonlinear Setting - Nonparametric
analysis of the nonlinear distortions and their impact on the best
linear approximation}

\author{J. Schoukens, M. Vaes, and R. Pintelon. Vrije Universiteit Brussel,
dept. ELEC, Pleinlaan 2, 1050 Brussel, Belgium\\ E-mail:
Johan.Schoukens@vub.ac.be\\}

\maketitle
\emph{This is a postprint copy of the following article under IEEE copyright: Linear System Identification in a Nonlinear Setting: Nonparametric Analysis of the Nonlinear Distortions and Their Impact on the Best Linear Approximation. J.Schoukens, M. Vaes, and R. Pintelon. IEEE Control Systems Magazine, Volume: 36, Issue: 3, June 2016, pp. 38 - 69. DOI: 10.1109/MCS.2016.2535918}\\

Linear system identification \cite{Ljung boek 1999}-\cite{Schoukens 2012  Exercises book}
is a basic step in modern control design approaches. Starting from
experimental data, a linear dynamic time-invariant model is identified
to describe the relationship between the reference signal and the
output of the system. At the same time, the power spectrum of the
unmodeled disturbances is identified to generate uncertainty bounds
on the estimated model.

Linear system identification is also used in other disciplines, for
example vibrational analysis of mechanical systems, where it is called
modal analysis \cite{Ewins Modal Testing 2000,Oomen (2014) Waferstage}.
Because linear time-invariant models are a basic model structure,
linear system identification is frequently used in electrical \cite{Verbeeck (1999) Identification Synchronous Machines}-\cite{Dedene Nele (2003) Global Synchronous Machine model MIMO},
electronic, chemical \cite{Rivera Daniel (2009) Constrained Multisine  plant-friendly identification},
civil \cite{Peeters Bart (2003)MSSP  Comparative modal analysis on bridge},
and also in biomedical applications \cite{Westwick boek 2003}. It
provides valuable information to the design engineers in all phases
of the design process.

Starting from the late 1960s, system identification tools have been
developed to obtain parametric models to describe the dynamic behavior
of systems. A formal framework is set up to study the theoretical
properties of the system identification algorithms \cite{Ljung boek 1999}-\cite{Pintelon 2012 book}.
The consistency (does the estimated model converge to the true system
as the amount of data grows?) and the efficiency (is the uncertainty
of the estimated model as small as possible?) are analyzed in detail.
Underlying all these results are the assumptions that the system to
be modeled is linear and time invariant.

It is clear that these assumptions are often (mostly?) not met in
real-life applications. Most systems are only linear to a first approximation.
Depending on the excitation level, the output is disturbed by nonlinear
distortions so that the linearity assumption no longer holds. This
immediately raises doubts about the validity of the results obtained
and validated by the linear system identification framework. The term
\emph{nonlinear distortions} indicates that nonlinear systems with
a (dominant) linear term are considered. The deviations from the linear
behavior are called nonlinear distortions. 

Moreover, because a linear model cannot capture the nonlinear distortions, it may be necessary to identify a nonlinear model to obtain results that are useful and reliable. Identification of nonlinear models requires more data and is more involved than linear identification.
Currently, identification of nonlinear systems is a hot research
topic, but the nonlinear identification framework has not yet reached
the same level of maturity linear identification theory has \cite{Westwick boek 2003}-\cite{Paduart NLSS 2010}.
Since the cost of a nonlinear approach is significantly higher, additional
information is needed to guarantee that there will be sufficient return
on the additional investment of time, money, and human resources that
is needed.

This article addresses the following problems:
\begin{itemize}
\item First, a nonlinearity analysis is made looking for the presence of
nonlinearities in an early phase of the identification process. The
level and the nature of the nonlinearities should be retrieved without
a significant increase in the amount of measured data.
\item Next it is studied if it is safe to use a linear system identification
approach, even if the presence of nonlinear distortions is detected.
The properties of the linear system identification approach under
these conditions are studied, and the reliability of the uncertainty
bounds is checked.
\item Eventually, tools are provided to check how much can be gained if
a nonlinear model were identified instead of a linear model.
\end{itemize}
Addressing these three questions forms the outline of this article.
The possibilities and pitfalls of using a linear identification framework
in the presence of nonlinear distortions will be discussed and illustrated
on lab-scale and industrial examples. 

In this article, the focus is on nonparametric and parametric black
box identification methods, however the results might also be useful
for physical modeling methods. Knowing the actual nonlinear distortion
level can help to choose the required level of detail that is needed
in the physical model. This will strongly influence the modeling effort.
Also, in this case, significant time can be saved if it is known
from experiments that the system behaves almost linearly. The converse
is also true. If the experiments show that some (sub-)systems are
highly nonlinear, it helps to focus the physical modeling effort on
these critical elements.

Three major steps are made to reach the main goals. First, a motivational
example is given, using linear system identification tools in the
presence of nonlinear distortions. This will give a first idea about
the possibilities and problems. Next, a nonparametric nonlinear distortion
analysis is proposed and illustrated on many real-life examples. It
includes experiment design, nonparametric preprocessing, and dealing
with closed-loop measurement conditions. In the first approach, open
loop measurement conditions are considered; the closed-loop measurement
conditons are postponed until the end of the article. To generalize the linear framework to include nonlinear effects,
a new paradigm is developed, representing nonlinear systems using
the best linear approximation (BLA) plus a nonlinear noise source.
First, an analysis of the impact of the user choices is made (choice of the excitation signal, the convergence criteria, and the approximation criterion). Next, a mathematical framework is introduced to give a sound theoretical basis for the description of nonlinear systems using linear models. The concept of the BLA is formally introduced, and an optimized measurement strategy to measure the frequency response function is developed. Again, these results are illustrated by some lab-scale and real-life examples. This is followed by a study of the impact of nonlinear distortions on the parametric linear identification framework.
At the end of the article, a short discussion about publicly available
software is given, followed by the conclusions.

This article is an extension of the keynote address that was given
at The 13th International Workshop on Advanced Motion Control AMC2014
\cite{Schoukens AMC2014}.

\section*{A motivational example\label{sec:A-motivational-example}}

\begin{figure}
\begin{centering}
\includegraphics[scale=0.1]{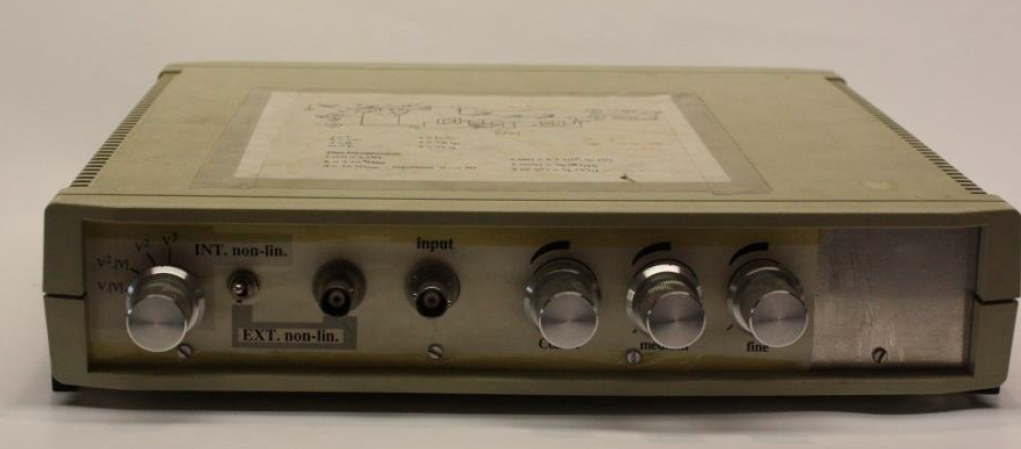}
\par\end{centering}
\begin{centering}
(a)
\par\end{centering}
\begin{centering}
\includegraphics[scale=0.5]{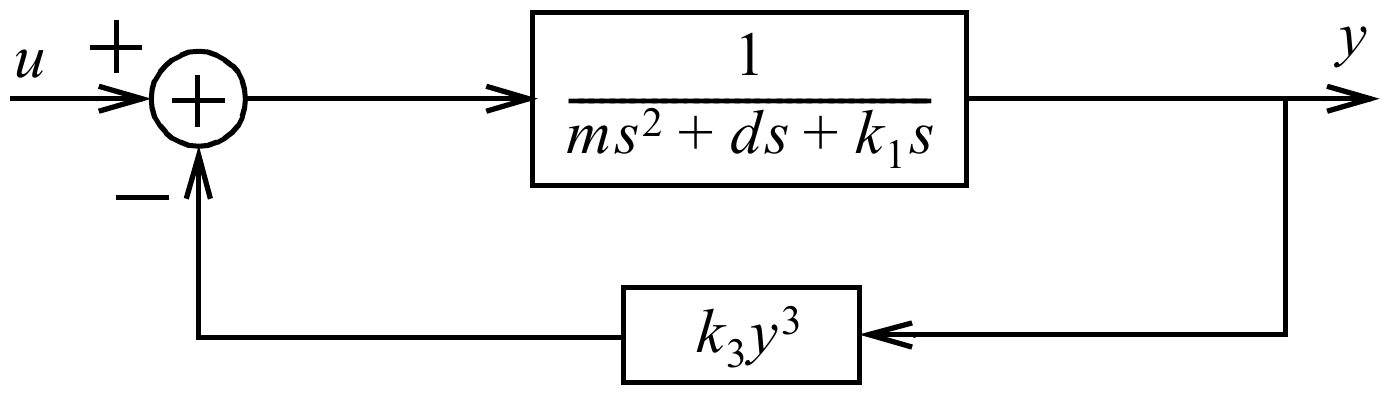}
\par\end{centering}
\begin{centering}
(b)
\par\end{centering}
\centering{}\caption{Forced Duffing oscillator (a): The electronic circuit mimics a nonlinear
mechanical system with a hardening spring. Such a system is sometimes
called a forced Duffing oscillator. The system is excited with an
input $u(t)$ (the applied force to the mechanical system). The output
of the system corresponds to the displacement $y(t)$. The schematic
representation of the system is given in (b) as a second-order system
with a nonlinear feedback.\label{fig:Loudspeaker Setup}}
\end{figure}

Consider the test setup in Figure \ref{fig:Loudspeaker Setup}. The
electronic circuit mimics a nonlinear mechanical system with a hardening
spring. Such a system is sometimes called a forced Duffing oscillator
\cite{Thompson and Stewart book Nonlinear Dynamics and Chaos,Ueda (1991) Forced Duffing oscillator}.
This class of nonlinear systems has a very rich behavior, including
regular and chaotic motions, and generation of subharmonics. The
system is excited with an input $u(t)$ (the applied force to the
mechanical system). The output of the system corresponds to the displacement
$y(t)$ .

\begin{figure}
\centering{}\includegraphics{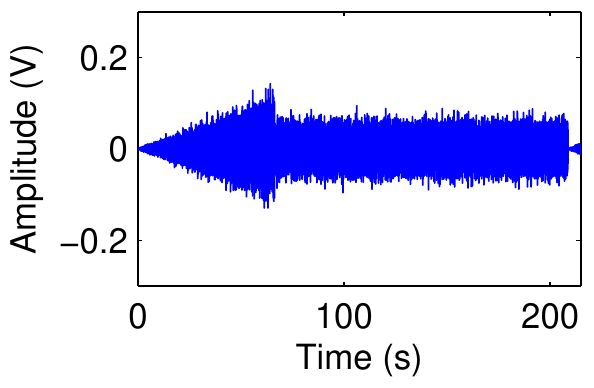}\caption{The system is excited with a low-pass signal, with a maximum excitation
frequency of 200 Hz. The excitation signal consists of two parts.
The tail part consists of 10 sub-experiments. Each of these is a realization
of a random signal, and will be used to estimate a linear model (Box-Jenkins
structure) to model the data. The arrow-like part will be used to
validate the estimated model. Observe that at the end of the arrow,
the excitation level is larger than the tail amplitude. This gives
the possibility to test the extrapolation capacity of the linear model.\label{fig:Motivating Exampe Excitation}}
\end{figure}

\begin{figure}
\begin{centering}
\includegraphics{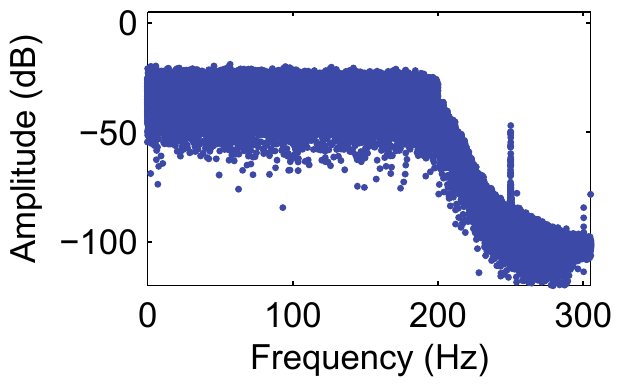}
\par\end{centering}
\begin{centering}
(a)
\par\end{centering}
\begin{centering}
\includegraphics{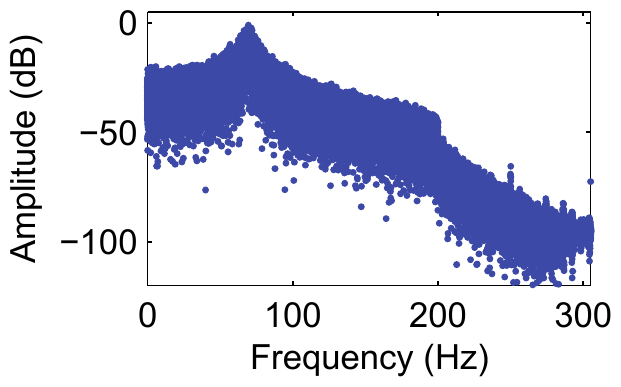}
\par\end{centering}
\centering{}(b)\caption{The amplitude spectrum of the input (a) and output (b) signal. The
spike at 250 Hz is a harmonic disturbance of the mains.\label{fig:Motivating Example: Spectrum of input and output signal}}
\end{figure}

This system can be schematically represented as a second-order system
with a nonlinear feedback. It is excited with a low-pass random excitation
with a maximum excitation frequency of 200 Hz, as shown in Figures
\ref{fig:Motivating Exampe Excitation} and \ref{fig:Motivating Example: Spectrum of input and output signal}. 

\subsection*{Modeling the nonlinear system using linear system identification
tools}

A linear approximating model will be estimated to describe the input-output
relation of the system from the flat tail part. 

\begin{figure}
\centering{}\includegraphics{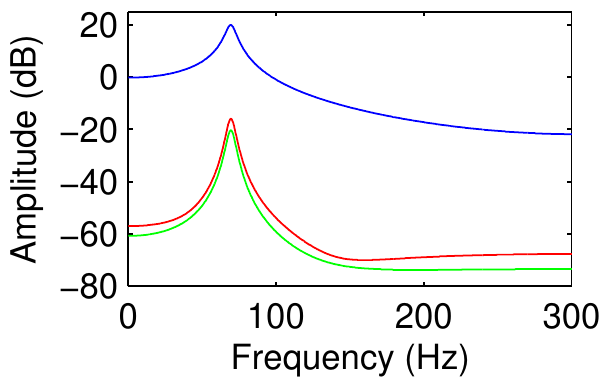}\caption{The amplitude of the estimated transfer function model is shown (blue).
Green line: the theoretic standard deviation of the estimated plant
model, calculated from the estimated noise model. Red line: the actual observed
standard deviation of the estimated plant model, estimated from the variations of the estimated plant model over the 10 subrecords.
It can be seen that the actually observed standard deviation is underestimated
with 4 dB by the theoretical results. This leads to too small error
bounds.\label{fig:Motivating Example: transfer function}}
\end{figure}

The tail is split in 10 sub-records with a length of 8692 points,
and each of these is used to identify a second-order discrete-time
plant model and a sixth-order noise model using the Box-Jenkins model
structure of the prediction-error method \cite{Ljung boek 1999,Soderstrom boek 1989}.
The estimated second-order plant transfer function, is shown in Figure
\ref{fig:Motivating Example: transfer function}. 

\begin{figure}
\centering{}\includegraphics{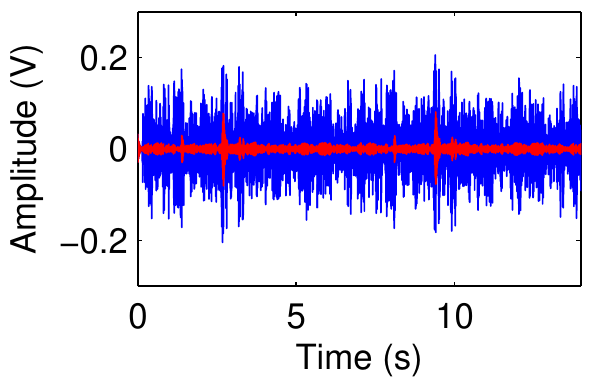}\caption{The output of the forced Duffing oscillator is simulated using an
estimated Box-Jenkins model (plant model order 2 poles and 2 zeros,
noise model order 6 poles and 6 zeros). The blue line is the measured
output, the red line is the simulation error. \label{fig:Motivating example: Simulation Error TD}}
\end{figure}

\begin{figure}
\centering{}\includegraphics{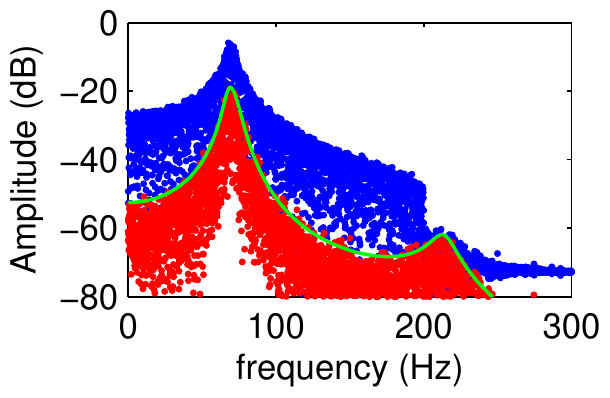}\caption{The output of the forced Duffing oscillator is simulated using an
estimated Box-Jenkins model (plant model order 2 poles and 2 zeros,
noise model order 6 poles and 6 zeros). The amplitude of the discrete
Fourier transform of the measured output and the simulation error
are shown. The blue dots are the measured output, the red dots are
the simulation error. The green line is the 95\% error level that
is calculated from the estimated noise model. \label{fig:Motivating Example: Simulation Error FD}}
\end{figure}

\begin{figure}
\begin{centering}
\includegraphics[scale=0.6]{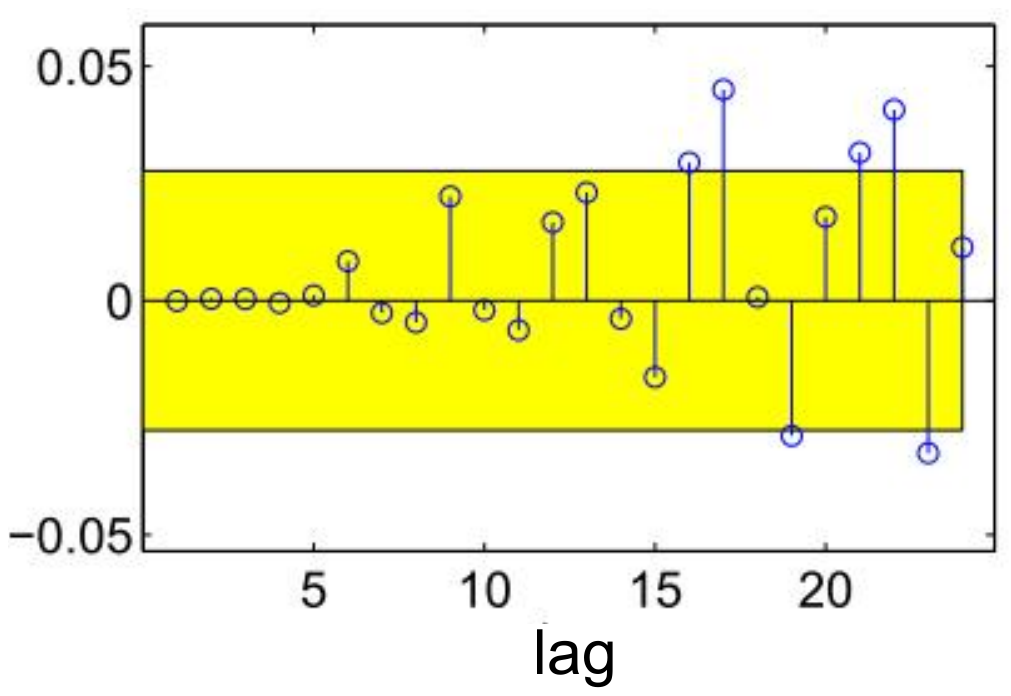}
\par\end{centering}
\begin{centering}
(a)
\par\end{centering}
\begin{centering}
\includegraphics[scale=0.6]{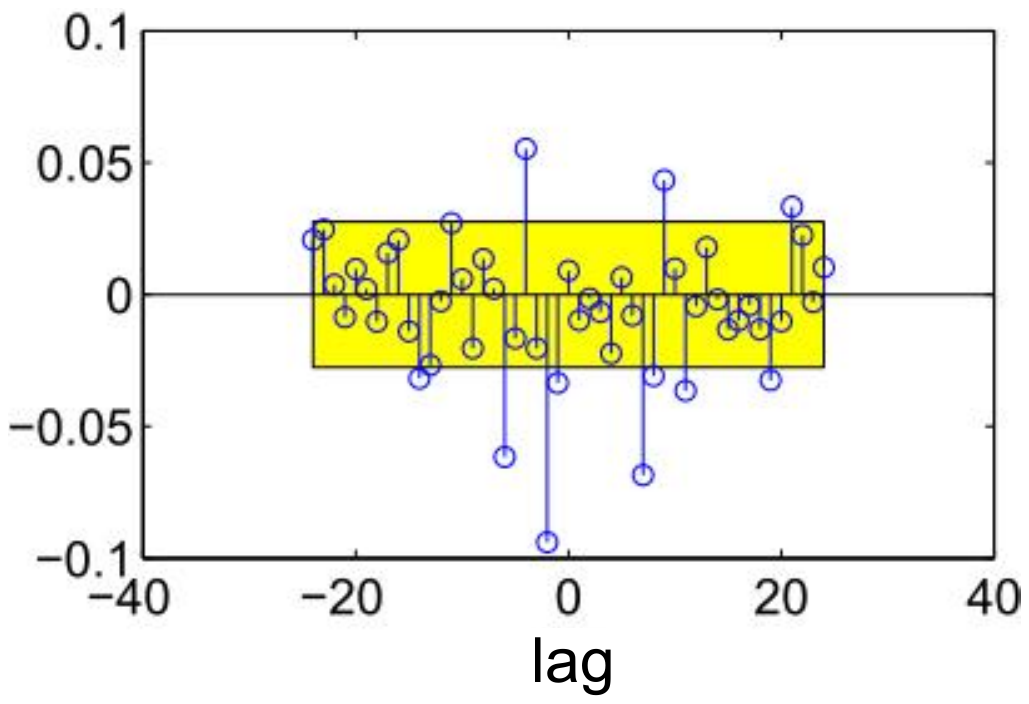}
\par\end{centering}
\centering{}(b)\caption{Analysis of the residuals of the Box-Jenkins fit for the last subexperiment.
(a): the auto-correlation of the output innovations (the residuals
that are whitened with the estimated noise model). A few points are
outside the 95\% interval. For a perfect fit, the innovations should
be white, 95\% of the blue dots should be in the yellow uncertainty
band. This points to a good, but not perfect model. This is also confirmed
by the cross-correlation between the input and the output innovations
shown in (b). Many points are outside the 95\% uncertainty interval.
Moreover, strong correlations for negative lags can be observed.
This points to a non-causal linear relation, it is possible to improve
the linear model by including also future input data. This noncausal
behavior is also discussed in \cite{Forssell and Ljung (2000) Projection method}.
\label{fig:Motivating Example: residue analysis}}
\end{figure}

Using this model, the output is 'simulated'. This is the identification
term to indicate that the ouput is calculated from the measured input.
The simulation error, which is the difference between this simulated
and measured output, is shown in Figure \ref{fig:Motivating example: Simulation Error TD}
(time domain) and Figure \ref{fig:Motivating Example: Simulation Error FD}
(frequency domain) for the last subrecord. The latter shows also the
95\% amplitude bound of the simulation error that is calculated from
the estimated sixth-order noise model. From these results it can be concluded
that a linear model gives still a reasonable approximation of the
output of the nonlinear system. Moreover, the power spectrum of the
errors is well captured by the noise model, even if the dominating
error is, in this case, the nonlinear distortions of the system.
That part of the nonlinear distortions that cannot be captured by
the linear model is added to the noise disturbances in the linear
identification framework. The whiteness test of the residuals in Figure
\ref{fig:Motivating Example: residue analysis} shows that the estimated
noise model describes well the power spectrum. But from the cross-correlation
test between the input and the residuals, it can be seen that there
are still some unexplained linear relations. Observe that the largest
spikes occur at negative lags, which indicates the need for noncausal
terms in the BLA \cite{Forssell and Ljung (2000) Projection method}.
Since the identification is not done under closed-loop conditions,
this behavior can only be due to the nonlinear nature of the system.

\subsection*{Validation of the linear model}

\begin{figure}
\centering{}\includegraphics{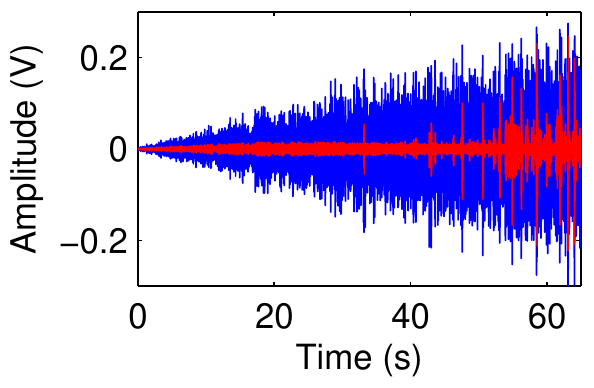}\caption{Validation of the model. The output of the system on a slowly growing
noise excitation is simulated. The model does well for small inputs,
but it fails for large inputs. The simulation error becomes very large
at the end of the amplitude sweep. Such a behavior points often to
nonlinear distortions.\label{fig:Motivating Example: validation test}}
\end{figure}

In a second step, the identified linear model is validated on the
arrow-like part of the data. This is a challenging test, because the
excitation level on part of the data exceeds that of the tail that
is used to estimate the linear model. From Figure \ref{fig:Motivating Example: validation test},
it is seen that the errors become very large once the excitation level
exceeds that of the tail part. The approximating linear model fails
completely under these conditions, because it cannot capture the underlying
nonlinear behavior of the system outside the domain where it was fitted
to the data. 

\subsection*{Analysis of the model uncertainty}

The estimation procedure resulted in the plant and noise model. From
this information it is possible to obtain also an estimate of the
uncertainty on the results. In Figure \ref{fig:Motivating Example: transfer function}
the estimated standard deviation of the transfer function is compared
with the sample standard deviation that is calculated from the repeated
estimates on the 10 subrecords. Both curves look very similar, but
the model-based estimated value (green) underestimates the actual
observed standard deviation (red) by 50\% or more. This is due to
the fact that the linear identification framework fails to estimate
precisely the uncertainty in the presence of nonlinear distortions.
The user should keep in mind that the confidence bounds are wrong, whenever
these are used during the design.

\subsection*{Conclusions}

The results from the motivational example show that even in the presence
of significant nonlinear distortions, it is still possible to obtain
a useful linear approximation with the classical linear identification
methodology. This model is only reliable under the conditions that
it is obtained. Changing the excitation, as was done in the validation
test, can lead to very large errors. Moreover, the uncertainty bounds
that are obtained from the linear identification framework are unreliable.
When the nonlinear distortions dominate the disturbing noise, significant
underestimation of the variances appears. This problem will be analyzed
in more detail later in this article in the section on the parametric
estimation of the BLA.

\subsection*{How to deal with nonlinear systems in system identification?}

From these observations, the reader could decide that in the presence
of nonlinear distortions it is better to build a complete nonlinear
model. But this choice is not without its own drawbacks. Nonlinear
identification is more involved and often more time consuming. This
leads to more experiments and longer development times. Moreover,
most engineers and designers are often very familiar with linear design
tools, but they are not trained in dealing with nonlinear systems.
In many cases, imperfect models with known error bounds are still
very useful to make a design that meets the requested specifications.
To follow this strategy, tools are needed to detect in an early phase
of the modeling process the presence of nonlinear distortions, and
to quantify their level. On the basis of this information, the design
engineer can decide wether a cheaper linear identification approach
can be made, or if the more expensive nonlinear identification framework
should be used. Using imperfect linear models is not a problem as
long as the user understands very well the validity of the linear
models, and knows what will be the impact of nonlinear distortions. The major goal of this article is to provide this background by discussing the three main topics that were formulated at the end of the 'introduction section': 1) Detection and characterization of nonlinear distortions; 2) Extending the linear framework to include the effect of nonlinear distortions; 3) Quantifying the potential gain by switching from a linear to a nonlinear identification framework.

\section*{Detection, qualification, and quantification of the nonlinear distortions\label{sec:Detection,-qualification,-and quantification of NL distortions}}

In this section tools will be presented that allow the user to detect
and analyze the presence of nonlinear distortions during the initial
tests. Without needing more experiments, the frequency response function
of the BLA, the power spectrum of the disturbing noise, and the level
of the nonlinear distortions will be obtained. All these results are
obtained from a nonparametric analysis, so that no user interaction
is needed. At the basis of the proposed solution is the use of well-designed
periodic excitations. The restriction to periodic signals is the price
to be paid to access all this information. The user can set the desired
frequency resolution and the desired power spectrum of the excitation
signal. The phase will be chosen randomly on $[0,2\pi)$. First,
the response of a nonlinear system to a periodic excitation is studied,
next it is explained how to design good periodic excitation signals.
Eventually, these signals will be used to make a nonparametric distortion
and disturbing noise analysis.

The nonlinear distortion analysis is initially made under open-loop
measurement conditions. The discussion of how to operate under closed-loop
conditions is postponed, because to do so the concept of 'BLA', which will be introduced later in this article, is needed

\setcounter{subsection}{0}

\subsection*{The response of a nonlinear system to a periodic excitation\label{subsec:Response NL system to periodic signal}}

A linear time-invariant system cannot transfer power from one frequency
to another. In contrast, a nonlinear system can transfer power from one frequency to another.
Understanding this power transfer mechanism is an essential tool
in the detection and analysis of nonlinear distortions\cite{Chua and Ng (1979)}.
Consider a cosine signal passed through a cubic static nonlinear system
$y=u^{3}$

\[
u(t)=2cos\omega t=e^{j\omega t}+e^{-j\omega t}.
\]

\begin{figure}
\begin{centering}
\includegraphics[scale=0.5]{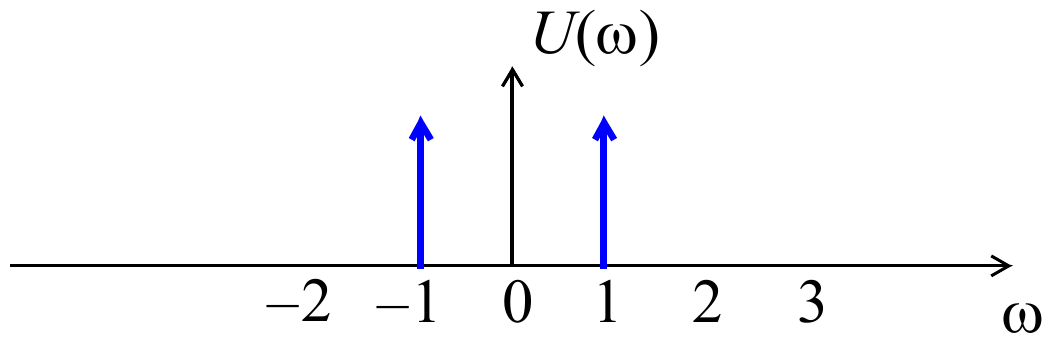}
\par\end{centering}
\begin{centering}
(a)
\par\end{centering}
\begin{centering}
\includegraphics[scale=0.5]{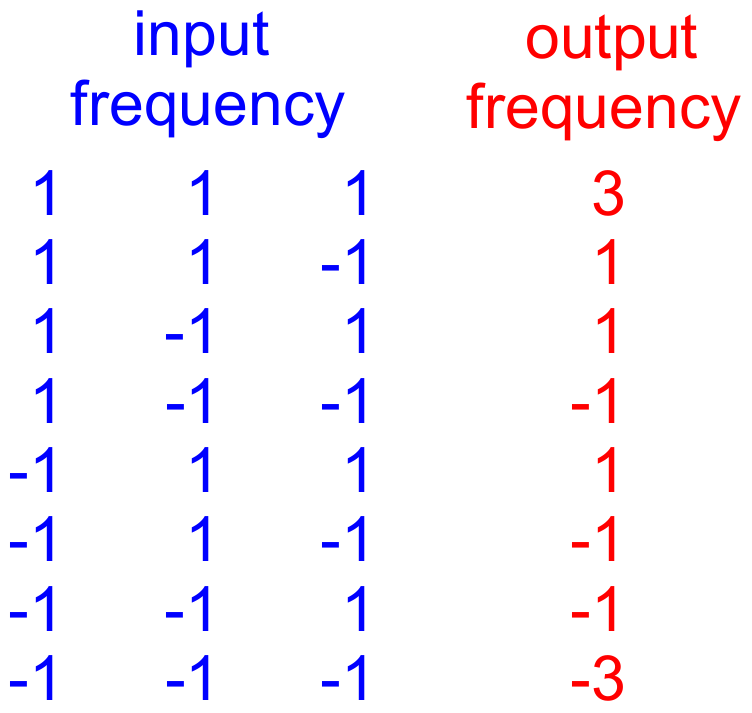}
\par\end{centering}
\begin{centering}
(b)
\par\end{centering}
\begin{centering}
\includegraphics[scale=0.5]{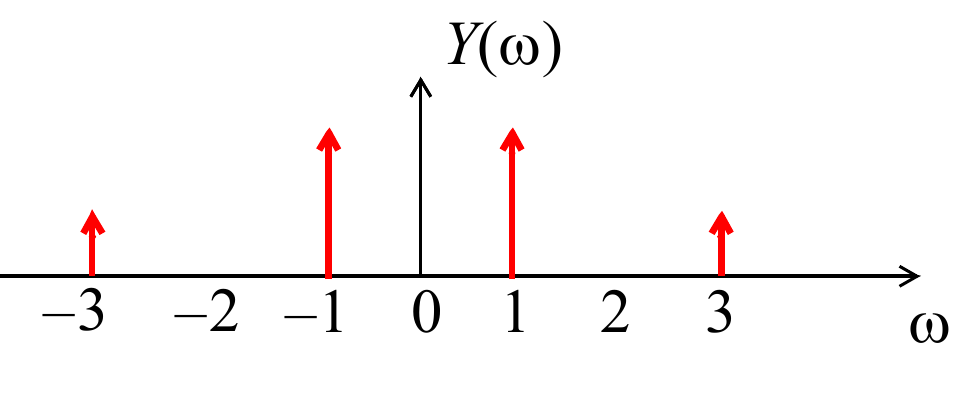}
\par\end{centering}
\centering{}(c)\caption{Spectrum of a sinosoid (a) that is passed through a cubic nonlinearity
$y=u^{3}$. The frequencies of the output spectrum (c) are obtained
by making the sum over each of all possible combinations of 3 frequencies
selected from the input frequencies (b), for $\omega=1$. Keep in
mind that both the positive and the negative frequencies should be
considered.\label{fig:NonLinear System: cos}}
\end{figure}

Remark that the signal has a positive and negative frequency, as shown
in Figure \ref{fig:NonLinear System: cos}. The steady-state output
of the nonlinear system is

\[
y(t)=u(t)^{3}=(e^{j\omega t}+e^{-j\omega t})(e^{j\omega t}+e^{-j\omega t})(e^{j\omega t}+e^{-j\omega t}).
\]

In the calculation of this product, terms of the form $e^{\pm j\omega t}e^{\pm j\omega t}e^{\pm j\omega t}=e^{j(\pm\omega\pm\omega\pm\omega)t}$
appear (Figure \ref{fig:NonLinear System: cos}(b) for $\omega=1$),
resulting eventually in the frequency components $-3,-1,1,3,$ as
shown in Figure \ref{fig:NonLinear System: cos}(c). 

This result can be generalized. Consider a nonlinear system $y=u^{\alpha}$,
excited at the frequencies $\pm\omega_{k},k=1,\ldots,F.$ The frequencies
at the output of such a system are given by making all possible combinations
of $\alpha$ frequencies, including repeated frequencies, selected
from the set of $2F$ excited frequencies 
\begin{equation}
\sum_{i=1}^{\alpha}\omega_{k_{i}},\textrm{ with}\;\omega_{k_{i}}\in\{-\omega_{F},\ldots,-\omega_{1},\omega_{1},\ldots,\omega_{F}\}.\label{eq:NLS output frequencies}
\end{equation}
Every static nonlinearity $y=f(u)$ can be approximated arbitrarily
well in least-squares sense, under some regularity conditions, by
a polynomial $y_{P}=\sum_{\alpha=1}^{P}a_{\alpha}u^{\alpha}$ 
\[
lim_{P\rightarrow\infty}E_{u}\{|y-y_{P}|^{2}\}=0,
\]
for some specified classes of inputs (see later in this article).
On each of the monomial terms $a_{\alpha}u^{\alpha}$ in the sum,
the previous analysis can be applied, and hence it is very simple
to know all the frequencies that can appear at the output of a static
nonlinear system. 

The result can be further generalized to dynamic nonlinear systems,
using Volterra series \cite{Schetzen 2006}. A formal development
is given in \cite{Pintelon 2012 book}, pages 74-75, and illustrated
in a set of Matlab exercises in \cite{Schoukens 2012  Exercises book}.
Under some regularity conditions, Volterra series can approximate
arbitrarily well fading-memory systems and discontinuous nonlinear
systems \cite{Boyd Fading memory paper 1985}. Because a periodic
input results in a periodic output with the same period, it is clear
that a Volterra series cannot be used to describe a chaotic system
because chaotic systems have no periodic output for a periodic input.

\subsection*{Design of a multisine for nonlinear detection and frequency response function
measurements\label{subsec:Multisine Design}}

The choice of the excitation signal is extremely important in a nonlinear
framework. The behavior of a nonlinear system depends on both the power spectrum and the amplitude distribution of the applied excitation signal
\cite{Pintelon 2012 book}, as shown in Figure \ref{fig:Excitation-signals-impact}.
In this article, signals with a Gaussian amplitude distribution will be used.\foreignlanguage{english}{}
\begin{figure}
\selectlanguage{english}%
\begin{centering}
\includegraphics[scale=0.45]{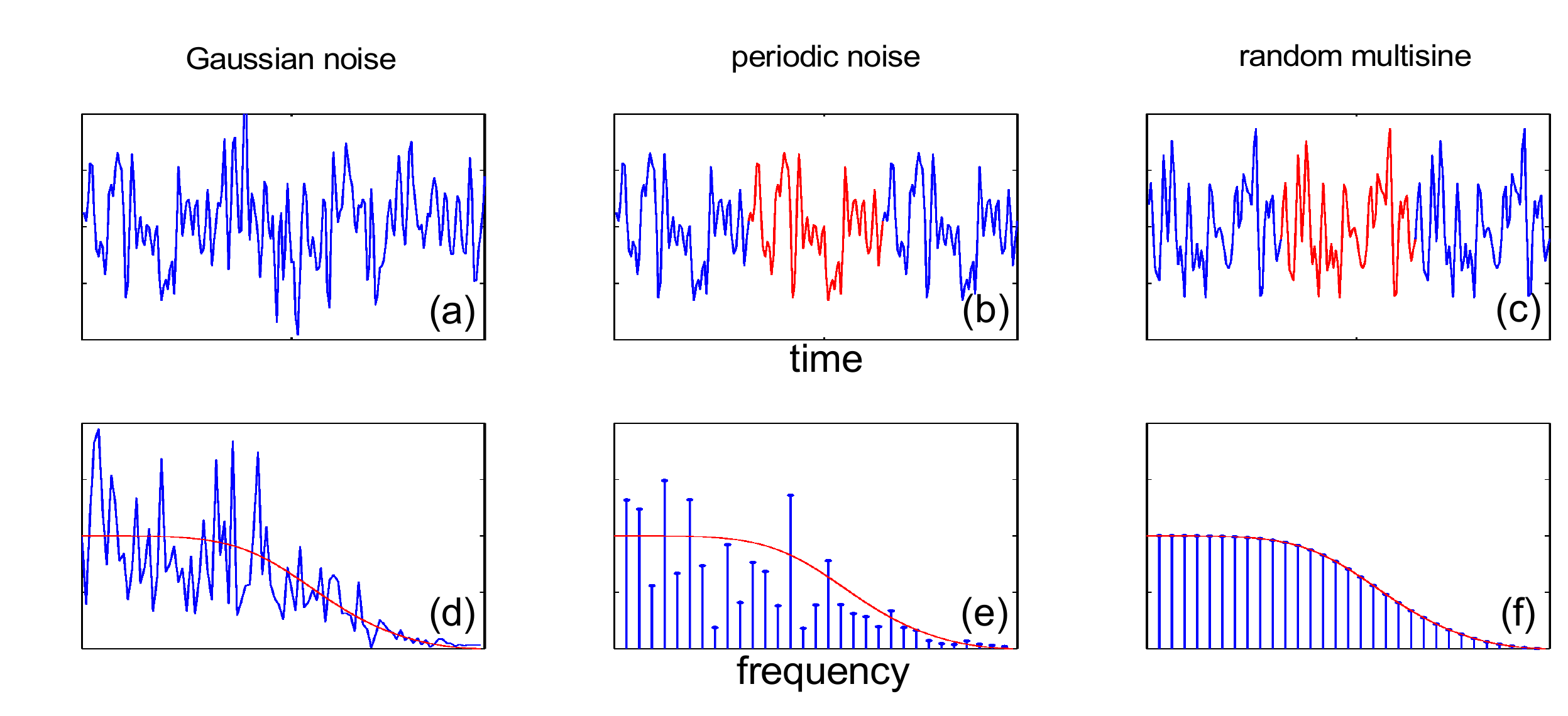}
\par\end{centering}
\caption{\foreignlanguage{british}{Examples of excitation signals in time and frequency domain: Gaussian noise (a,d), periodically repeated Gaussian noise (b,e), random-phase multisine (c,f). In the frequency domain, the amplitude spectrum of the actual realization (blue) and the power spectrum (red) is shown.\label{fig:Examples-of-excitions}}}
\selectlanguage{british}%
\end{figure}

Gaussian random noise excitations (Figure \ref{fig:Examples-of-excitions}(a,d))
are very popular among practicing engineers, because they seem to
be simple to design. However, in this article periodic excitations
are used because these signals offer significant advantages for making
a nonparametric nonlinear distortion analysis. It will be shown later
in the article, in the section on Riemann-equivalent excitation signals,
that the results obtained with the periodic excitation can also be
transferred to random Gaussian noise excitations after proper normalization
\cite{Schoukens RiemanEquivalence}. 

One possibility to generate a periodic signal is to periodically
repeat a finite segment of a random noise sequence (Figure \ref{fig:Examples-of-excitions}(b,e)).
However, using a random-phase multisine (Figure \ref{fig:Examples-of-excitions}(c,f))
\cite{Pintelon 2012 book,Schoukens 2012  Exercises book,Schoukens RiemanEquivalence}
gives a much better control over the amplitude spectrum of the excitation, resulting in lower uncertainties on the measured frequency response function (FRF).
Consider the signal 
\begin{eqnarray}
u_{0}\left(t\right) & = & \frac{1}{\sqrt{N}}\sum_{k=-N/2+1}^{N/2-1}U_{k}e^{j\left(2\pi kf_{0}t+\varphi_{k}\right)},\label{eq:random phase multisine}\\
 & = & \frac{2}{\sqrt{N}}\sum_{k=-N/2+1}^{N/2-1}U_{k}cos\left(2\pi kf_{0}t+\varphi_{k}\right),
\end{eqnarray}
where $\varphi_{-k}=-\varphi_{k}$ and $U_{-k}=U_{k}$, $U_{0}=0$,
and $f_{0}=f_{s}/N=1/T$ . The sample frequency to generate the signal
is $f_{s}$, and $T$ is the period of the multisine. The phases $\varphi_{k}$
will be selected independently such that $E\{e^{j\varphi_{k}}\}=0$,
for example by selecting a uniform distribution on the interval $[0,2\pi)$.
The amplitudes $U_{k}$ are chosen to follow the desired amplitude
spectrum (Figure \ref{fig:Examples-of-excitions}(f)). See \cite{Schoukens RiemanEquivalence}
for a detailed discussion about the user choices and the properties
of these signals. The major advantage of the random-phase multisine
is that it still has (asymptotically for sufficiently large $N$)
all the nice properties of Gaussian noise, while it also has the advantages
of a deterministic signal: the amplitude spectrum does not show dips
at the excited frequencies (see Figure \ref{fig:Examples-of-excitions}f))
as the two other signals do (see Figure \ref{fig:Examples-of-excitions}(d) and (e)).
At those dips, the measurements are very sensitive to all nonlinear
distortions and disturbing noise. 

\emph{Remark}

Initially, multisine excitations were introduced for the frequency
response function (FRF) measurement of linear dynamic systems \cite{Schoukens (2002) Survey excitations signals}.
To maximize the signal-to-noise ratio (SNR) of the measurements, an
intensive search for compact signals was made. For a given amplitude
spectrum, the phases were chosen such that the peak value of the signal
is minimized \cite{Guillaume (2002) Crest-factor minimization}. Alternatively,
well-designed binary signals could be used \cite{Tan and Godfrey (2002) Binary excitations overview}.
Although these compact signals are superior for linear measurements,
they are not so well suited to measure the FRF in the presence of
nonlinear distortions. It will be explained in this article (see Figure
\ref{fig:Excitation-signals-impact}), that the linearized measurements
depend strongly on the amplitude distribution of the excitation. The
specially designed multisines with a minimized peak factor have an
amplitude distribution that is close to that of a sine excitation
(a high probability to be close to the extreme values, a low probability
to be around zero, as shown in Figure \ref{fig:Excitation-signals-impact}).
Random-phase multisines are asymptotically (with growing number of
frequencies) Gaussian distributed, which is often preferred in applications.
Moreover, it will be possible to make explicit statements on the properties
of the linear approximation and the remaining errors for the latter
case. For that reason, the focus will be from here on random-phase
multisines and random Gaussian excitations. More information on the
impact of the amplitude distribution on the linear approximation can
be found in \cite{Wong Hin Kwan (2012) IEEE I=000026M paper 1  distribution,Wong Hin Kwan (2013) IEEE I=000026M paper 2  PRBS}.

\emph{User guidelines}:
\begin{itemize}
\item Use random-phase multisine excitations.
\item The spectral resolution $f_{0}$ of the multisine should be chosen
high enough so that no sharp resonances are missed \cite{Geerardyn Egon 2013 Multisine design}.
Since $f_{0}=1/T$, it sets immediately the period length $T$ of
the multisine. A high-frequency resolution requires a long measurement
time because at least one, and preferably a few, periods should be
measured.
\item The amplitude spectrum should be chosen such that the frequency band
of interest is covered. The signal amplitude should be scaled such
that it also covers the input amplitude range of interest.
\end{itemize}
In the next section, it will be shown that nonlinear distortions can
be easily detected by putting some amplitudes $U_{k}$ in \eqref{eq:random phase multisine}
equal to zero for a well-selected set of frequencies.

A detailed step-by-step procedure of how to generate and process
periodic excitations is given in Chapter 2 of \cite{Schoukens 2012  Exercises book}.

\subsection*{Riemann-equivalent excitation signals\label{subsec:Riemann-equivalent-excitation}}

The goal is to characterize a nonlinear system for Gaussian excitation
signals, using random-phase multisines. The design of the amplitude
spectrum of the multisine should be such that the equivalence between
the random-phase multisine and the Gaussian random noise with respect
to the nonlinear behavior is guaranteed. To do so, the equivalence
class $E_{S_{U}}$ is defined that collects all signals that are (asymptotically)
Gaussian distributed, and have asymptotically, for $N\rightarrow\infty$,
the same power on each finite frequency interval. This is defined
precisely in the next definition.
\begin{defn}
\emph{Riemann-equivalence class $E_{S_{U}}$ of excitation signals}.
Consider a power spectrum $S_{U}(\Omega)$ that is piecewise continuous,
with a finite number of discontinuities. A random signal belongs to
the equivalence class if:
\end{defn}
\begin{enumerate}
\item It is a Gaussian noise excitation with power spectrum $S_{U}(\Omega),$
\item It is a random multisine or random-phase multisine such that
\[
\frac{\mathrm{1}}{N}\sum_{k=k_{\omega_{\mathrm{1}}}}^{k_{\omega_{\mathrm{2}}}}\{E(\left|U(k)\right|^{\mathrm{2}})\}=\frac{\mathrm{1}}{\mathrm{2}\pi}\int_{\omega_{\mathrm{1}}}^{\omega_{\mathrm{2}}}S_{U}(\nu)d\nu+O(N^{-\mathrm{1}})\quad\textrm{for all }k_{\omega_{i}},
\]
\end{enumerate}
with
\[
k_{\omega_{i}}=\mathrm{floor}(\frac{\omega_{i}}{\mathrm{2}\pi f_{s}}N)\mathrm{\;and\;}\mathrm{0}<\omega_{\mathrm{1}},\omega_{\mathrm{2}}<\pi f_{s}.
\]

Using the Riemann equivalence, it is possible to use periodic random-phase
multisines to characterize the properties of the nonlinear system
excited with filtered Gaussian noise. This will be explained in the
next section.

\subsection*{Detection, separation, and characterization of the nonlinear distortions
and the disturbing noise\label{subsec:Detection, Separation NLD section}}

This article presents only the basic principles of the nonlinear
distortion analysis; see \cite{Pintelon 2012 book} for a more detailed discussion.

\subsubsection*{Detection and characterization of the nonlinear distortions}

\begin{figure}
\begin{centering}
\includegraphics[scale=0.4]{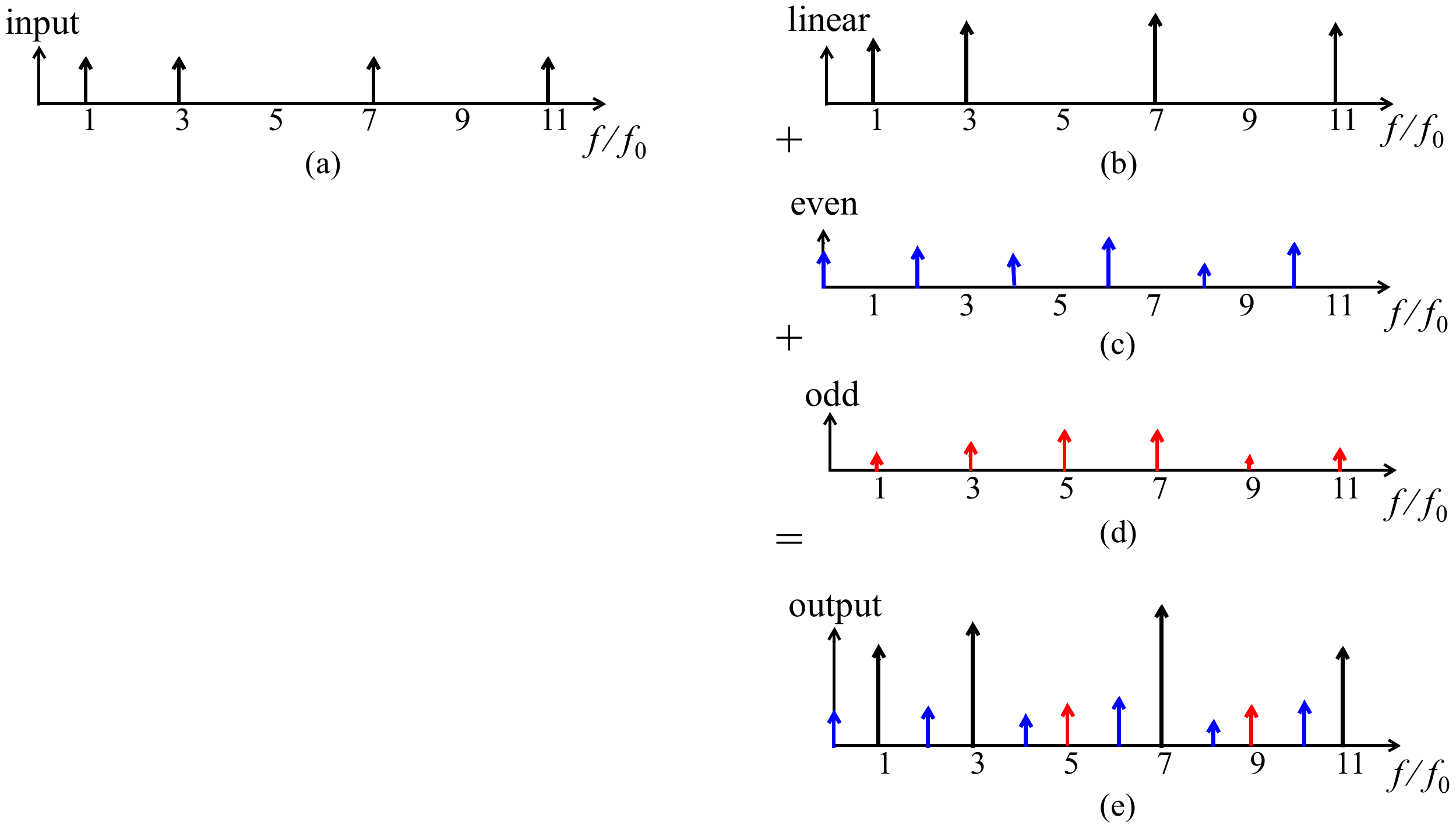}
\par\end{centering}
\caption{\foreignlanguage{british}{Design of a multisine excitation for a nonlinear analysis. (a): Selection of the excited frequencies at the input (left side); At the output (right side), from top to bottom: linear (b), even (c), odd (d) contributions, and total output (e).
\label{fig:Multisine NonLin detection}}}
\end{figure}

The basic idea illustrated in Figure \ref{fig:Multisine NonLin detection}
is very simple and starts from (see Figure \ref{fig:Multisine NonLin detection})
a multisine \eqref{eq:random phase multisine} that excites a well-selected
set of odd frequencies (odd frequencies correspond to odd values of
$k$ in \eqref{eq:random phase multisine}). This excitation signal
is applied to the nonlinear system under test. Even nonlinearities
show up at the even frequencies because an even number of odd frequencies
is added together. Odd nonlinearities are present only at the odd
frequencies because an odd number of odd frequencies is added together.
At the odd frequencies that are not excited at the input, the odd
nonlinear distortions become visible at the output because the linear
part of the model does not contribute to the output at these frequencies
(for example, frequencies 5 and 9 in Figure \ref{fig:Multisine NonLin detection}).
By using a different color for each of these contributions, it becomes
easy to recognize these in an amplitude spectrum plot of the output
signal. 

\selectlanguage{english}%

\selectlanguage{british}%

\subsubsection*{Disturbing noise characterization}

\begin{figure}
\centering{}\includegraphics[scale=0.55]{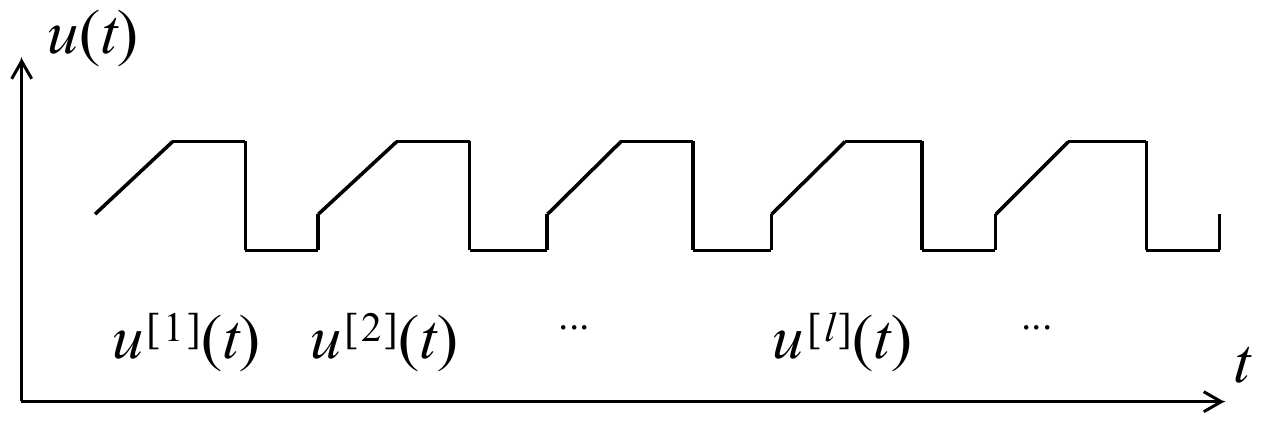}\caption{Calculation of the sample mean and variance of a periodic signal.\label{fig:Periodic Signal Variance Mean}}
\end{figure}

In the next step, the disturbing noise analysis is made. By analyzing
the variations of the periodic input and output signals over the measurements
of the repeated periods, the sample mean and the sample (co-)variance
of the input and the output disturbing noise can be calculated, as
a function of the frequency. Although the disturbing noise varies
from one period to the other, the nonlinear distortions do not, so
they remain exactly the same. This results eventually in the following
simple procedure: consider the periodic signal $u(t)$ in Figure
\ref{fig:Periodic Signal Variance Mean}. The periodic signal is measured
over $P$ periods. For each subrecord, corresponding to a period,
the discrete Fourier transform is calculated using the fast Fourier
transform (FFT) algorithm, resulting in the FFT spectra of each period
$U^{[l]}(k),Y^{[l]}(k),$ for $l=1,\ldots,P$. Because an integer
number of periods is measured, there will be no leakage in the results.
The sample means $\hat{U}(k),\hat{Y}(k)$ and noise (co)variances
$\hat{\sigma}_{U}^{2}(k),\hat{\sigma}_{Y}^{2}(k),\hat{\sigma}_{YU}^{2}(k)$
at frequency $k$ are then given by
\[
\hat{U}(k)=\frac{1}{P}\sum_{l=1}^{P}U^{[l]}(k)\quad\hat{Y}(k)=\frac{1}{P}\sum_{l=1}^{P}Y^{[l]}(k),
\]

and 

\begin{equation}
\begin{array}{c}
\hat{\sigma}_{U}^{2}(k)=\frac{1}{P-1}\sum_{l=1}^{P}|U^{[l]}(k)-\hat{U}(k)|^{2},\\
\hat{\sigma}_{Y}^{2}(k)=\frac{1}{P-1}\sum_{l=1}^{P}|Y^{[l]}(k)-\hat{Y}(k)(k)|^{2},\\
\hat{\sigma}_{YU}^{2}(k)=\frac{1}{P-1}\sum_{l=1}^{P}(Y(k)-\hat{Y}(k))(U(k)-\hat{U}(k))^{H}.
\end{array}\label{eq:Sample variances}
\end{equation}
In \eqref{eq:Sample variances}, $(.)^{H}$ denotes the complex conjugate.
The variance of the estimated mean values $\hat{U}(k)$ and $\hat{Y}(k)$
is $\hat{\sigma}_{U}^{2}(k)/P$ and $\hat{\sigma}_{Y}^{2}(k)/P,$
respectively. Adding together all this information in one figure
results in a full nonparametric analysis of the system with information
about the system (the FRF), the even and odd nonlinear distortions,
and the power spectrum of the disturbing noise. Note that no interaction
with the user is needed during the processing. This makes the method
well suited to be implemented in standard measurement procedures.

\subsubsection*{Combining multiple realizations of the random input}

This measurement can be repeated over $M$ realizations of the random-phase
multisine by generating each time a new multisine excitation with
another random-phase realization. The results can then be averaged
over these realizations to obtain more reliable estimates of the distortion
and noise levels. At the same time, the standard deviation of the
FRF, due to the nonlinear distortions and the disturbing noise, will
be reduced by $\sqrt{M}$.

In \cite{Pintelon (2011) MSSP SISO LPM en NL analysis,Pintelon (2011) MSSP  MIMO LPM and NL analysis},
a detailed analysis is given of how these ideas can be generalized
to deal with initial transient effects in SISO and MIMO FRF measurements.

\subsection*{User guidelines}
\begin{itemize}
\item Design a random multisine excitation following the guidelines specified
earlier in this article.
\item Excite the system with the multisine and measure $P\geq2$ periods
of the steady-state response.
\item Repeat this procedure for $M$ successive realizations of the random-phase
multisine.
\item Choose $P,M$ such that within the available measurement time the
number of repetitions $M$ is as large as possible. This advice can
be refined, depending on the prior knowledge of the user:

\begin{itemize}
\item No prior knowledge available: select $P=2$, and $M$ as large as
possible.
\item Maximize the nonlinear detection ability: $M=2$, and $P$ as large
as possible.
\item If it is known that the nonlinear distortions dominate: $P=1$, and
$M$ as large as possible (the disturbing noise level will not be
estimated in this case).
\end{itemize}
\end{itemize}

\section*{Characterizing nonlinear distortions: experimental illustrations\label{sec:Nonlinear distortions: Experiments}}

In this section, a series of experimental illustrations are presented.
The first example is the forced Duffing oscillator that was already
used in the motivating example. The second and third examples are
industrial applications (air path characterization of a diesel engine,
and a ground vibration test of an F-16 fighter). 

\setcounter{subsection}{0}

\subsubsection*{Characterization of a forced Duffing oscillator\label{subsec:Nonlinear distortions: Experiments Duffing oscilator}}

The nonlinear analysis method is experimentally illustrated on the
electronic circuit (see Figure \ref{fig:Loudspeaker Setup}, top)
\cite{Pintelon 2012 book,Paduart NLSS 2010,Anna vgl NN SVM P}. Although
this is a nonlinear feedback system, it behaves as a fading-memory
system \cite{Boyd Fading memory paper 1985} for sufficiently small
input amplitudes, and hence the proposed method can be applied. 

The following settings were used to make the measurements: sample
frequency is about 1220 Hz, the period length is 4096 samples, the
frequency resolution is $f_{0}\approx0.30$ Hz, and the maximum excited
frequency is 200 Hz. Only the odd frequencies are excited, and in
each block of 5 consecutive odd frequencies, the amplitude of one
randomly selected frequency is put to zero so that it can be used
as a nonlinear detection line. All the excited frequencies have the
same amplitude. For each realization, 3 periods of the output were
measured. The first period is dropped to avoid initial transient effects. 

\begin{figure}
\centering{}\includegraphics[scale=0.6]{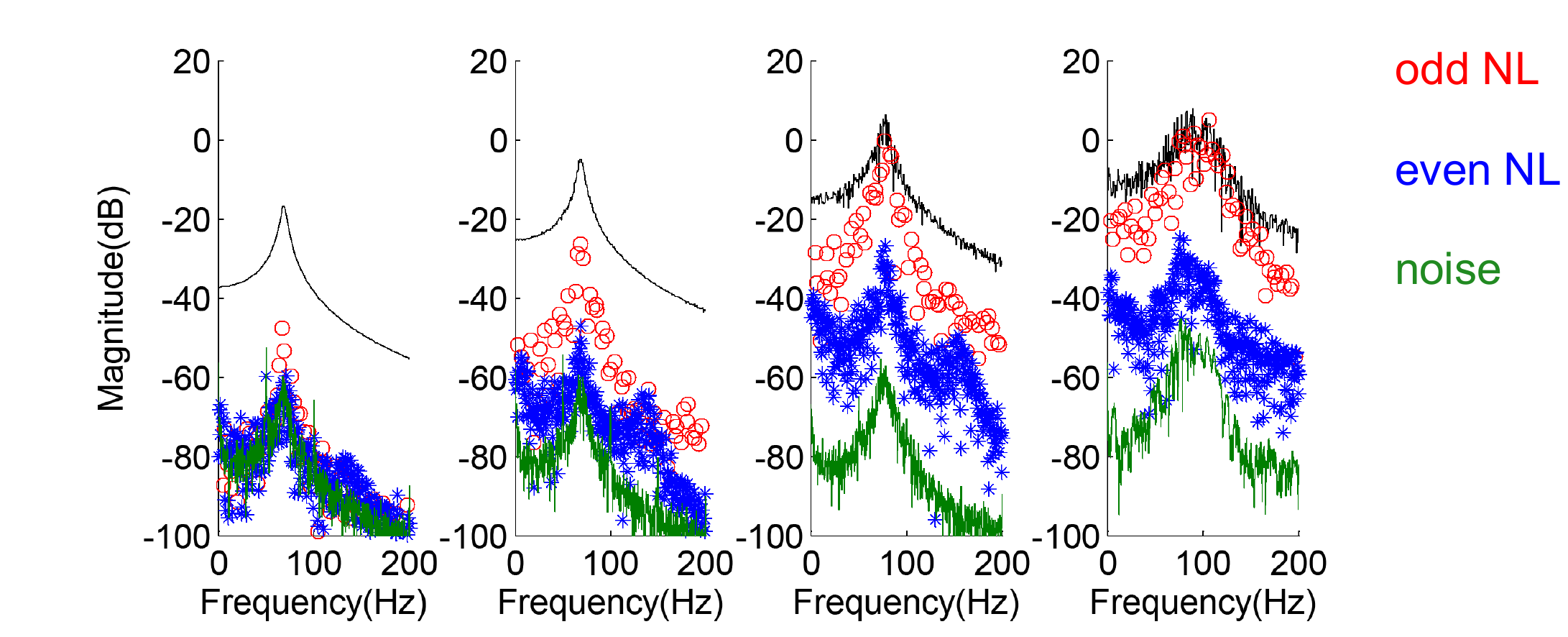}\caption{Nonparametric analysis of the nonlinear distortions on a forced Duffing
oscillator. The system is excited at a well-selected set of frequencies
as explained in the section: Design of a multisine for nonlinear detection and frequency response function measurements.
The nonlinearities become visible at the unexcited frequencies. Black
dots: output at the excited frequencies; Red bullets: odd nonlinearities;
Blue stars: even nonlinearities; Green line: disturbing noise level.
The excitation level is growing from the left to the right figure.
Observe that the level of the nonlinear distortions grows with the
excitation level, the disturbing noise level remains almost constant.\label{fig: Forced Duffing oscilator NonLinDist}}
\end{figure}

In Figure \ref{fig: Forced Duffing oscilator NonLinDist}, the evolution
of the nonlinear distortions as a function of the frequency is shown
for different excitation levels. These distortion levels can be compared
to the output at the excited frequencies to obtain an idea about the
relative distortion levels and the SNR. It can be seen that, for a
low excitation level, the presence of odd nonlinear distortions is
detected around the resonance frequency. When the excitation level
grows, the odd nonlinear distortions grow faster than the even ones,
while the observed disturbing noise level remains almost the same.
This figure is very informative for the designer. For small excitation
levels (left side of the figure), the nonlinear distortions are 30
dB below the linear contributions. In that case, a linear model can
be used if a moderate precision is sufficient. For higher excitation
levels (right side of the figure), it is clear that the nonlinear
distortions can no longer be neglected since the nonlinear distortions
are as large as the linear contributions. In that case, a full nonlinear
model will be needed. All this information is directly available from
a simple nonparametric nonlinear analysis that requests no user interaction.
It can be easily implemented in a measurement instrument. The figure
shows also the level of the disturbing noise. This level remains constant
in the first three experiments, but grows significantly in the last
one. This might be due to a nonlinear mixing of the process noise
and the signals in the loop, leading to signal-dependant noise levels.

\subsection*{Characterization of the air path of a diesel engine}

\begin{figure}
\centering{}\includegraphics[scale=0.6]{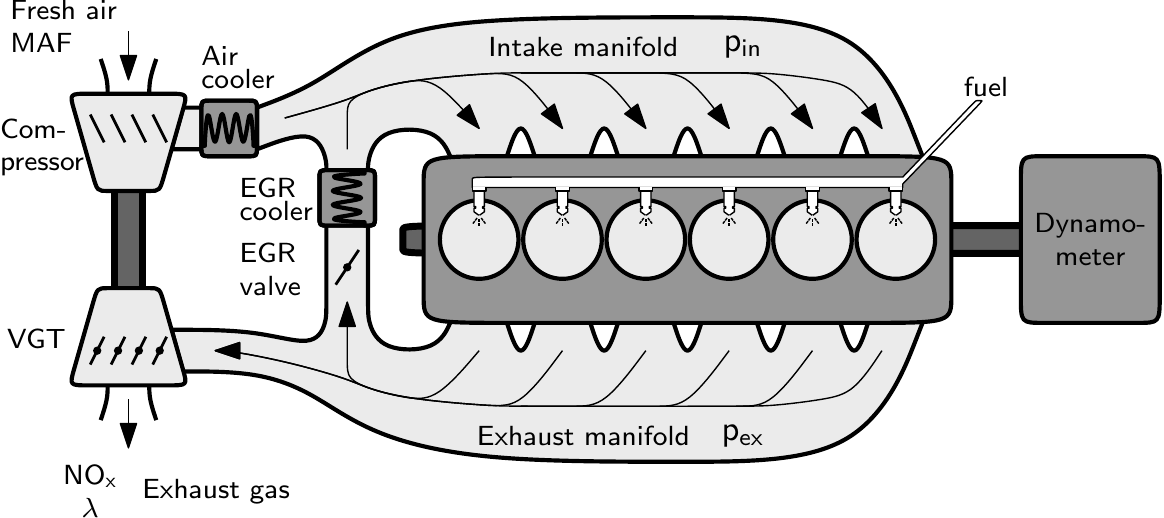}\caption{Nonparametric measurement of the frequency response function and
the nonlinear distortions analysis of the air path of a turbocharged
diesel engine around a fixed operating point\cite{Criens (2014) PhD,Criens (2010) Diesel paper,Criens et al. (2015) IJPT journal InPress}.
\label{fig:Diesel Engine}}
\end{figure}

The results and figures in this section are taken from \cite{Criens (2014) PhD}.
The goal of the thesis was the design of a control system for heavy-duty
diesel engines that is capable of combining a low fuel consumption
with low emissions of nitrogen oxides (NOx) and particulate matter
(PM) \cite{Criens (2010) Diesel paper,Criens et al. (2015) IJPT journal InPress}.
In addition, these properties should be maintained when disturbances
are present. The control design for the diesel engine air path is
considered. A feature of the control system to be designed was that
the required design effort is low. Air path control is particularly
interesting. It is challenging and time consuming to calibrate using
current control methods. Moreover, the air path contains a variety
of sensors and actuators, which means that within the current hardware
constraints, several control layouts are possible. The main actuators
in the air path are the variable geometry turbine (VGT) and exhaust gas recirculation
(EGR) valve (see Figure \ref{fig:Diesel Engine}). The VGT and EGR
valve are used as inputs. The $NO_{x}$ emissions, air-fuel equivalence
ratio $\lambda$, and pressure difference $\Delta P$ between the
intake and exhaust manifold are the considered outputs. The nonlinear
engine behavior is reduced to a series of linear submodels, where
each submodel describes the engine behavior in a part of the speed-load
operating range. Since the controllers are designed using the local
linear properties, they do not make full use of the actual nonlinear
system description. A nonlinear distortion analysis provides the necessary
information to verify the validity of this approach.

\begin{figure}
\centering{}\includegraphics[scale=0.55]{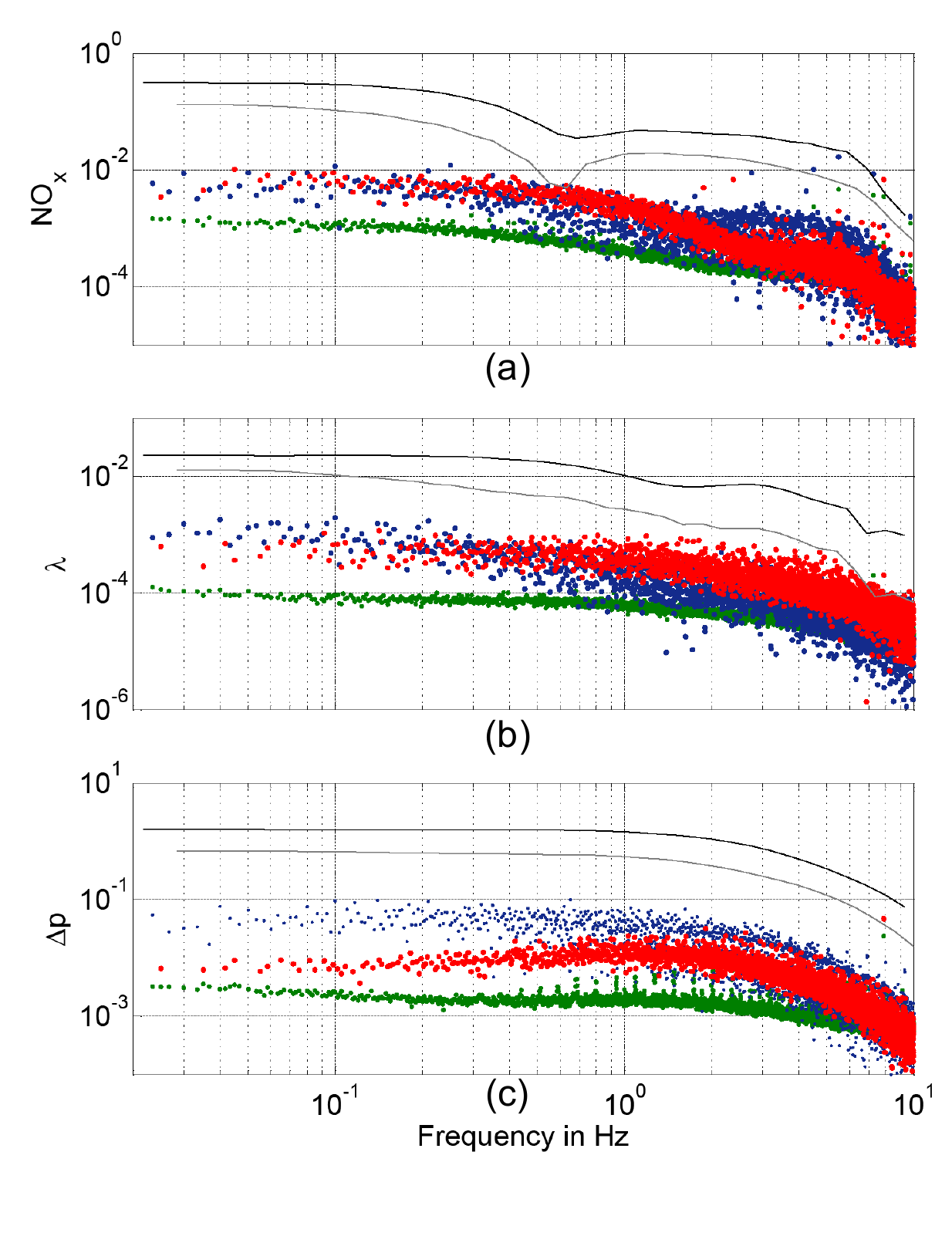}\caption{Measurement of the frequency response function and the nonlinear
distortion levels of the diesel engine for the operating point 1455
RPM and 120mg/injection \cite{Criens (2010) Diesel paper,Criens (2014) PhD,Criens et al. (2015) IJPT journal InPress}.
The figures show respectively the $NO_{x}$ emissions (a), the engine-out
air-fuel equivalence ratio $\lambda$ (b), and the pressure difference
between intake and exhaust manifold $\triangle p$ (c). The black
and grey lines show the FRF from the VGT and the EGR, respectively.
The blue, red, and green dots show, respectively, the odd and even
nonlinearities, and the disturbing noise level. From these figures,
it can be concluded that the nonlinear distortions are well above
the disturbing noise level. It can be seen that the nonlinear distortions
are about a factor 10 below the linear contributions for the actual
settings. The $\triangle p$ signal is  dominated by the even nonlinearities.
\label{fig:Diesel Engine results}}
\end{figure}

The FRF and the nonlinear distortion levels of the diesel engine are
measured at the operating point 1455 RPM and 120mg/injection. To separate
both transfer functions (respectively from the VGT and the EGR inputs),
two zippered multisines are created \cite{Rivera Daniel (2009) Constrained Multisine  plant-friendly identification,Verbeeck (1999) Identification Synchronous Machines}
so that nonoverlapping frequency grids are used for both inputs.
This allows the two FRF's to be measured in a single experiment. The
inputs were normalized on the maximum of the corresponding actuator
position, and the amplitudes were set to 0.7\%. The results are given
in Figure \ref{fig:Diesel Engine results} (a),(b),(c) showing, respectively,
the $NO_{x}$ emissions, the engine-out air-fuel equivalence ratio,
and the pressure difference between intake and exhaust manifold $y_{\triangle p}$.
The nonlinear distortions were scaled with the normalized input levels,
so that they can be plotted in the FRF figures \cite{Criens (2014) PhD}.
From Figure \ref{fig:Diesel Engine results}, it can be concluded
that the nonlinear distortions are well above the disturbing noise
level. The level of the nonlinearities is about a factor 10 below
the linear contributions for the actual settings. There is no clear
dominance of the even or odd nonlinearities in the first two figures,
however for the $\triangle p$ signal, the even nonlinearities dominate.
On the basis of these results, the control engineer can decide that
a linear control design can still be used, keeping in mind that model
errors up to 10\% are present. The nonlinear distortion levels are
useful to set uncertainty bounds on the FRF that can be used in robust
control design.

\subsection*{Ground vibration test on an air fighter\label{subsec:Nonlinear distortions: Experiments Ground vibration tests}}

Ground vibration testing (GVT) is an essential step in the development
of a new aircraft. Also, after each structural modification, new
GVT should be done. From these tests, the dynamical characteristics
of the air plane are obtained. These are necessary to update the finite
element models that are used, for example, during a flutter analysis.
These tests should be conducted in a very short time period because
the test is made in the critical path of the development program.
An introduction to the state-of-the-art of GVT can be found in \cite{Govers (2014) ISMA ground vibration tests},
which states that the major goal of the GVT is to measure the eigenfrequencies,
the mode shapes, the generalized mass and damping matrices, and FRFs.
Also, the structural nonlinear behavior must be studied. Measurement
and excitation strategies are developed to minimize the required total
measurement time (for example 9 days to test the Airbus A350XWB).
The excitation signals should meet level constraints, and also the hardware limitations
should be respected. At the same time, good SNRs should be obtained. 

\begin{figure}
\begin{centering}
\par\end{centering}
\begin{centering}
\includegraphics[scale=0.4]{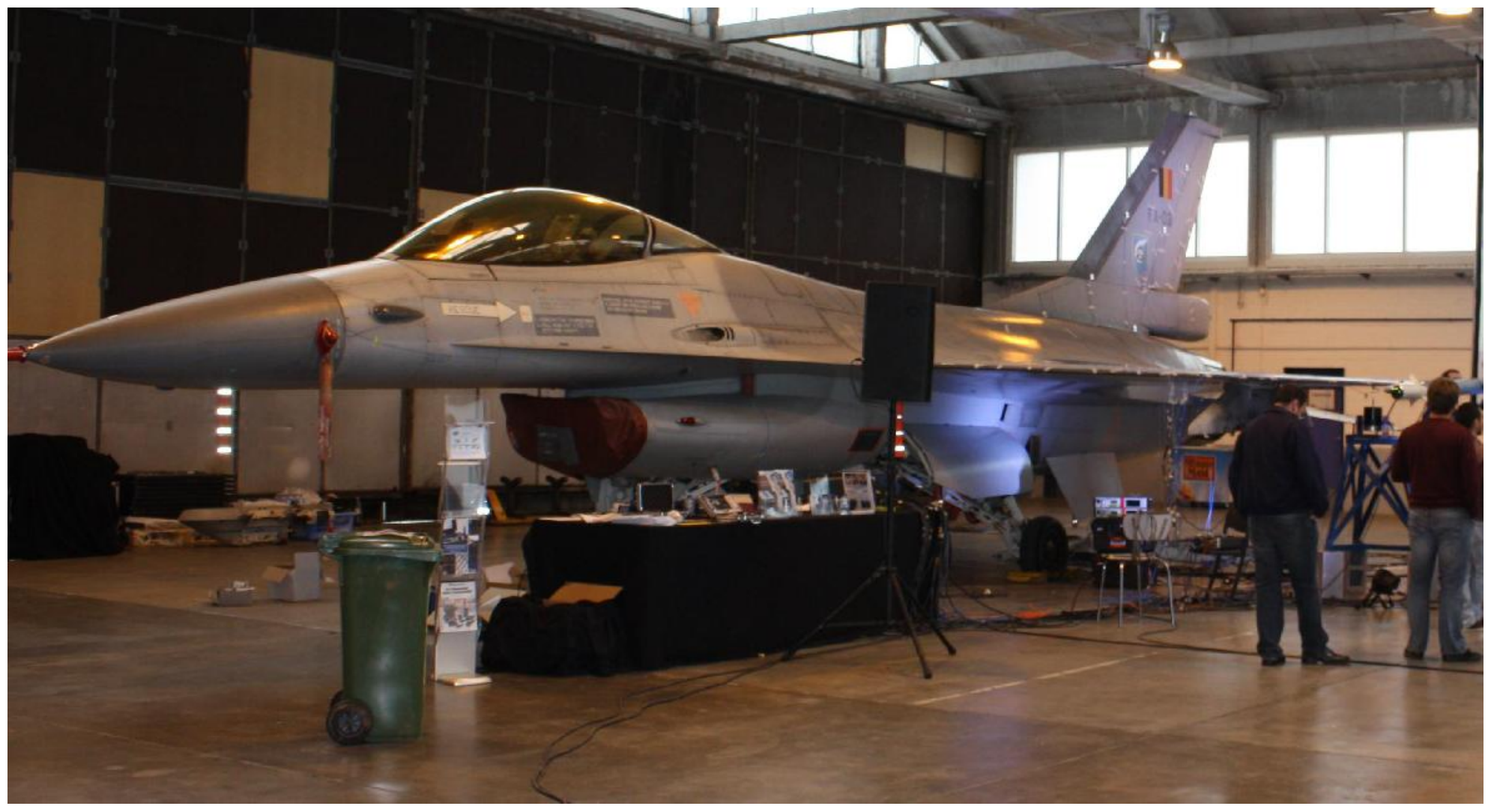}
\par\end{centering}
\begin{centering}
(a)
\par\end{centering}
\begin{centering}
\includegraphics[scale=0.27]{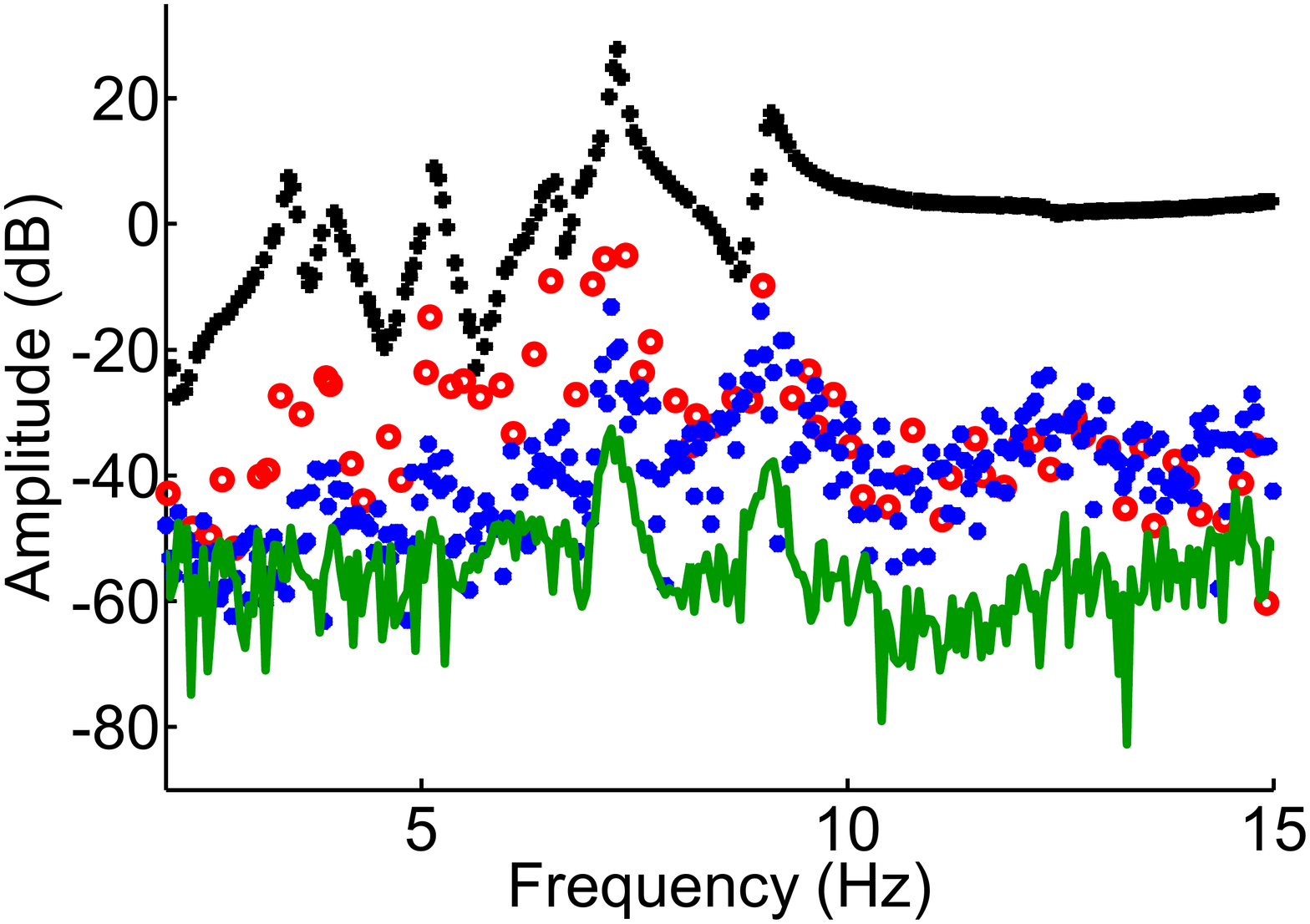}
\par\end{centering}
\centering{}(b)\caption{Ground vibration test on the General Dynamics F16 fighter jet (a).
The right wing is excited with a shaker, and the accelerations are
measured at 140 places. In figure (b) the measured acceleration for
a measurement point close to the right tip, near the missile connections,
is shown. Black: output at the excited frequencies, Red: odd nonlinear
distortions, Blue: even nonlinear distortions, Green: disturbing noise
level. These measurements show that the level of the nonlinear distortions
is well above the disturbing noise level. \label{fig:F16-fighter NL dist}}
\end{figure}

The measurement strategy that was presented earlier in this article
allows the user to meet all these customer requirements, and go even
beyond these expectations. This is illustrated on a GVT of an General
Dynamics (now Lockheed Martin) F-16 Fighting Falcon \ref{fig:F16-fighter NL dist}.
During the measurement campaign, shakers were put at the right and
left wing tip, and the accelerations are measured at 140 places. The
results shown here focus on the wing-to-payload mounting interfaces.
For large-amplitude vibrations, friction and gaps may be triggered
in these connections and markedly impact the dynamic behavior of the
complete structure. The right wing shaker is used, and the accelerations
at a point close to the mounting interface are analyzed.

The data were measured with a sample frequency $f_{s}=200$ Hz. A
multisine with a period length of about 41 s ($f_{0}=0.0244$ Hz)
is used. Only the odd frequencies between 1Hz and 60 Hz are excited,
and in each group of 4 odd frequencies one line is not excited. It
is used as a detection line for the odd nonlinear distortions. The
results for the frequency band between 3 and 11 Hz are shown in Figure
\ref{fig:F16-fighter NL dist} for the intermediate excitation level.
These measurements show  that the nonlinear distortion levels are
far above the noise level. So, the uncertainty on the linear measurements
(FRF, damping estimates, ...) are completely dominated by the nonlinear
behavior and hence they should be used with care. The resonance around
7 Hz corresponds to the first mode of the wings , which also excites
the wing-to-payload mounting interface at the tip of the wing. The
nonlinear distortions are largest around this resonance and are dominated
by an odd linear behavior. Later on in this article, it will be explained
that this will result in an excitation-dependent resonance frequency
and damping. Both will shift by a changing excitation level. 

Observe that the disturbing noise levels are at -40 to -60 dB which
is very good for mechanical measurements. This illustrates that the
proposed measurement strategy meets all the formulated expectations
for a good GVT. In a single experiment, it is possible to measure
the mode-shapes and the resonance frequencies, together with a full
nonlinear signature of the nonlinear behavior of the tested structure.
The measured FRFs will be discussed later (see Figure \ref{fig:F16-fighter-FRF}).

\section*{Nonlinear distortions characterization: alternative methods}

In the first part of this article, a nonlinear distortion analysis
method has been presented that strongly relies on the use of random-phase
multisines with a well-designed frequency grid. Alternative approaches to detect the presence of nonlinear distortions are described in the survey article
\cite{Vanhoenacker (2002) ISMA nonlinear tests}, and \cite{Vanhoenacker (2003) PhD},
with a focus on mechanical applications. Amongst others, the following
methods are discussed:\emph{ superposition principle} and \emph{homogeneity
principle} \cite{Worden and Tomlinson (2001) book}; \emph{overlaid
Bode plot} and \emph{Nyquist plot distortions} \cite{Silva and Maia (1997) book};
\emph{coherence function measurements} \cite{Bendat and Piersol (2010) book,Pintelon 2012 book}\emph{;
bispectral analysis} \cite{Choi and Chang (1984). Bispectral NL analysis,Chang and Stearman (1985) Bispectral NL analysis}\emph{;
Hilbert transform} \cite{Tomlinson (1987) Hilbert transform}\emph{;
correlation methods\cite{Enqvist (2007) correlation nonlinearity tests}},
\cite{McCormack (1994) correlation NL periodic signals}.

This article discusses two alternative methods in more detail: the
higher-order sinusoidal input describing functions (HOSIDFs), and
the swept sine test. There are three reasons for this choice: i) these
methods can be considered as special cases of the previously presented
framework in which the multisine signal is replaced by a single (swept)
sine excitation; 2) The HOSIDFs are an elegant and practical useful
generalization of the concepts that are presented in this article;
3) The swept sine analysis provides additional nonparametric information
about the nonlinear distortion in mechanical vibrating systems.

\setcounter{subsection}{0}

\subsection*{\emph{Higher-order sinusoidal input describing functions} }

The HOSIDFs are a generalization of the sinusoidal input describing
function \cite{Gelb (1969)}, and describe the gain and phase relation
of a system between the input at the fundamental frequency $f_{0}$
and the output at the harmonics $kf_{0}$, using a sinusoidal input
signal \cite{Nuij (2006) MSSP  HOSIDF}

\[
G_{k}(f_{0},a)=Y(kf_{0})/U_{s}^{k}(f_{0}),
\]
where $a$ indicates the amplitude of the excitation signal. The method
can be used under feedback conditions \cite{Nuij (2006) CEP  HOSIDF feedback}.
The HOSIDFs give a simple description of complex nonlinear behaviors
of mechanical systems, for example, the transition from stick to sliding
in precision mechatronic systems \cite{Nuij (2008) HOSIDF Stick/sliding}. 

\begin{figure}
\centering{}\includegraphics[scale=0.65]{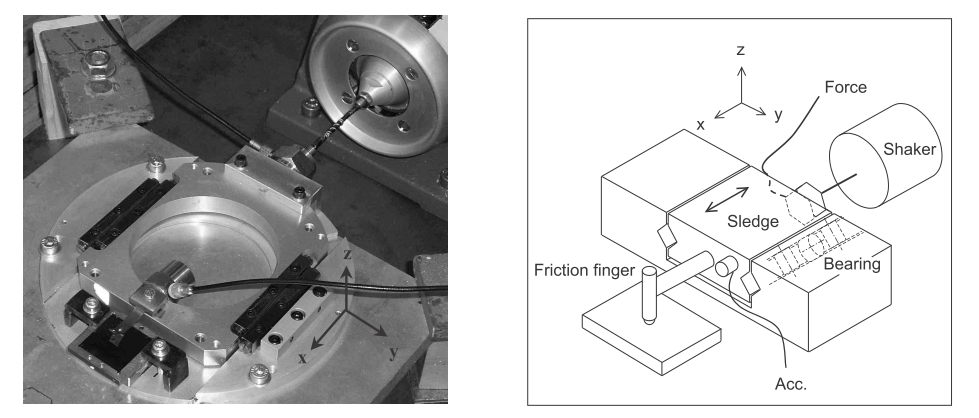}\caption{Experimental setup to analyze stick/sliding in a linear bearing with
friction \cite{Nuij (2006) CEP  HOSIDF feedback}\label{fig:Nuij Setup StickSlip}
.}
\end{figure}

An electromechanical shaker drives a sledge that is prone to dry friction
mainly created by the dry friction finger, resulting in a stick/slip
behavior (see Figure \ref{fig:Nuij Setup StickSlip}) \cite{Nuij (2008) HOSIDF Stick/sliding,Nuij (2006) CEP  HOSIDF feedback,Nuij (2006) MSSP  HOSIDF}.
The driving current of the shaker is used as an input, and the measured
acceleration is the output of the system. 

\begin{figure}
\begin{centering}
\includegraphics{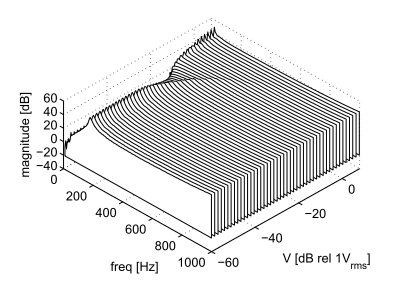}
\par\end{centering}
\begin{centering}
(a)
\par\end{centering}
\begin{centering}
\includegraphics{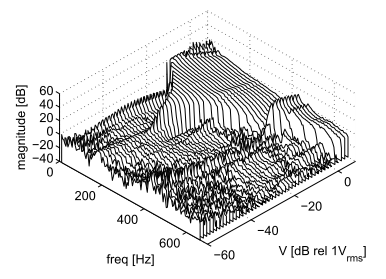}
\par\end{centering}
\begin{centering}
(b)
\par\end{centering}
\caption{Magnitude and phase of the first-order (a) and third-order (b) HOSIDFs
for the system shown in the previous figure \cite{Nuij (2006) CEP  HOSIDF feedback,Nuij (2006) MSSP  HOSIDF,Nuij (2008) HOSIDF Stick/sliding}
\label{fig:Nuij HOSIDF 1 and 3}.}
\end{figure}

The amplitudes of the first- and third-order HOSIDFs are shown in
Figure \ref{fig:Nuij HOSIDF 1 and 3}. As long as the system is in
the stick phase, it behaves as a linear system with a large stiffness.
Once the sledge starts to move, nonlinear distortions become visible
in the measured acceleration, resulting in a large increase of the
third-order HOSIDFs. This makes it possible to detect very clearly
the transition from stick to slip for varying excitation conditions
(frequency and amplitude of the sine excitation). These results show
that the HOSIDFs are a versatile tool that provides intuitive insight
in the behavior of a nonlinear system that is directly accessible
for the design engineer. It complements the multi-frequency tests
that were explained before in the section: Detection, separation,
and characterization of the nonlinear distortions and the disturbing
noise.

\subsection*{\emph{Swept sine test} }

The swept sine test works well for mechanical systems with isolated
resonance modes, and the sensors positioned close to the nonlinear
component. The system is excited with a swept sine (this is a sine
with constant amplitude, the frequency varies linearly with time),
and the presence of nonlinear distortions is looked for either by
\cite{Noel (2014) NL analysis using swept sine}: i) searching for
anomalies in the envelope of the response, ii) by plotting the acceleration
against the relative displacement or relative velocity, or iii) by
making a time-frequency analysis using short-time Fourier transforms
or a wavelet analysis. From these measurements, it is also possible
to make a first estimate of the function describing the local nonlinear
component.

The sweep rate should be kept sufficiently low, such that the structure
gets enough time to built up the full resonance power when passing
through a resonance. If only the acceleration signal is used in the
analysis, sharp resonances might be missed or strongly underestimated
\cite{Roy (2012) Sweep rate impact on modal analysis}. As a rule
of thumb, the maximum sweep rate is proportional to $\omega_{3dB}^{2}$.
This problem disappears when the FRF is estimated from the input-output
measurements \cite{Gloth (2012) Sweep rate analytical study}, although
even in this case the sweep rate should remain low enough to have
a good frequency resolution. As mentioned before, the frequency resolution
is the inverse of the measurement time. An increasing sweep rate decreases
the measurement time required to cover a given frequency band, and
so the frequency resolution drops. In some standards, for example,
the standard for space engineering testing \cite{Esa Standard (2012)},
the users are advised to use a logarithmic sweep rate between 2 or
4 octaves/min, independent of the structure. It is clear that such
a setting can become critical if the damping is too low.

\begin{figure}
\begin{centering}
\includegraphics[scale=0.7]{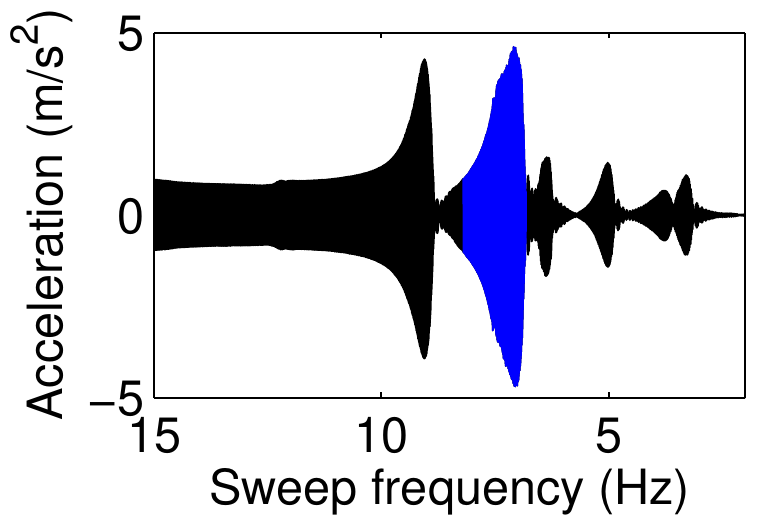}
\par\end{centering}
\begin{centering}
(a)
\par\end{centering}
\begin{centering}
\includegraphics[scale=0.5]{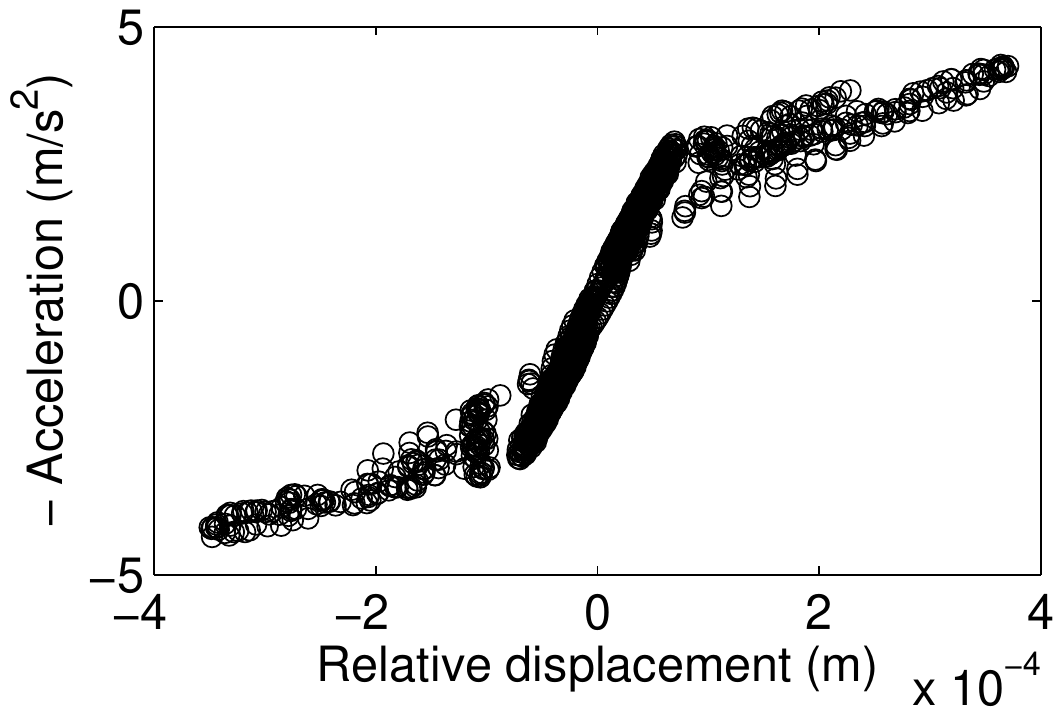}
\par\end{centering}
\centering{}(b)\caption{Ground vibration test on the General Dynamics F16 fighter (see Figure
\ref{fig:F16-fighter NL dist}) using a swept sine excitation. The
accelerations on both sides of the bolted missile connection to the
wing tip are measured. In (a), the measured acceleration is shown.
In (b), the acceleration is plotted with respect to the relative displacement
between the two sensors. \label{fig:Ground-vibration-test Swept Sine}}
\end{figure}

These ideas are illustrated on the fighter measurements in Figure
\ref{fig:Ground-vibration-test Swept Sine}. A swept sine excitation,
sweeping from 2 Hz to 15 Hz with a constant sweep rate of 0.05 Hz/s
is applied to the wing. Figure \ref{fig:Ground-vibration-test Swept Sine}(a) shows the measured
acceleration of the wing tip against the instantaneously swept sine
frequency. The resonances that were already  visible in Figure \ref{fig:F16-fighter NL dist}
and also in Figure \ref{fig:F16-fighter-FRF} (which will be discussed
later) show up also in Figure \ref{fig:F16-fighter NL dist}. In
the plot the crossing of the instantaneous frequency through the resonance
at 7 Hz is highlighted in blue. Observe that this blue section is
asymmetric, which is a strong indicator of the presence of a nonlinear
resonance. This part of the signal is further analyzed in Figure \ref{fig:F16-fighter NL dist}(b),
plotting the measured acceleration versus the relative displacement
of two sensors put on the left and right side of the bolted connection.
It is shown in \cite{Noel (2014) NL analysis using swept sine} that
such a plot gives a good indication of the shape of the local stiffness.
A detailed description and illustration on an aerospace structure
is given in \cite{Noel (2014) NL analysis using swept sine}. The
key idea is to discard all the inertia and force contributions that
are not directly related to the nonlinear component, as they are generally
unknown or not measured. In Figure \ref{fig:Ground-vibration-test Swept Sine}(b)
a softening spring behavior is  observed (the acceleration is proportional
to the force). This will be later confirmed by the FRF measurements
shown in Figure \ref{fig:F16-fighter-FRF}.

\begin{figure}
\centering{}\includegraphics[scale=0.5]{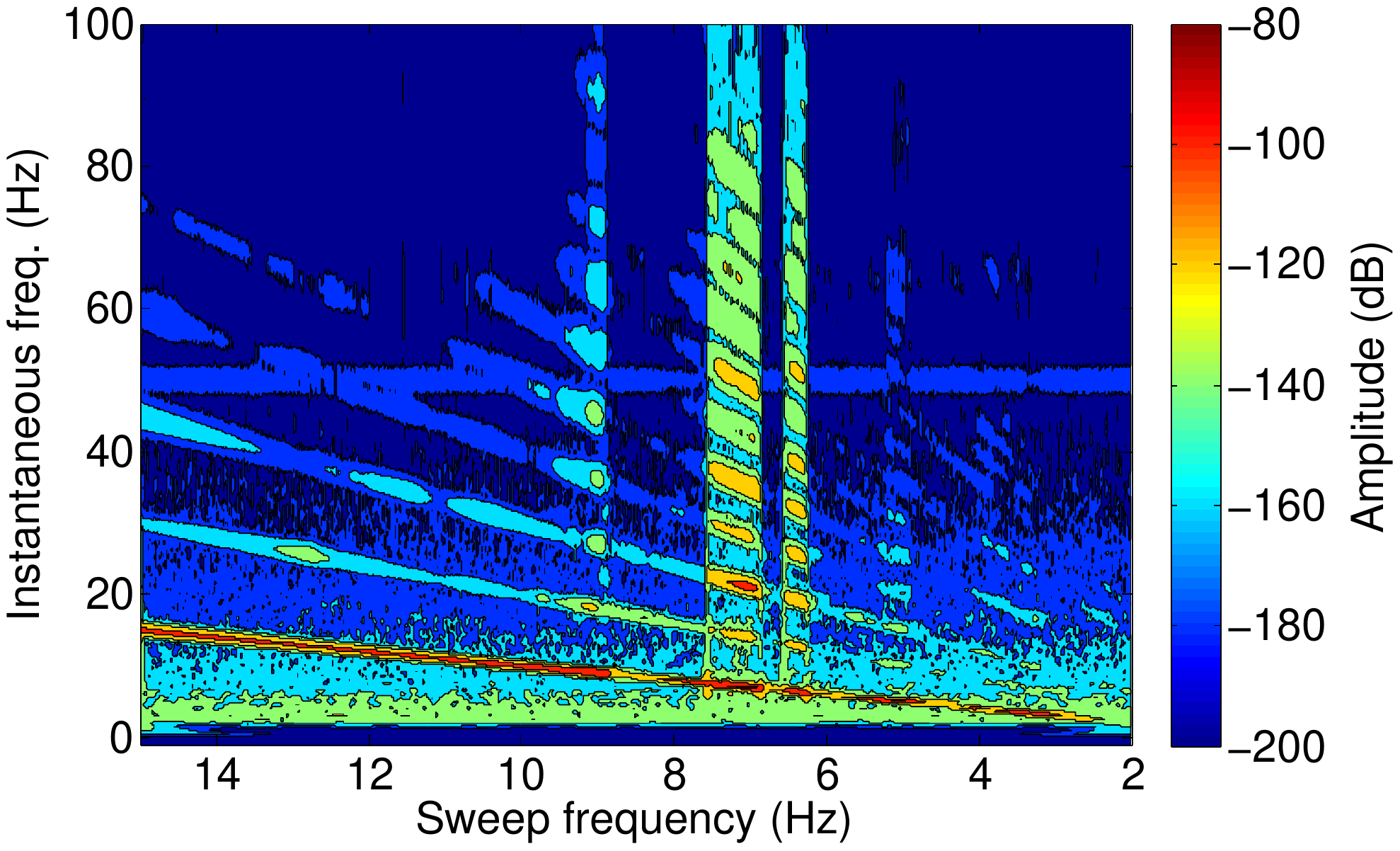}\caption{Time-frequency analysis of the measured acceleration signal at the
tip of the wing \cite{Noel (2014) NL analysis using swept sine}.
\label{fig:Ground-vibration-test Time-Freq analysis}}
\end{figure}

In Figure \ref{fig:Ground-vibration-test Time-Freq analysis}, a time-frequency
analysis is made of the acceleration signal and plotted as a function
of the instantaneous frequency (which replaces the time axis). The
decreasing red line corresponds with the instantaneous swept sine
frequency applied to the fighter. Some harmonic frequencies are  visible
at the integer multiples of this frequency. Observe also that around
the resonance frequency, the intensity and the number of higher harmonics
grows very fast. This points again to the presence of a strong nonlinear
behavior in the resonances. 

This analysis complements well the multisine method that was explained
before. It is applicable whenever local nonlinearities are present,
and it is possible to put sensors on both sides of the nonlinear structure.

\section*{System identification in the presence of nonlinear distortions: selection
of a linear or nonlinear modeling approach?}

Using the nonparametric test procedure described in in the previous
sections, the user gets a clear view on the presence and the behavior
of the nonlinear distortions. The procedure is formalized in a set
of user guidelines:
\begin{itemize}
\item Design a random-phase multisine to detect the presence of nonlinear
distortions following the guidelines of the multisine design section.
To do so, the even frequencies and a set of randomly selected odd
frequencies should be put to zero. The bandwidth, power spectrum,
and peak amplitude should be similar to the signals that will be later
on applied to the model. See \cite{Schoukens RiemanEquivalence} for
a detailed discussion.
\item Make a series of (steady-state) measurements with varying amplitudes
or offsets of the excitation signal that cover the amplitude range
of interest, and make the nonlinear analysis. More advanced signal
processing methods can be used to remove transient effects \cite{Pintelon MSSP part I,Pintelon (2011) MSSP SISO LPM en NL analysis,Pintelon (2011) MSSP  MIMO LPM and NL analysis}.
\item If the nonlinear distortions are smaller than the specified level
of accuracy of the model to be built, a linear design might be sufficient.
This will lead to the BLA of the nonlinear system. Otherwise, a
more involved nonlinear model will be needed. The BLA will be studied
in detail later in this article.
\item Be aware that the BLA varies in general as a function of the power
spectrum and amplitude distribution of the excitation signal. For
that reason, the excitation signals during the experiments should
match as well as possible the signals that will be applied later
on to the model as explained in the first bullet above.
\item Detailed step-by-step instructions for a nonparametric nonlinear distortion analysis are given in Section
6.1 of \cite{Schoukens 2012  Exercises book}, including a set of
routines to prepare the experiments and process the data. 
\end{itemize}

\section*{Approximation of nonlinear systems: user choices}

Once a nonparametric nonlinear distortion analysis is made, the user
has to decide, on the basis of this information, if a linear model
will be sufficient to meet the modeling goals, or if it is instead
necessary to use a nonlinear model. To make this choice, it is important
to understand the behavior of the linear modeling framework in the
presence of nonlinear distortions. Some of the theoretical properties  that are obtained under the linear assumptions will no longer hold. The asymptotic properties of the linear model that is estimated from a nonlinear system need to be verified, and the physical interpretation of the noise model needs to be modified. For a formal mathematical framework, see "A mathematical framework for nonlinear systems". Within this framework, it is possible to give
a precise definition and interpretation of the BLA that will be identified
under these settings.

Describing a system with a model that is too simple results in model
errors. These model errors depend upon some choices that are implicitly
or explicitly made by the user. To address these issues and to understand
the results, it is necessary to line up the user choices that are
present in each identification strategy. It is dangerous if the user
is not aware of these choices, or if their impact is not well understood.
The impact of the selected approximation criterion, related convergence
criterion, and the chosen excitation signal are discussed below.

\setcounter{subsection}{0}

\subsection*{Approximation methods}

\begin{figure}
\centering{}\includegraphics[scale=0.6]{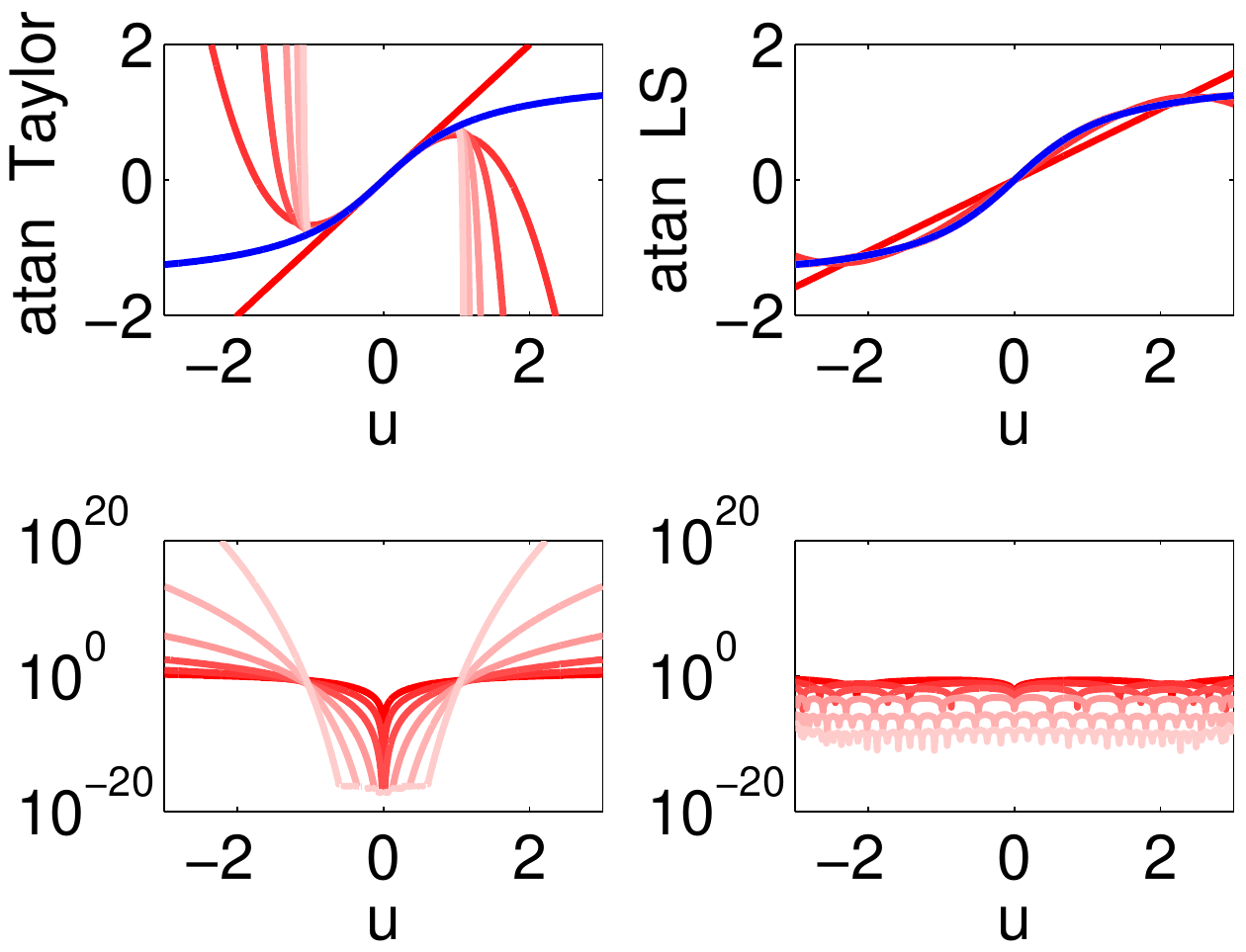}\caption{Illustration of the impact of the approximation criterion on the approximation
errors. A static nonlinear system (blue line) is approximated using
two different approaches. That is, the comparison is between a Taylor
series of order 1,3,5,7,9,11, and a polynomial model of order 1,3,5,7,9,11.
The polynomial is fit using the least squares. The errors are shown
in the bottom figures. Observe that the Taylor series approximation
gives a much better fit around the origin, but fails to converge for
an input $|u|>1$. The convergence of the least squares fit on the
right side is much slower, but the approximation converges everywhere
on the interval {[}-3,3{]}.\label{fig:Approximation method Taylor and LS}}
\end{figure}

The quality of the fit of a model to a system, or to the data that
describe this system, can be expressed by defining a distance between
the model and the data. This distance is called the approximation
criterion or the cost function. The sum of the absolute or the squared
errors are two popular choices. A first possibility to find an 'optimal'
approximating model is to minimize the selected cost function with
respect to the model parameters for the given data set. If the model
can exactly describe a system, and the data are free of measurement
error, the choice of the approximation criterion is not so critical
as long as it becomes zero with exact model parameters. The choice
of the cost function that minimizes the impact of disturbing noise
(the combined effect of measurement noise and process noise), still
assuming that the system is in the model set, is the topic of system
identification theory. This section is focussed on the alternative
situation. It starts from exact data, but the model is too simple
to give an exact description of the system. This results in model
errors, and the choice of the approximation criterion will significantly
impact on the behavior of the resulting model errors. The ideas are
presented by analyzing a simple example. In the right side of Figure
\ref{fig:Approximation method Taylor and LS}, the atan function is
approximated by a polynomial $atan(x)\approx\sum_{k=0}^{n}a_{k}x^{k}$.
The polynomial coefficients $a_{k}$ are obtained as the minimizers
of the squared differences $e_{n}(x)=atan(x)-\sum_{k=0}^{n}a_{k}x^{k}$
\begin{equation}
\hat{a}=\arg\min_{a}\sum_{x\in D}e_{n}(x)^{2}.\label{eq:Approximation LS fit polynomial}
\end{equation}

In this expression, the sum over $x$ stands for the sum over all
data points in the data set $D$. 

Alternative approaches to obtain an approximating polynomial representation
exist, for example, using a Taylor series approximation. In that
case, no explicit cost function interpretation is made. The polynomial
coefficients are calculated from the values of the function's derivatives
at a single point, in this case $x=0$. In the left side of Figure
\ref{fig:Approximation method Taylor and LS}, the successive Taylor
approximations are shown for growing orders $n$. 

Both approximations are  different from each other, and the model
errors have a completely different behavior. While the Taylor approximation
converges very fast around zero, it fails to converge outside the
interval $[-1,1]$. The least squares approximation converges over
the full interval $[-3,3]$ but at a slower rate. The width of the
interval can be made arbitrarily large.

In this article, the least squares model fitting approach is followed.

\subsection*{Convergence criteria}

\begin{figure}
\centering{}\includegraphics[scale=0.75]{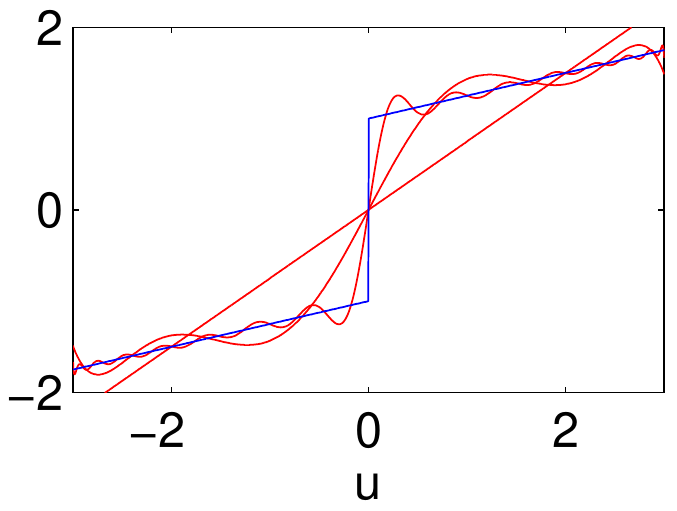}\caption{Least-squares approximation of a discontinuous function with continuous
basis functions.\label{fig:Approximation discon function} }
\end{figure}

In the previous example, a polynomial approximation of the $atan(x)$
function is made. Using the least-squares cost function, the error
can made arbitrarily small in a given interval by increasing the complexity
of the model. In Figure \ref{fig:Approximation discon function},
a discontinuous function is approximated using polynomials of different
degrees. Again, it can be seen that the error can be made arbitrarily
small for all inputs, except at the discontinuity at $u=0$ where
the error converges to half the discontinuity. This prohibits uniform
convergence that is characterized by a decrease of the maximum error
in the interval. More formaly, $\textrm{for all }\varepsilon,\textrm{ there exists a value }N\mathrm{\,\textrm{such that}\:sup_{\mathit{x}\in\mathit{D}}|\mathit{e}_{\mathit{n}}(\mathit{x})|}<\varepsilon\;\mathrm{for}\;n>N.$ 

To include also discontinuous functions in the framework, the convergence
criterion should be weakened to point-wise convergence, which can
be obtained by using the convergence in the mean-square sense. Mean-square
convergence requires that

\[
\lim_{n\rightarrow\infty}\sum_{x\in D}e_{n}(x)^{2}=0,
\]
which guarantees that the approximation converges everywhere excepted
for some isolated points where the function is discontinuous. For
continuous functions, uniform convergence is retrieved. It is clear
that this concept matches very well with the least-squares model-fitting
approach.

In this article mean-square convergence will be used.

\subsection*{Impact of the choice of the excitation signal}

\begin{figure}
\centering{}\includegraphics[scale=0.5]{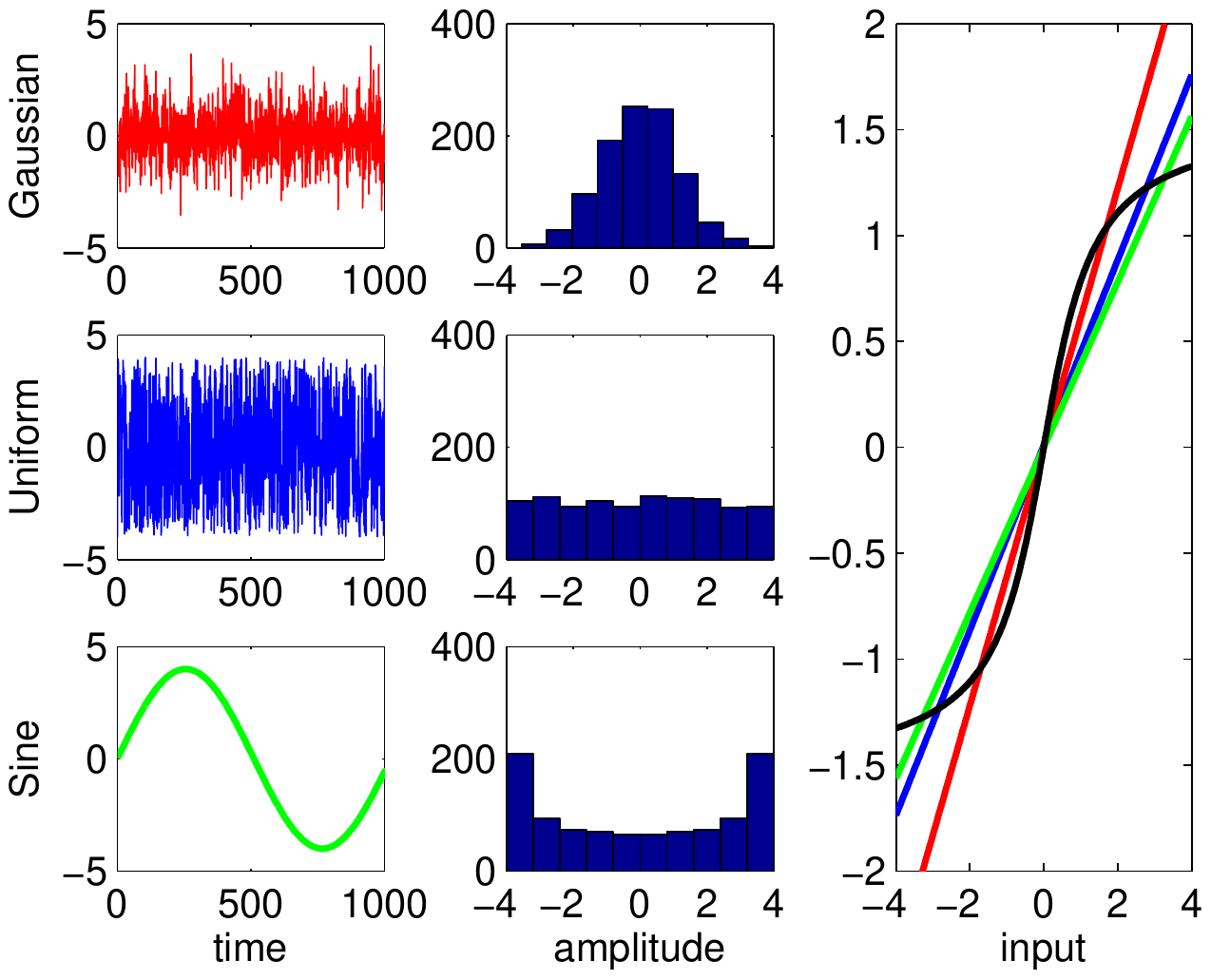}\caption{The best linear approximation (BLA) of a static nonlinear system (the
black line in the right figure) depends on the amplitude distribution
of the excitation signal. The BLA for a Gaussian (red), uniform (blue),
and sine (green) excitation (left column) are shown. The histogram
(for 1024 samples) for each of the excitation signals is shown in
the middle column. Since most of the probability mass of a Gaussian
distribution is around the origin (see Gaussian histogram), the Gaussian
excitation results in the best fit in that domain. A sine excites
mostly the extreme values (see the histogram of the sine excitation),
and it results in a fit that better approximates the nonlinear function
for these extreme values. This comes at a cost of larger approximation
errors around the origin. The behavior of the uniform distribution
is in between these two extreme distributions, and this is also true
for the corresponding fit (blue line). \label{fig:Excitation-signals-impact}}
\end{figure}

The actual fit of the model, in the absence of model errors and noise-free
data, will not depend on the characteristics of the excitation signal.
This changes drastically when the model is not rich enough to capture
the full system behavior. In that case, errors will be present, and
during the fit these will be pushed to those parts of the input domain
that are not so well excited because that reduces the cost in \eqref{eq:Approximation LS fit polynomial}.
This makes the results dependent on the choice of the excitation,
which is illustrated in Figure \ref{fig:Excitation-signals-impact}
where $atan(u)$ is linearly approximated using the model $y=au$.
Figure \ref{fig:Excitation-signals-impact} shows that the BLA
depends on the amplitude distribution of the excitation signal. The
results for a Gaussian, uniform, and sine excitation (see left column)
are shown in the right figure. The histogram for each of the excitation
signals is shown in the middle figures. Since most of the probability
mass of a Gaussian distribution is around the origin (see Gaussian
histogram), the Gaussian excitation results in the best fit in that
sub-domain. A sine excites mostly the extreme values (see the histogram
of the sine excitation), and it results in a fit that better approximates
the nonlinear function for these extreme values. This comes at a cost
of larger approximation errors around the origin. The behavior of
the uniform distribution is in between these two extreme distributions,
and this is also true for the corresponding fit (blue line).

This example shows  that the experiment design in the presence of
model errors will be even more important than in classical system
identification where no model errors are considered. If the user is
aware that model errors will be present, care should be taken that
at least a part of the experiment consists of signals that mimic very
well the signals that will applied later on to the model. The remaining
part of the experiment can be used to obtain a sufficiently rich excitation
so that the uncertainty remains small enough. Such an approach is
illustrated on the identification of an industrial clutch in \cite{Dhammika excitation signals CEP}.
To identify the BLA for a nonlinear dynamic system, not only is the
power spectrum of the excitation important (to cover the frequency
band of interest), also the amplitude distribution plays a crucial
role (to excite those amplitude regions that will be used in later
applications). In \cite{Dhammika excitation signals CEP}, a mixture
of random-phase multisines and impulsive excitations is used to cover
the later use of the model.

\section*{A new paradigm: replacing the nonlinear system by a linear system
plus a noise source\label{subsec:Approximating NLS:A-new-paradime}}

In the previous section, the impact of some user choices (approximation
method, convergence criteria, excitation signal) on the behavior of
model errors is discussed. In this section, these results are used
to approximate a nonlinear system with a linear model. First, the
user choices are  specified. Next, the BLA will be introduced, and
the properties of the model errors are discussed. This leads eventually
to a new paradigm to deal with nonlinear systems in a linear setting.

\setcounter{subsection}{0}

\subsection*{User choices}

Since the approximation of a nonlinear system by a linear model creates
model errors, the user choices that were discussed before should
be carefully made. As explained before, the linear approximation is
tuned by minimizing the mean-square error between the measured and
modeled output. As a direct result, the output error will be uncorrelated
with input. The excitation will be restricted to random signals with
a Gaussian distribution. These include filtered Gaussian noise and
random-phase multisines \eqref{eq:random phase multisine} with a
sufficiently large number of components (in practice $F>10$ works
in many applications). 

\subsection*{The best linear approximation $G_{BLA}$}

The linear system that fits best the data is called the BLA, given
either by its impulse response $g_{BLA}(t)$, or its FRF $,G(\omega).$
More formally, $G_{BLA}$ is defined as \cite{Pintelon 2012 book},
\cite{Enqvist 2005 Thesis}-\cite{Bussgang},\cite{Schoukens RiemanEquivalence}
\begin{equation}
G_{BLA}\left(q\right)=\arg\min_{G}E\left\{ \left|y_{0}\left(t\right)-G\left(q\right)u\left(t\right)\right|^{2}\right\} ,\label{eq: 2 def Gbla}
\end{equation}
where $q$ is the shift operator for a discrete-time model. Similar
expressions can be given for continuous-time models. All expected
values $E\left\{ \right\} $ in this article are taken with respect
to the random input $u\left(t\right)$. In most applications, the
dc-value of the input and output signal should be removed to obtain
a model that is valid around a given setpoint. 

\subsection*{A nonlinear 'noise' source}

The difference between the output of the nonlinear system and that
of the BLA $y_{s}(t)=y(t)-G_{BLA}(q)u(t)$ is called the stochastic
nonlinear contribution or nonlinear noise. Although this name might
be misleading (the error is deterministic for a given input signal),
it is still prefered to call it a stochastic contribution because
it looks very similar to a noise disturbance for a random excitation
\cite{schoukens dobrowiecki NL dist,Schoukens Automatica 2005 plenary,Schoukens 2012  Exercises book,Pintelon 2012 book}. 

\subsection*{A new paradigm}

\begin{figure}
\centering{}\includegraphics[scale=0.4]{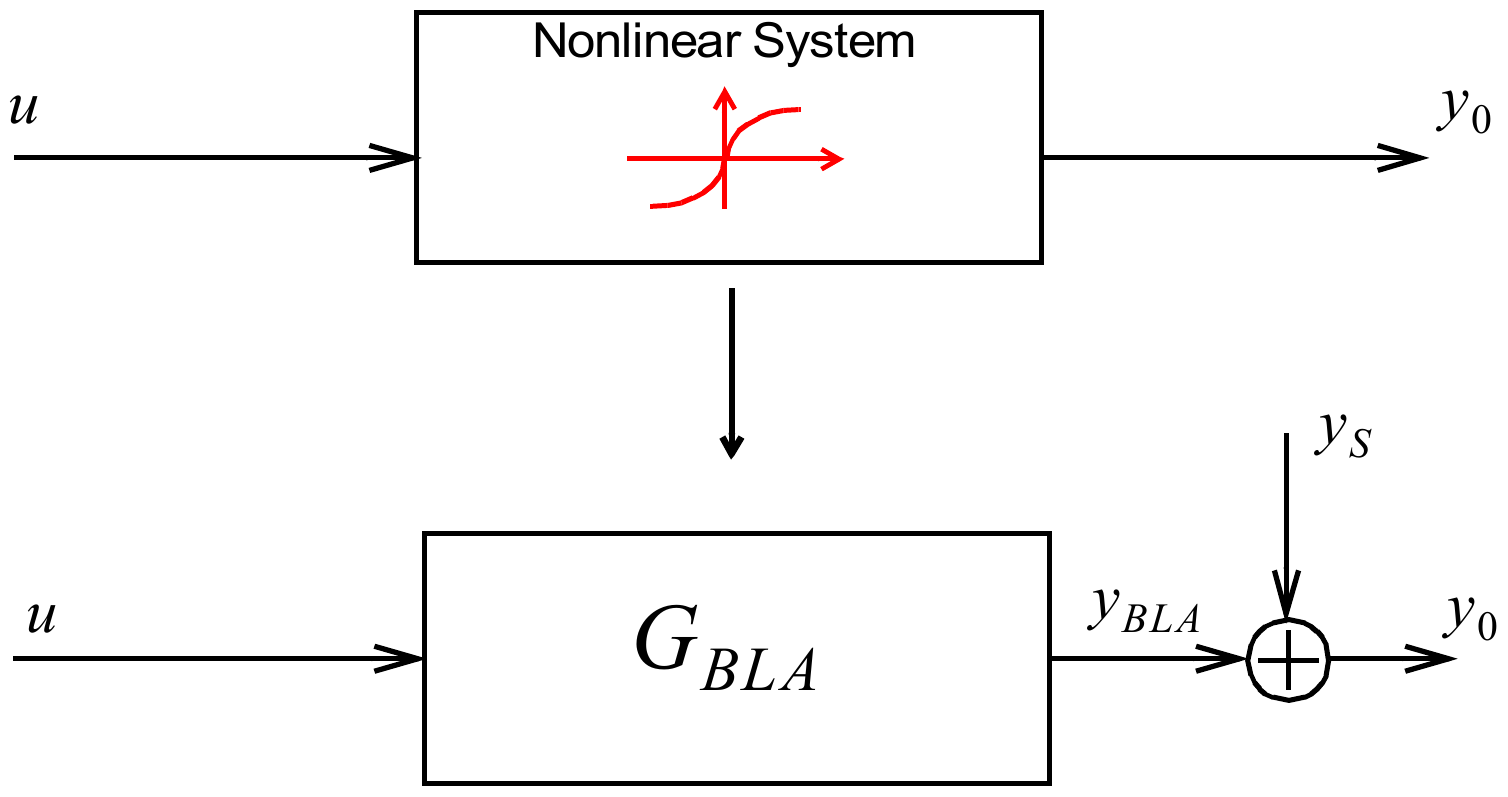}\caption{A new paradigm: the nonlinear system (top figure) is replaced by the
best linear approximation $G_{BLA}$ plus an error term $y_{S}(t)$
(bottom figure).\label{fig:A-new-paradigm:}}
\end{figure}

By combining these results, the output of a nonlinear system that
is driven by a random excitation (or a Riemann-equivalent signal
\cite{Schoukens RiemanEquivalence}) is split in two classes of contributions,
being the coherent contributions $Y_{BLA}$ and the noncoherent contributions
$Y_{S}$ (see Figure \ref{fig:A-new-paradigm:}). The linear part
of the system contributes to the coherent output only, while the nonlinear
distortions contribute to both the coherent and noncoherent output.
\begin{itemize}
\item \emph{Coherent output}: The relation between the input $U_{0}(k)$
and the coherent (non)linear contributions $Y_{BLA}(k)$ is given
by 
\[
Y_{BLA}(k)=G_{BLA}(k)U(k)+T(k),
\]
where $T(k)$ models the transient effects and leakage errors \cite{Schoukens 2006 leakage,Pintelon 2012 book}.
From now on it is assumed, without loss of generality, that steady-state conditions apply,
such that the transient terms can be neglected in what follows. The
transfer function $G_{BLA}(k)$ depends on the power spectrum of the
Gaussian random excitation. Changing the Gaussian distribution to
an alternative such as a uniform distribution, can change the BLA.
From a spectral point of view the phase of $Y_{BLA}(k)$ is equal
to the phase of the input plus the phase of the transfer function
$G_{BLA}(k)$. Since $G_{BLA}$ is an expected value over the random
input, its actual value will not depend upon the actual realization
of the random input. 
\item \emph{Noncoherent output}: The noncoherent output $y_{S}$ accounts
for the difference between the output of the BLA and the actual nonlinear
output. For random excitations, it is very difficult for an untrained
user to distinguish the nonlinear noise $y_{S}\left(t\right)$~from
the additive disturbing output noise $v\left(t\right)$ (Figure \ref{fig:A-new-paradigm:}).
The nonlinear noise $y_{S}\left(t\right)$ is uncorrelated with $u\left(t\right)$
because they are the residuals of the solution of a least-squares
problem. However, $u\left(t\right)$ and $y_{S}\left(t\right)$ are
mutually dependent since there exists a nonlinear relation between
both signals, namely 
\[
y_{S}\left(t\right)=y_{0}(t)-G_{BLA}\left(q\right)u\left(t\right).
\]
\end{itemize}
Combining both results, the noise-free output $y_{0}\left(t\right)$
can be written as the sum $y_{BLA}(t)+y_{S}(t)$ (see Figure \ref{fig:A-new-paradigm:})
\cite{schoukens dobrowiecki NL dist,Schoukens Automatica 2005 plenary,Pintelon 2012 book} 

\begin{eqnarray}
y(t) & = & y_{0}(t)+v(t),\nonumber \\
y_{0}\left(t\right) & = & G_{BLA}\left(q\right)u\left(t\right)+y_{S}\left(t\right).\label{eq:}
\end{eqnarray}
In the frequency domain the relation between the FFT spectra becomes
\begin{align}
Y\left(k\right)= & Y_{0}\left(k\right)+V\left(k\right)\nonumber \\
= & G_{BLA}\left(k\right)U\left(k\right)+Y_{S}\left(k\right)+V\left(k\right),\label{eq:5}
\end{align}
disregarding again the transients $T\left(k\right)$~representing
the initial transients and leakage errors. The phase of $Y_{S}(k)$
will depend upon the phase of the input $U(l)$, for some values $l\neq k$.
This was not so for $Y_{BLA}(k)$, whose phase depends only on the
input phase $\angle U(k)$. Since the phases are stochastic variables,
$Y_{S}(k)$ will also be a stochastic value with respect to the random
input.

The power spectra of $Y_{S}$ and $V$ can be measured using the nonparametric nonlinear detection methods that were explained before.
In \cite{Schoukens (2010) Ys --> Ybla} a rationale is given that
shows that the level of the stochastic nonlinearities (noncoherent
output) is also a good indicator for the level of the nonlinear coherent
output for the considered class of excitations (random-phase multisines
and Gaussian noise).

For the specified user choices (mean-square error, random Gaussian
excitation), the asymptotic properties of $G_{BLA}$ and $Y_{s}$
are well known, assuming that the number of frequencies $N$ in the
multisine \eqref{eq:random phase multisine} grows to infinity. A
detailed discussion is given in Section 3.4 of \cite{Pintelon 2012 book},
here only a brief summary of the most important properties is given.

The BLA $G_{BLA}$ is shown to be smooth; it does not depend on
$N$, and it is the same for all Riemann-equivalent excitations. Only
the odd nonlinearities contribute to $G_{BLA}$.

The stochastic nonlinearities $Y_{S}$ have a smooth power spectrum.
They are zero-mean circular complex normally distributed, and they
have the same power spectrum for all Riemann-equivalent excitations.
Both the even and the odd nonlinearities contribute to $Y_{S}$.

\subsection*{A toy example\label{subsec:BLA FRF toy example}}

Consider a static nonlinear system 
\[
y=\sum_{k=1}^{n}a_{k}u^{k}.
\]
For (filtered) Gaussian noise excitations, it is known from Bussgang's
theorem \cite{Bussgang}, that the BLA is also static $y_{BLA}=a_{BLA}u$.
The least-squares estimate is
\[
\hat{a}_{BLA}=\frac{\sum y(t)u(t)}{\sum u(t)^{2}},
\]
which converges for large data sets to 
\[
a_{BLA}=\sum_{k=1}^{n}a_{k}\mu_{k+1}/\mu_{2},
\]
with $\mu_{\alpha}$ the moment of order $\alpha.$ This simple example
shows that the BLA depends on the higher-order moments of the excitation.
Observe that this result depends only on the amplitude of the Gaussian
noise (all higher-order moments of a zero-mean Gaussian distribution
are set by its variance), and not on its power spectrum. For zero-mean
excitations, only the odd degrees will contribute to the estimate. 

\begin{figure}
\centering{}\includegraphics[scale=0.6]{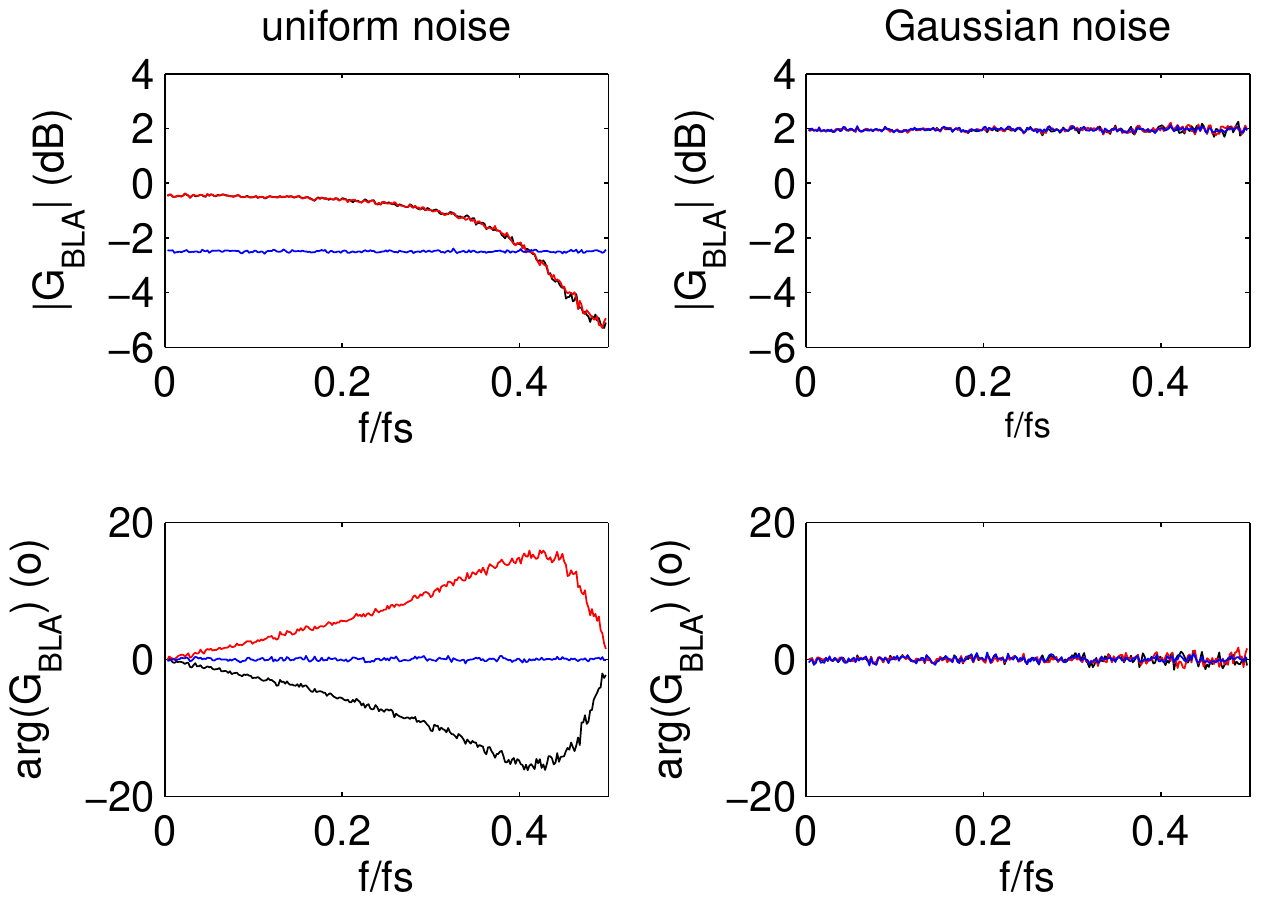}\caption{Illustration of the dependency of the best linear approximation (BLA)
of a static nonlinear system on the distribution of the excitation
signal. Results are shown for a (filtered) uniform and a (filtered)
Gaussian excitation on a normalized frequency axis, whith $f$ the
frequency, and $f_{s}$ the sample frequency. The amplitude and phase
of the estimated $\hat{G}_{BLA}$ are shown for the following noise filters: Blue: white noise, Black:
$u(t)=e(t)+0.5e(t-1)$, Red: $u(t)=0.5e(t)+e(t-1)$.\label{fig:Enqvist example}}
\end{figure}

This result changes when the excitation is no longer Gaussian distributed.
A nice illustration is given in Exercise 83.b of \cite{Schoukens 2012  Exercises book},
taken from \cite{Enqvist 2005 Thesis}. The BLA of a cubic static
nonlinear system $y=u^{3}$ is estimated for 6 different situations:
$u$ is zero-mean white Gaussian or uniformly distributed noise, $u(t)=e(t)+0.5e(t-1)$,
or $u(t)=0.5e(t)+e(t-1)$, with $e(t)$ being zero-mean white Gaussian
or uniformly distributed noise. The resulting FRF of the BLA is given
in Figure \ref{fig:Enqvist example}. From the right figures, it is
seen that for Gaussian noise $G_{BLA}$ has a constant amplitude and
phase which corresponds to a static system. In the left figures, the
excitation is generated starting from a uniformly distributed noise
generator. For white noise, $G_{BLA}$ is still a constant. However,
for the filtered uniform noise, a frequency-dependent FRF is retrieved
that depends on the actual filter that is applied. In this example,
short filters were used. If the impulse response becomes longer, the
distribution of the filtered signal will converge to a Gaussian distribution
and the dependency on the distribution of $e$ will disappear \cite{Wong Hin Kwan (2012) IEEE I=000026M paper 1  distribution}.
This allows well-selected pseudo-random binary excitations to be used
in many practical applications to measure $G_{BLA}$ \cite{Wong Hin Kwan (2013) IEEE I=000026M paper 2  PRBS}.
Since in some industrial applications, binary excitations are the
only feasible excitation (for example, opening or closing a valve),
this might become an attractive practical extension.

\section*{Nonparametric identification of the best linear approximation\label{sec: BLA FRF measurement}}

In this section it is explained how to measure the FRF of the BLA.
An optimal strategy (choice excitation signals, reduction of the impact of the nonlinear distortions) to obtain the best FRF measurement within a given time is proposed.
First, the error sources in the $G_{BLA}$ measurement will be discussed,
and it will be shown how to reduce them. Next, experimental illustrations
will be shown.

\begin{figure}
\centering{}\includegraphics[scale=0.4]{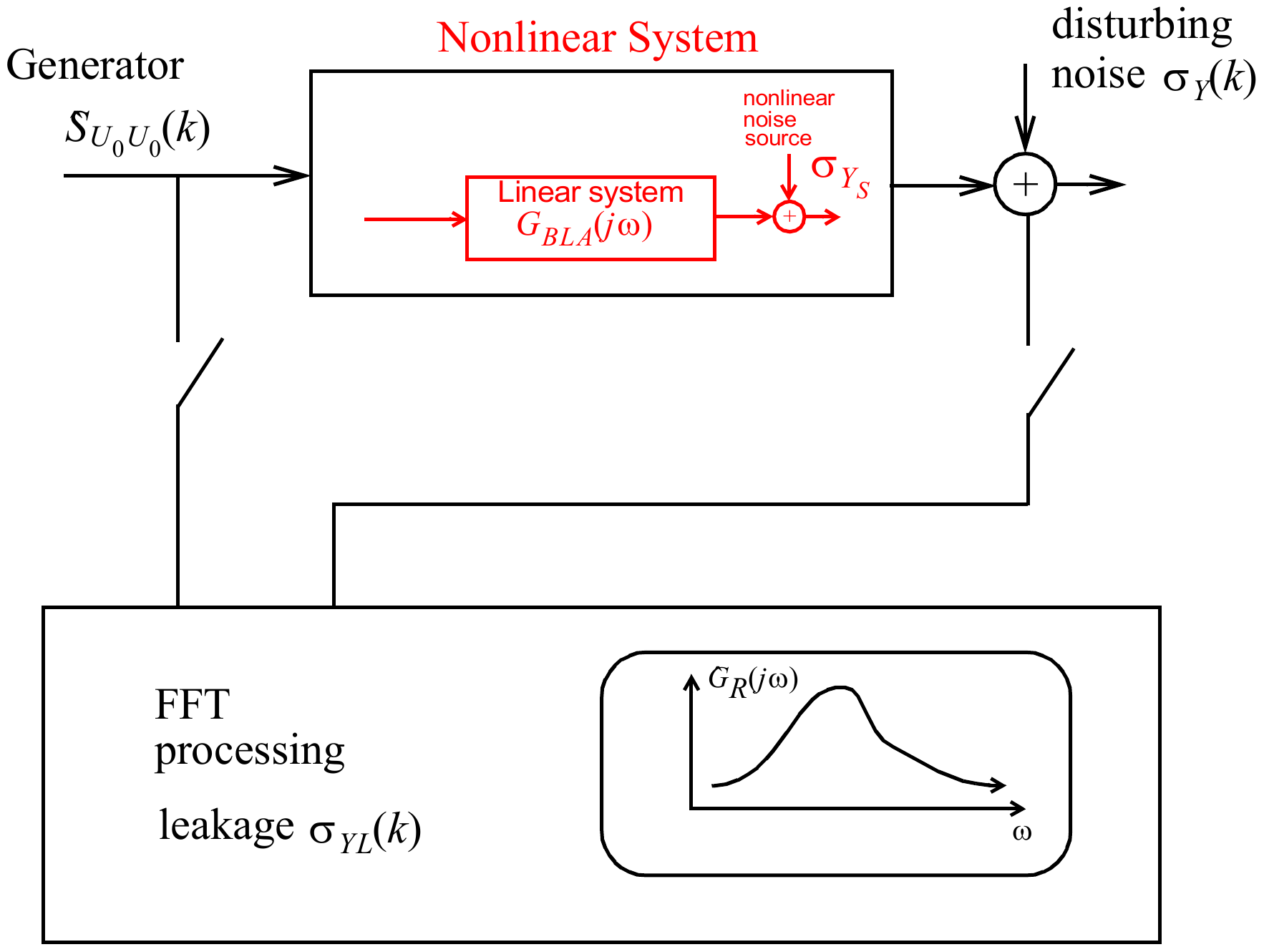}\caption{Error sources in a frequency response function (FRF) measurement: leakage error with standard deviation
$\sigma_{YL}$, disturbing noise (process noise, measurement noise, environmental noise) with standard deviation
$\sigma_{Y}$,  and the nonlinear noise source (stochastic nonlinear contributions) with standard deviation
$\sigma_{YS}$. The variance of the FRF at frequency $f_{k}$ is $\sigma_{G_{BLA}}^{2}(k)=\frac{\sigma_{YL}^{2}(k)+\sigma_{Y}^{2}(k)+\sigma_{YS}^{2}(k)}{S_{UU}}$.\label{fig:FRF BLA noise sources}}
\end{figure}

It is shown \cite{Schoukens variance FRF BLA 2012} that all the results
for nonparametric FRF measurements, developed for the linear framework,
hold also for the nonlinear situation
\[
\hat{G}_{BLA}(k)=\frac{\hat{S}_{YU}(k)}{\hat{S}_{UU}(k)},
\]

and

\begin{equation}
\sigma_{G_{BLA}}^{2}(k)=\frac{\sigma_{disturbances}^{2}(k)}{\hat{S}_{UU}(k)},\label{eq:BLA variance}
\end{equation}
with $\hat{S}_{UU}(k)$ and $\hat{S}_{YU}(k)$ the sample auto and
cross-power spectrum obtained from the finite set of repeated measurements.
In Figure \ref{fig:FRF BLA noise sources}, it can be seen that there
are three contributions to the noise variance that show up in the
numerator of the variance expression
\begin{equation}
\sigma_{G_{BLA}}^{2}(k)=\frac{\sigma_{disturbances}^{2}(k)}{\hat{S}_{UU}(k)}=\frac{\sigma_{YL}^{2}(k)+\sigma_{Y}^{2}(k)+\sigma_{YS}^{2}(k)}{\hat{S}_{UU}},\label{eq:BLA variance 3 contributions}
\end{equation}
with $\sigma_{YL}^{2},\sigma_{Y}^{2},\sigma_{YS}^{2}$ being respectively
the variance of the leakage error, the disturbing noise, and the stochastic
nonlinearities. In the next section, it will be shown how these different
contributions to the variance can be reduced to minimize the variance. 

\setcounter{subsection}{0}

\subsection*{Reduction of the errors on the FRF in the presence of nonlinear distortions}

Two possibilities to reduce the variance \eqref{eq:BLA variance}
on the FRF measurement of $G_{BLA}$ are discussed. The first one
is to avoid dips in the power spectrum estimate $\hat{S}_{UU}$ of
the input. These (very) small values of $\hat{S}_{UU}(k)$ result
in a much higher noise sensitivity since $\hat{S}_{UU}(k)$ is in
the denominator of \eqref{eq:BLA variance}. The second possibility
is to reduce the variance $\sigma_{disturbances}^{2}(k)$ as much
as possible. Both possibilities are discussed below.

\subsubsection*{Avoiding dips in the observed input power spectrum \foreignlanguage{english}{\textmd{$\hat{S}_{UU}$}}}

\begin{figure}
\centering{}\includegraphics[scale=0.6]{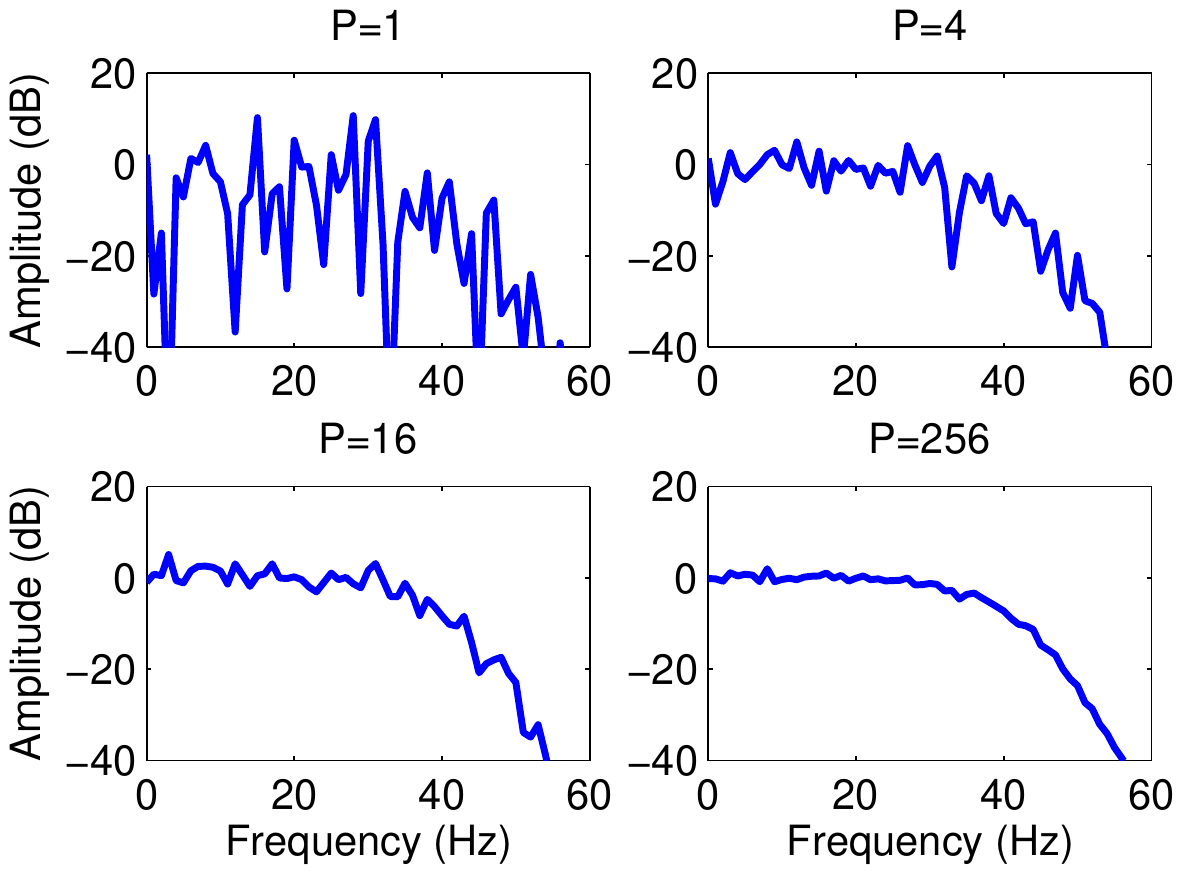}\caption{Power spectrum estimate $\hat{S}_{UU}$ of an observed random noise
sequence, averaged over $P$ realizations\label{fig:BLA FRF power spectrum input}.}
\end{figure}

The observed power spectrum $\hat{S}_{UU}(k)=\frac{1}{P}\sum_{l=1}^{P}|U^{[l]}(k)|^{2}$,
obtained from $P$ realizations of the random input, can be significantly
different from the power spectrum $S_{U_{0}U_{0}}(k)$, especially
for small values of $P$. This can be seen in Figure \ref{fig:BLA FRF power spectrum input}.
For small values of $P$, large dips can be observed with a loss of
20 dB or more. To reduce the loss to 1 dB (10\%) with a probability
of 95\%, at least $P=64$ realizations should be averaged (see \cite{Pintelon 2012 book},
Table 2-1 p. 58). For that reason, it is better to avoid random excitations
if possible, and use, for example, random-phase multisines (see Figure
\ref{fig:Examples-of-excitions}) that do not face this problem.

\subsubsection*{Reducing the noise contributions}

The reduction of the three noise contributions in \eqref{eq:BLA variance 3 contributions}
is discussed below.

\emph{Disturbing noise }$\sigma_{Y}^{2}(k)$ : The only possibility
to reduce the disturbing noise level is to be careful during the measurement
setup. Using shielded cables, low-noise signal conditioners, reducing
the environmental noise, for example, can all contribute to keep
the disturbing noise as small as possible. 

\emph{Leakage errors} $\sigma_{_{YL}}^{2}(k)$ : A random-phase multisine
eliminates also the leakage errors in the FFT-processing of the results,
so that $\sigma_{Y_{L}}=0$. This is a second reason why the use of
random-phase multisines is strongly advocated.

\emph{Nonlinear noise source} $\sigma_{Y_{S}}^{2}(k)$ : Although
nonlinear distortions are intrinsically linked with a nonlinear system,
it is still possible to partially eliminate their impact on the FRF
measurement by using an odd excitation. This can be either a random
noise source with a symmetric amplitude distribution (for example,
zero-mean Gaussian noise) or a well-designed multisine. An odd multisine
does excite only the odd frequencies, and the FRF is only measured
at those frequencies. Keep in mind that this doubles the required
measurement time for a given frequency resolution because the even
frequencies cannot be used in that case for the FRF measurement, so
that one frequency out of two is not in use. However, this comes with
the advantage that the even nonlinear distortions are no longer influencing
the odd frequencies, as explained in the section on nonlinear distortions detection.
For systems with dominating even nonlinearities, a large reduction
of the nonlinear noise variance on the FRF measurement will be obtained. 

\subsubsection*{Experimental illustration of the noise reduction}

\begin{figure}
\centering{}\includegraphics[scale=0.4]{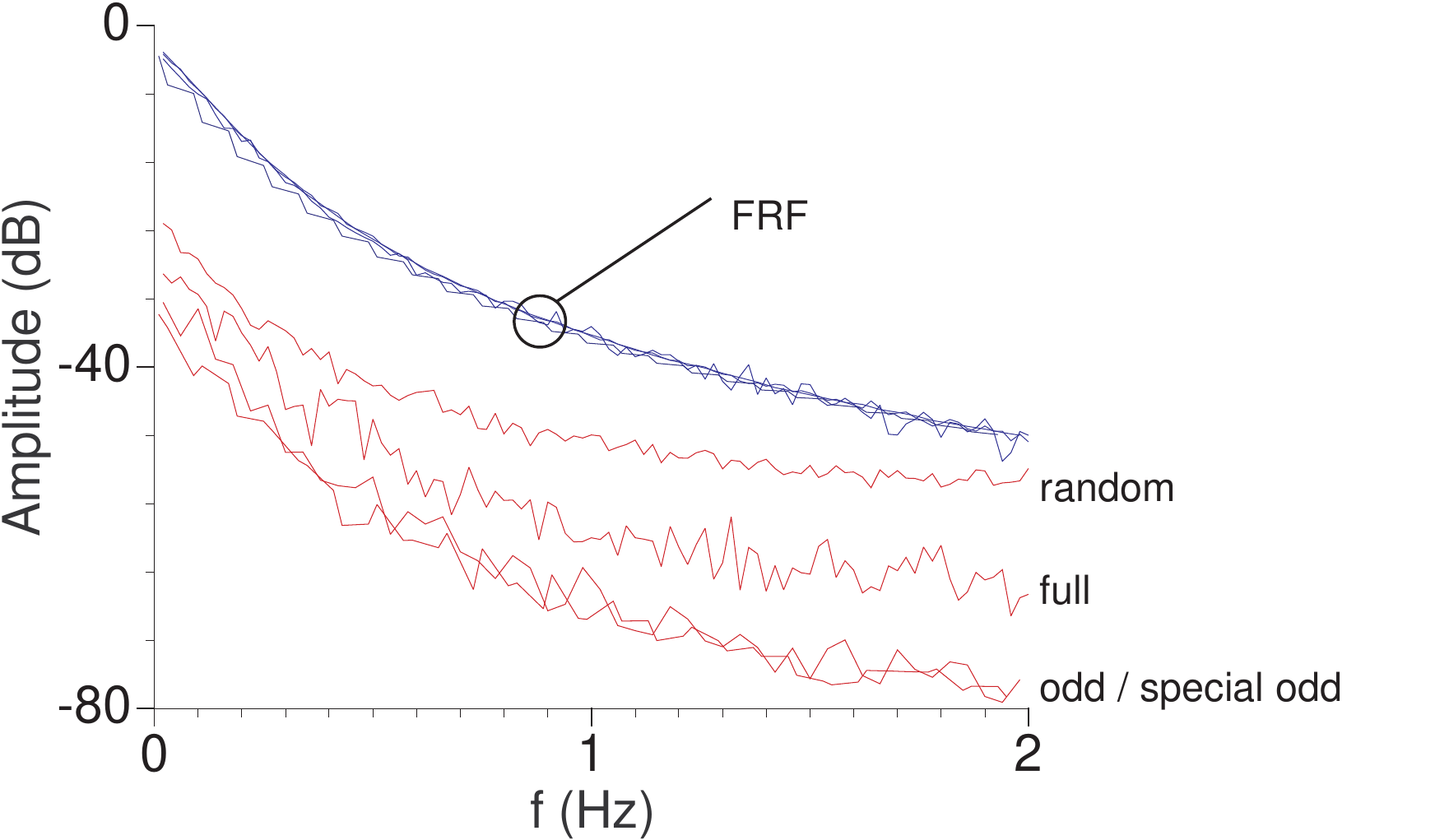}\caption{Illustration of the noise reduction in the frequency response function
(FRF) measurement of $G_{BLA}$ on a hot-air device. The FRF of the
best linear approximation is measured using random noise excitations,
a multisine that excites all frequencies, and two multisines that
excite only the odd frequencies. The blue lines show the measured
FRF, the red lines show the standard deviation.\label{fig:BLA FRF Hairdryer experiment}.}
\end{figure}

The huge gain that can be obtained by following the above guidelines
for excitation signal design are illustrated on a hair dryer setup.
The current that is driving the heating element is shaped with the
excitation signal around a given set point using a thyristor. The
firing angle was selected such that dominating even nonlinearities
show up in the thyristor characteristic. Three different classes of
Riemann-equivalent excitation signals were designed. The total available
measurement time was the same for each of these signals. Care was
taken so that the frequency resolution of the FRF measurement was
equal for all the excitations (an odd multisine has only half the
frequency resolution of a full multisine that excites all the frequencies).
The results are shown in Figure \ref{fig:BLA FRF Hairdryer experiment}.
The FRF for the three classes of excitations coincide well, as is
expected for Riemann-equivalent signals \cite{Schoukens RiemanEquivalence}.
However, the standard deviations are very different. The random noise
excitation is the worst; this is due to the presence of dips in the
input power spectrum. By using a multisine, the dips are eliminated
and this results in a reduced standard deviation. However, the best
results are obtained by the odd multisines, because these eliminate
completely the impact of the (dominant) even nonlinear distortions,
and reduce the standard deviation even more. Using an odd multisine
excitation eventually reduces the standard deviation almost 20 dB
(factor 10) with respect to the random excitation. Such a reduction
corresponds to a reduction in measurement time of a factor 100.

\subsubsection*{User guidelines}
\begin{itemize}
\item All the nonparametric expressions of the linear theory can be used
to measure the FRF of the BLA. So there is no need to change the
measurement equipment to deal with the nonlinear situation.
\item Use odd random-phase multisine excitations, designed following the
guidelines of the section on nonlinear detection, to measure the
FRF $\hat{G}_{BLA}(k)$ and its variance $\hat{\sigma}_{G}^{2}(k)$. 
\item Averaging over multiple realizations reduces the impact of the disturbing
noise and the stochastic nonlinearities $Y_{S}$. It results in a
smoother estimate. However, it does not reduce the systematic contributions
of the nonlinear distortions to the BLA. The latter cannot be reduced
using averaging techniques: averaging smooths the result, but the
nonlinear dependency of $G_{BLA}$ on the input characteristics will
not be reduced.
\item Use the available measurement time to maximize the number of realizations of the random-phase multisine by keeping the number of repeated periods
$P$ per realization small (for example, $P=3$). This advice can be refined, depending on the prior knowledge of the user:1)
No prior knowledge available: select $P=2$, and $M$ as large as
possible. 2) Maximize the nonlinear detection ability: $M=2$, and
$P$ as large as possible. 3) If it is known that the nonlinear distortions dominate:
$P=1$, and $M$ as large as possible (the disturbing noise level
will not be estimated in this case). \emph{If even nonlinearities
dominate} : use an odd multisine, exciting only the odd frequencies in the frequency band of interest. \emph{If
odd nonlinearities dominate}: use a full multisine, exciting all frequencies in the frequency band of interest.
\item A generalization to the measurement of the BLA of a MIMO system is
discussed in \cite{Dobrowiecki 2007 automatica}.
\end{itemize}

\subsection*{FRF measurement of the best linear approximation: experimental illustrations}

The measurement of the FRF of the BLA is illustrated on a labscale
(the forced Duffing oscillator) and on two real-life examples: ii)
A ground vibration test on a fighter jet, and iii) a MIMO measurement
on an industrial robot.

\subsubsection*{The forced Duffing oscillator}

\begin{figure}
\begin{centering}
\includegraphics[scale=0.5]{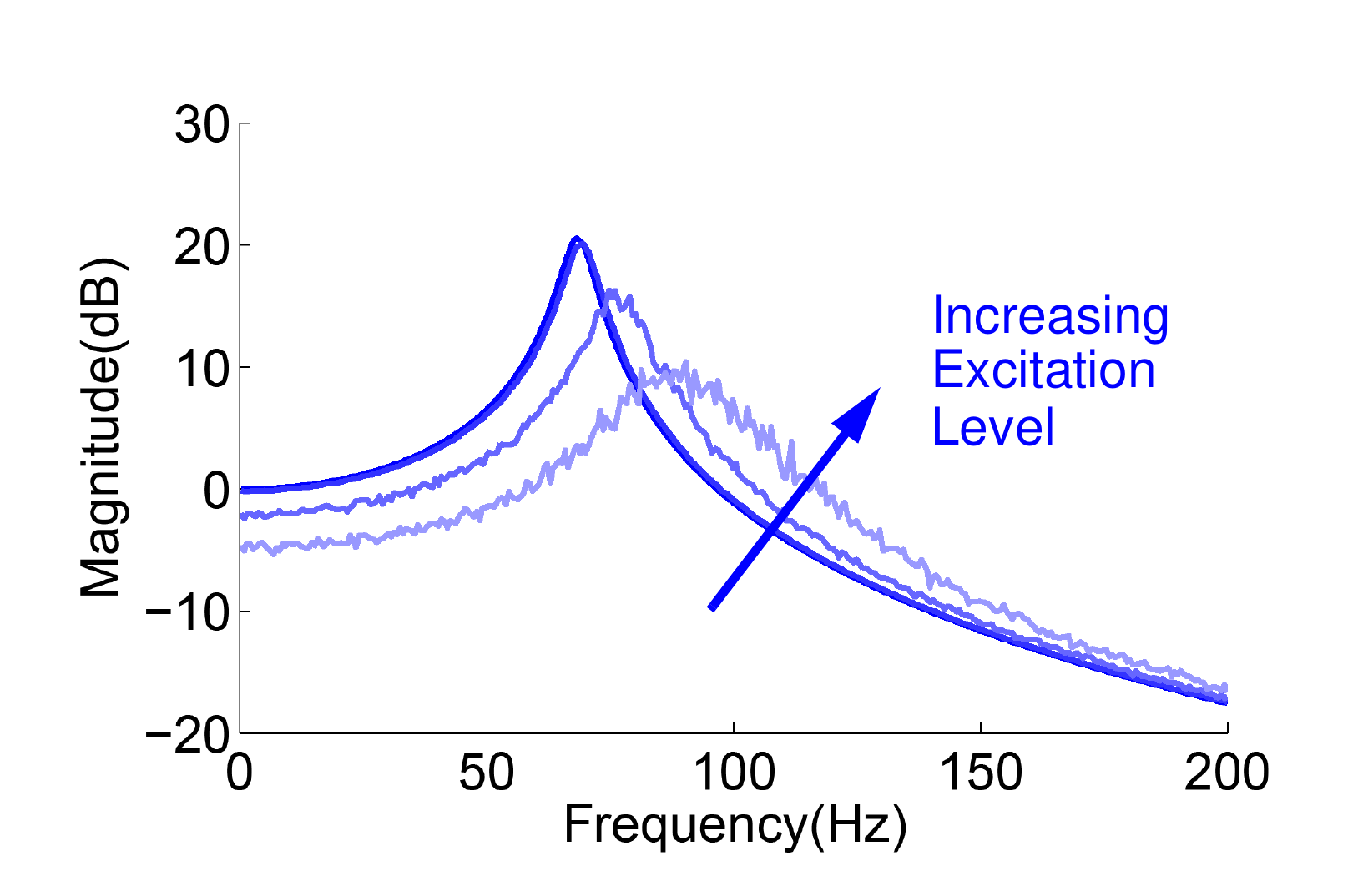}
\par\end{centering}
\caption{Measured frequency response function of a resonating system with
a hardening spring. The resonance frequency shifts to the right for
an increasing excitation level. This is the typical behavior for a
hardening spring. The shift is due to the systematic nonlinear contributions.
These create a shift in the dynamics of the best linear approximation.
The apparent noisy behavior, induced by the stochastic nonlinearities,
grows with the excitation level. \label{fig:Nonlinear-spring-FRF}}
\end{figure}

In this example, the measurements on the forced Duffing oscillator,
that are discussed earlier, are further processed. The FRF is measured
for 4 different excitation levels and shown in Figure \ref{fig:Nonlinear-spring-FRF}.
The FRF is averaged over 50 realizations of the input signal to obtain
a smoother result. Observe that the resonance frequency shifts to
the right for increasing ecitation levels and that the measurements
become more noisy. Both effects are completely due to the nonlinear
distortions. The level of the distortions that corresponds to these
measurements can be seen in Figure \ref{fig: Forced Duffing oscilator NonLinDist}. 

\subsubsection*{Ground vibration test on a fighter jet}

\begin{figure}
\begin{centering}
\includegraphics[scale=0.2]{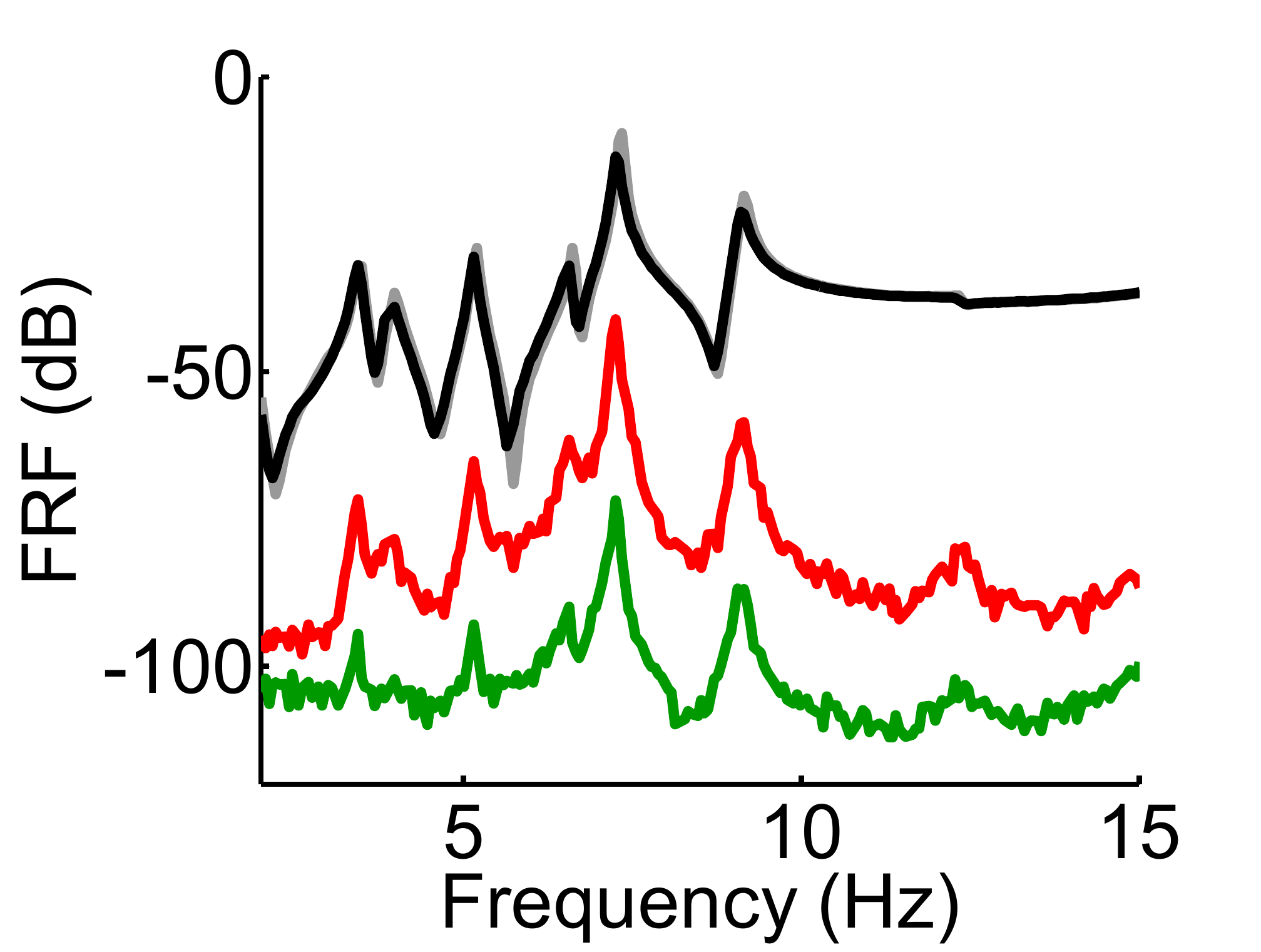}
\par\end{centering}
\begin{centering}
(a)
\par\end{centering}
\begin{centering}
\includegraphics[scale=0.2]{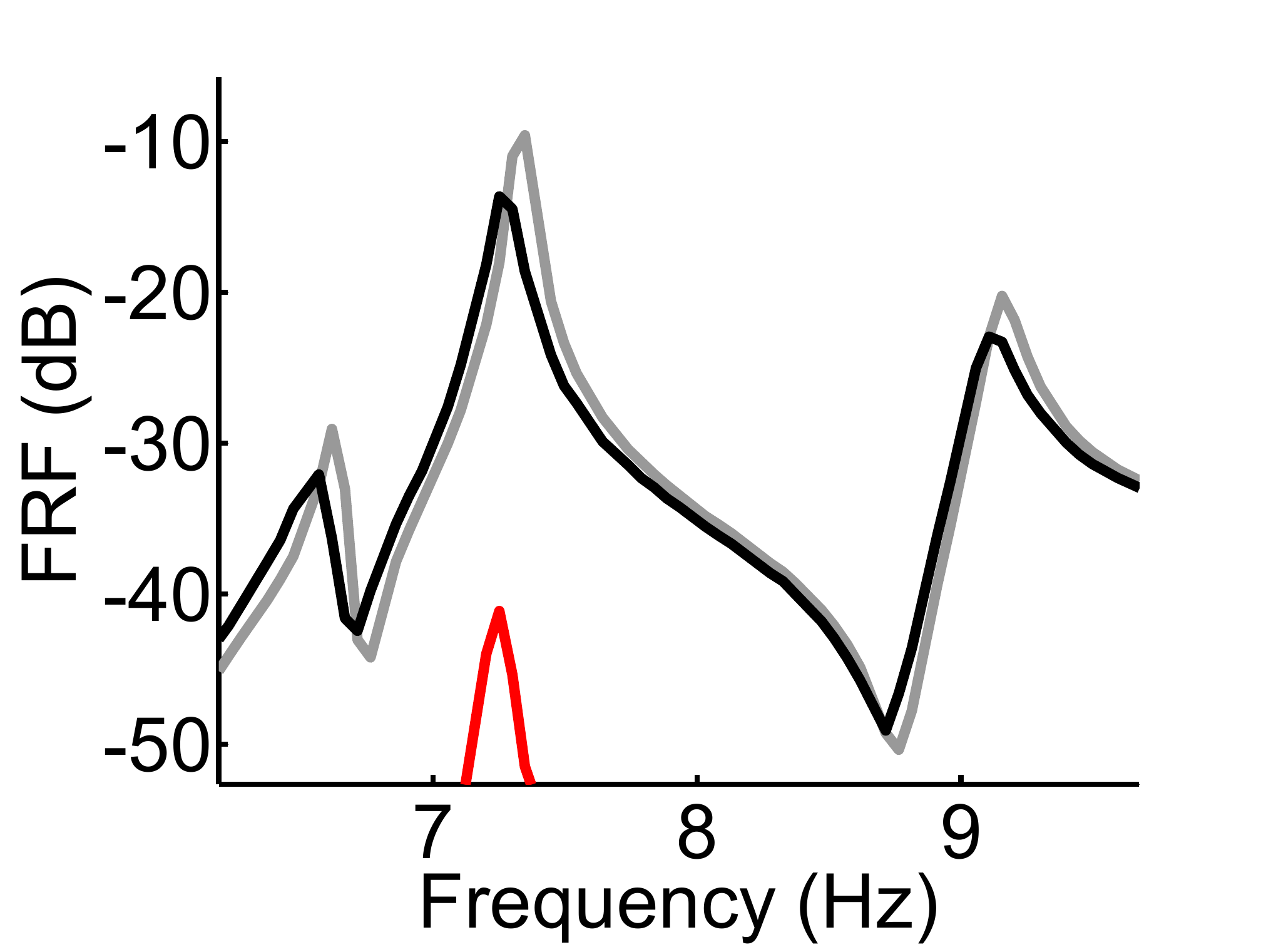}
\par\end{centering}
\centering{}(b)\caption{Measured frequency response function of $G_{BLA}$ of the F16-fighter
(see Figure \ref{fig:F16-fighter NL dist}) at a small (grey line)
and medium (black line) excitation levels. A specific challenge encountered
with fighter aircraft is the modeling of the wing-to-payload mounting
interfaces (for example the missile on the tip of the wing in Figure
\ref{fig:F16-fighter NL dist}(a)). For large amplitudes of vibration,
friction and gaps may be triggered in these connections, resulting
in a nonlinear behavior. Figure (b) shows a zoom around 7 Hz of the
measurements shown in Figure (a). The levels of the total variance
(red), and the disturbing noise (green) are given. Observe that the
resonance frequency shifts to the left for an increasing excitation
level. This corresponds to a softening spring. This behavior originates
from the bolted connections mentioned before. In  Figure (a) it is
seen that the nonlinearities are largest around the resonance frequencies,
and they are well above the disturbing noise level. \label{fig:F16-fighter-FRF}}
\end{figure}

In the second example, the measurements of the ground vibration test,
which was discussed earlier, are processed. In this example, some
additional signal processing was done to provide additional information.
This leads to an alternative method to detect and analyze the presence
of nonlinear distortions, called the robust method \cite{Dhaen Tom 2005,Pintelon 2012 book,Schoukens 2012  Exercises book}.
First, for each realization, the FRF is averaged over the successive
periods. This provides not only an averaged FRF, it gives also an
estimate of the disturbing noise variance $\sigma_{Y}^{2}(k)$ as
a function of the frequency. This estimate is not affected by the
nonlinear distortions because these do not vary over the periods.
Next, the FRFs and variance estimates are further averaged over the
realizations, and again the variance is calculated. This results in
a reduction of the impact of the stochastic nonlinearities on the
FRF measurement, it also allows the variance of disturbing noise plus
the stochastic nonlinearities $\sigma_{Y}^{2}(k)+\sigma_{Y_{S}}^{2}(k)$
to be estimated. This value, called the total variance, is shown in
Figure \ref{fig:F16-fighter-FRF}(a), together with the amplitude
of the output. Since the total variance $\sigma_{Y}^{2}+\sigma_{Y_{S}}^{2}$
is much larger than the noise variance $\sigma_{Y}^{2}(k)$, it follows
that it is also an excellent measurement of the nonlinear distortion
level in this case. 

This nonlinear analysis method is an alternative approach to measure
the level of the nonlinear distortions and the noise. Its major advantage
is that no detection lines are imposed on the excitation. This not
only increases the resolution of the measurement (the even lines are
also used to measure the FRF), it also relaxes the constraints on
the input signal because it is no longer necessary to impose the zero
lines. Nonlinear actuators are no longer a problem. Thus this method
is also called the robust method. The major disadvantage is that it
is no longer possible to make a distinction between the even and the
odd nonlinearities.

The FRF of $G_{BLA}$is shown in Figures\ref{fig:F16-fighter-FRF}(b).
Again, it can be observed that it varies with the excitation level.
This time the resonance is shifting to the left, which corresponds
to a softening stiffness, and is in agreement with the fact that,
in this case, the nonlinearities are due to a bolted connection between
the wing tip and the missile.

\subsubsection*{Multiple-input multiple-output FRF measurements on an industrial
robot}

\begin{figure}
\begin{centering}
\includegraphics[scale=0.12]{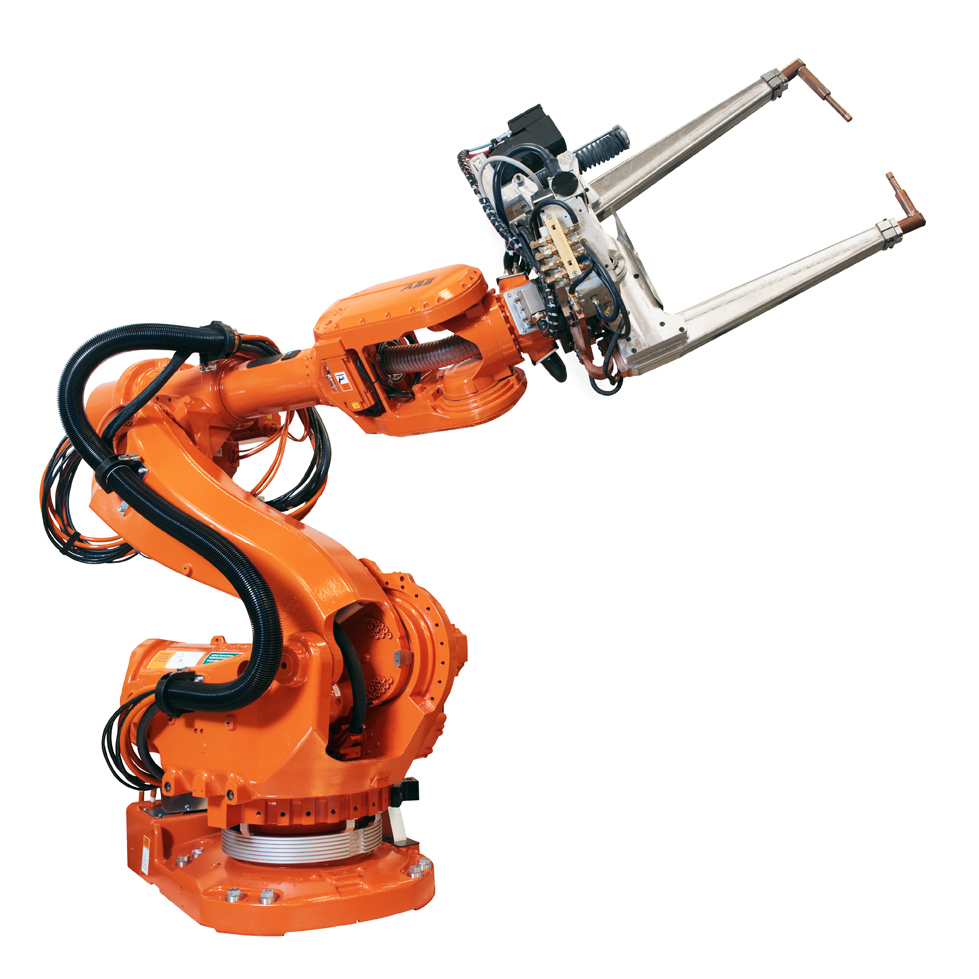}
\par\end{centering}
\begin{centering}
(a)
\par\end{centering}
\begin{centering}
\includegraphics[scale=0.5]{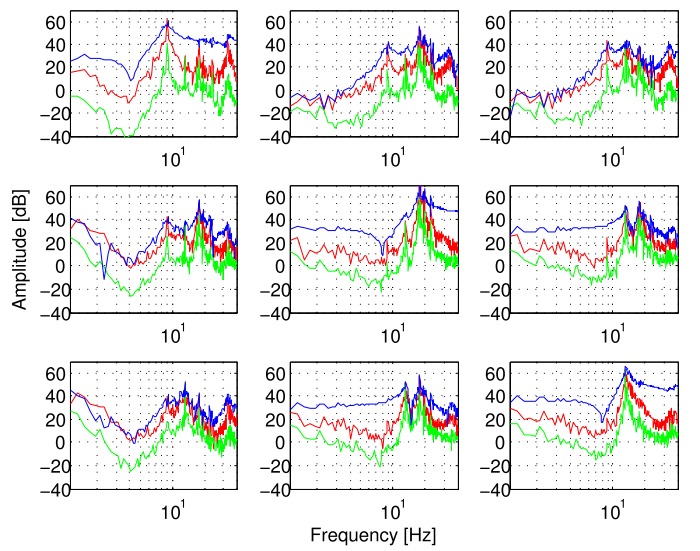}
\par\end{centering}
\begin{centering}
(b)
\par\end{centering}
\centering{}\caption{Nonparametric analysis of an industrial robot, Figure (a), with 6
degrees of freedom using orthogonal multisine excitations. The multi-input
multi-output frequency response function (FRF) for the 6 degrees of
freedom is measured. The FRFs between 3 motor torques and 3 motor
accelerations are shown in Figure (b). Blue line: frequency response
function, Red line: level of the total variance (nonlinear distortions
+ disturbing noise), Green line: level of the disturbing noise. The
nonlinear distortions dominate, the red line is everywhere well above
the green line \cite{Wernholt (2008) PhD,Wernholt (2008) I=000026M Robots ABB}.
\label{fig:Robot-ABB}}
\end{figure}

In this example, the MIMO FRF of an industrial robot with 6 degrees
of freedom is measured (see Figure \ref{fig:Robot-ABB}) \cite{Wernholt (2008) I=000026M Robots ABB}.
The figure shows a selected set of the measured FRFs between three
motor torques and three motor accelerations. When designing excitations
for MIMO measurements, additional design aspects come into the scope
besides those that were already discussed before. In a MIMO FRF measurement,
a set of linear equations with a dimension $n_{u}\times n_{u}$ ($n_{u}$
is the number of inputs) should be solved. The condition number of
this matrix affects very strongly the noise sensitivity. Using orthogonal
multisine excitations \cite{Pintelon (2011) MSSP  MIMO LPM and NL analysis,Dobrowiecki 2007 automatica},
it is possible to obtain a condition number equal to 1, while at the
same time it is still possible to make the nonlinear distortion analysis.
In this case, the total variance and the noise variance is shown.
Using these signals resulted in a significant improvement of the FRF measurements.
The settings for these MIMO measurements are as follows. The measurements
are averaged over 9 realizations of an odd random-phase multisine.
From each realization, a set of $n_{u}=6$ orthogonal multisines is
created and is used as the input for a single MIMO experiment. $P=2$
periods are measured in steady state for each realization. The period
length of the random-phase multisine is 10 s, and 195 odd frequencies
are excited in the frequency band from 1-40 Hz. The sample frequency
was $f_{s}=2$ kHz. More information can be found in \cite{Wernholt (2008) I=000026M Robots ABB},
\cite{Wernholt (2008) PhD}.

\section*{Parametric identification of the best linear approximation\label{sec:BLA parametric model}}

In many applications, a parametric transfer function model or state-space
representation of the system is needed, together with an estimate
of the model uncertainty. 

\setcounter{subsection}{0}

\subsection*{Plant model estimation}

Starting from $\hat{G}_{BLA}(k)$ and $\hat{\sigma}_{G}^{2}(k)$,
it is possible to obtain such a parametric model by minimizing the
following weighted least-squares cost function that comes from the
linear system identification theory \cite{schoukens dobrowiecki NL dist,Pintelon 2012 book}

\begin{equation}
V\left(\theta\right)=\frac{1}{F}\sum_{k=1}^{F}\frac{\left|\hat{G}_{BLA}(k)-G\left(\Omega_{k},\theta\right)\right|^{2}}{\hat{\sigma}_{G}^{2}\left(k\right)},\label{eq:Cost function}
\end{equation}
where $\Omega_{k}$ is the continuous or discrete-time frequency variable.
It can be shown that the minimizer $G(\Omega_{k},\hat{\theta)}$ of
the cost function \eqref{eq:Cost function} is a consistent estimator
for $G_{BLA}$ (the estimate converges to the exact value as the number
of data points tends to infinity) if the BLA is in the model set. 

An alternative is to use the results of the prediction error framework
\cite{Ljung boek 1999}-\cite{Pintelon 2012 book}. In that case,
a parametric model is used

\[
\hat{\sigma}_{Y}^{2}\left(k\right)=\lambda\left|H\left(\Omega_{k},\theta\right)\right|^{2},
\]
 and the cost function is formulated directly on the input output
data leading to

\[
V_{pe}\left(\theta\right)=\frac{1}{N}\sum_{t=1}^{N}(H^{-1}(q,\theta)(y(t)-G(q,\theta)u(t))^{2}.
\]

In the general problem, the plant and noise model parameters are estimated.
The reader is referred to \cite{Ljung boek 1999,Soderstrom boek 1989}
to learn how to choose the plant and noise model structure. In the
motivational example, a Box-Jenkins model structure was used. In
that case, the plant and noise model have no common parameters. A
simplified approach would be to put the noise model $H(q,\theta)=1$.
Under open-loop conditions, this will still lead to consistent estimates
for $G_{BLA},$ provided that $G_{BLA}$ is in the model set. However,
no information on the distortion levels will be available, and the
uncertainty on the estimated plant model will be larger. Under closed-loop
conditions, the estimate will become also biased.

\subsection*{Variance estimate of the plant model\label{subsec:variance parametric models}}

The linear system identification theory provides also a theoretical
estimate of the variance of the estimated model, starting from the
assumption that the disturbing noise is independent of the excitation
signal $u(t)$. This assumption does not hold for $y_{s}(t).$ As
mentioned before, the stochastic nonlinearity is uncorrelated with
the input, but still dependent on it. A detailed study shows that
this dependency between the input and the nonlinear distortions will
lead to a much higher variance than what is predicted by the linear
theory \cite{Schoukens variability parametric bla 2010}. 

The worst-case situation occurs for a static nonlinear system $y=u^{n}$,
as was studied in the toy example of the previous section. In that
case, an analytical analysis can be made made for a zero-mean white
Gaussian noise excitation. The BLA $y=a_{BLA}u$ was given by $a_{BLA}=\mu_{n+1}/\mu_{2}$.
The ratio between the actual variance $\sigma_{a_{BLA}}^{2}$ that
will be observed by repeated experiments, and the variance $\sigma_{a_{BLA,ind}}^{2}$
that is obtained from the linear identification theory assuming independent
noise, is \foreignlanguage{english}{
\[
\frac{\sigma_{a_{BLA}}^{2}}{\sigma_{a_{BLA,ind}}^{2}}=2n+1.
\]
} This shows that the underestimation of the variance by the linear
framework grows linearly with the nonlinear degree $n$. This leads
to far too optimistic uncertainty bounds; underestimation of the
actual variance with a factor 7 (about 8 dB) or more occurs. 

\begin{figure}
\begin{centering}
\includegraphics[scale=0.6]{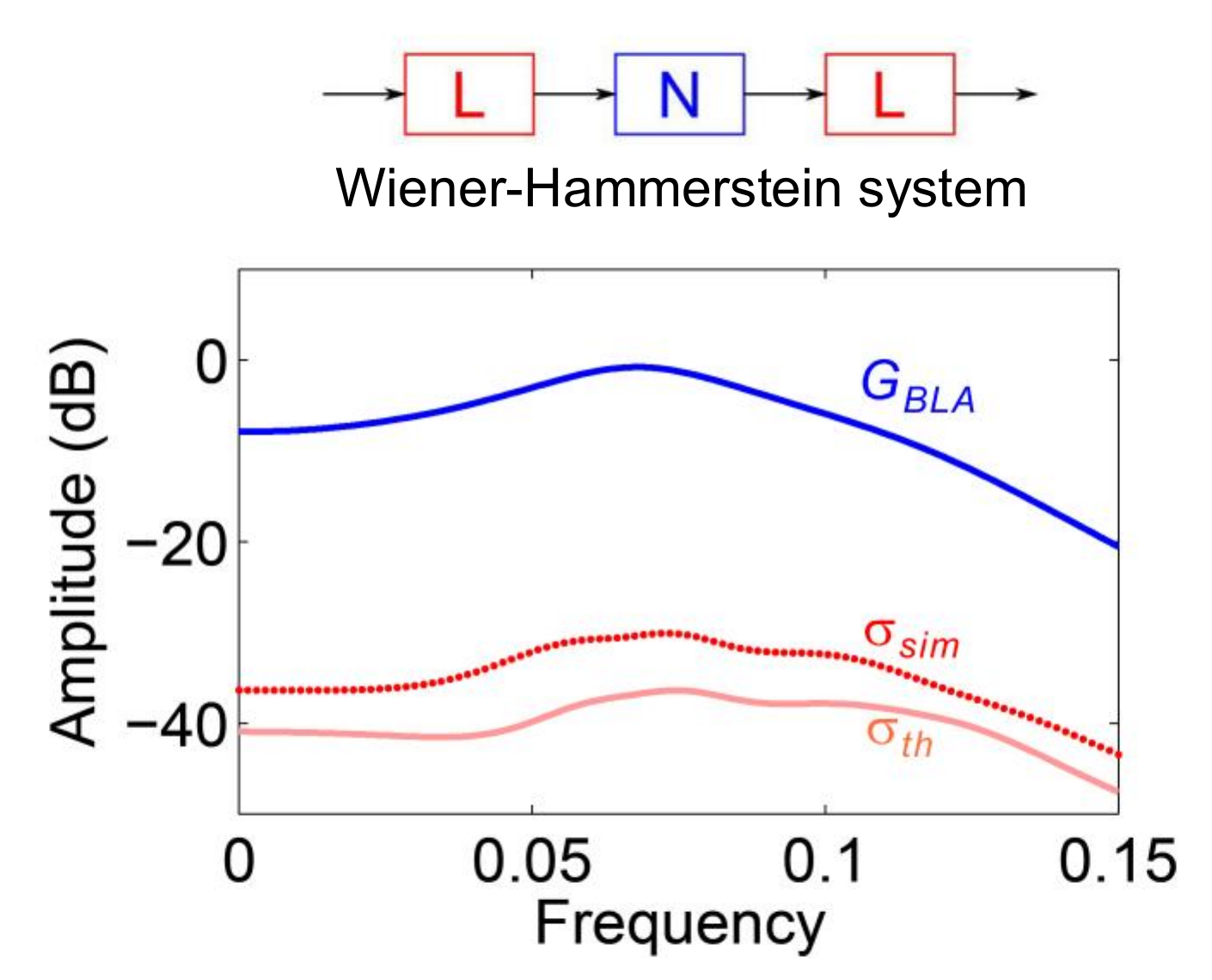}
\par\end{centering}
\caption{Parametric identification of a Wiener-Hammerstein system. The theoretical
$\sigma_{th}$ and the actually observed $\sigma_{sim}$ standard
deviation are shown. \label{fig:Parametric-identification-of-BLA}}
\end{figure}

In Figure \ref{fig:Parametric-identification-of-BLA}, the underestimation
effect is shown on the identification of a Wiener-Hammerstein system.
Such a system consists of the cascade of a linear dynamic system,
a static nonlinear system, and a linear dynamic system. It can be
shown, for Gaussian excitations, 
\[
G_{BLA}(k)=\alpha G_{1}(k)G_{2}(k),
\]
with $G_{1},G_{2}$ the transfer functions of the first and second
linear system, and $\alpha$ a constant that depends on the nonlinear
system and the properties of the excitation signal. From Figure \ref{fig:Parametric-identification-of-BLA},
it can be seen that the actual observed error level in the simulations
$\sigma_{sim}$ is significantly larger than the expected level $\sigma_{th}$
from the linear system identification theory. 

Similar results can also be observed in the experimental results in
Figure \ref{fig:Motivating Example: transfer function}, where it
is  visible that the actual variance is about 4 dB larger than what
is predicted by the linear framework.

\subsection*{User guidelines}
\begin{itemize}
\item Measure the FRF $\hat{G}_{BLA}(k)$ and its variance $\hat{\sigma}_{G}^{2}(k)$
following the guidelines for nonparametric measurements of the BLA,
and estimate the parametric model. Take care: while the uncertainty
bounds of the linear theory could be safely used for nonparametric
models, they are NOT valid for the parametric model. There exists,
for the time being, no simple theory to provide better error bounds. 
\item The BLA can also be directly estimated from the raw input-output
data in the time or frequency domain, using the classical linear framework. 
\item A detailed step-by-step procedure explaining how to identify a parametric
estimate of the BLA is given in Chapter 7 of \cite{Schoukens 2012  Exercises book}.
\end{itemize}

\section*{Nonlinear distortion analysis under closed-loop conditions\emph{\label{sec:Nonlinear-distortion-analysis under closed loop conditions}}}

The nonparametric nonlinear distortion analysis method that was proposed earlier in this article has to be used
under open-loop measurement conditions. To prevent unstable behavior
or saturation, the measurements on a dynamic system are often made
under closed-loop conditions. In other situations, the interaction
between the system and the actuator creates closed-loop effects, especially
when the input impedance of the system is not very large with respect
to the output impedance of the actuator. Because it is the typical
situation for mechanical systems, the open-loop nonlinear distortion
analysis and the concept of BLA need to be generalized to these closed-loop
measurement conditions. The impact of closed-loop conditions on the
measurement of the FRF of a linear system is discussed in "Measuring the FRF of a linear system uncer closed-loop conditions".  Without special precautions, the closed-loop effect will create systematic errors on the FRF-measurement: either the FRF of the feedforward, the FRF of the inverse feedback, or a combination of both is measured. For that reason, more advanced measurement techniques are needed under feedback conditions.

Also the the detection and characterization of the nonlinear distortions
needs to be changed. Due to the presence of a feedback loop, the nonlinear
distortions at the output of the system will also influence the input.
This destroys the special input design that was proposed and illustrated
in the previous section. The input signal spectrum should be zero
at the detection lines (no excitation), but under closed-loop conditions,
the nonlinear distortions will now excite these frequencies. Two strategies
are proposed to deal with this situation. First, a simple correction
method is proposed and illustrated on the measurement of the open-loop
characteristics of an operational amplifier. Next, the nonlinear distortion
concepts will be extended to include more formally the closed-loop
situation.

The presence of the feedback also makes it impossible to impose the
specially designed multisine signals with the detection lines put
equal to zero (those frequencies that were not excited) because the
feedback signal will be added to it. To deal with that situation,
two solutions are proposed. For large SNRs, a correction algorithm
is proposed to compensate for the non ideal excitation signal. Only
the signals in the loop ($u,y$) are needed. For lower SNRs, an indirect
method is proposed that requests the availability of the reference
signal $r$.

\begin{figure}
\centering{}\includegraphics[scale=0.6]{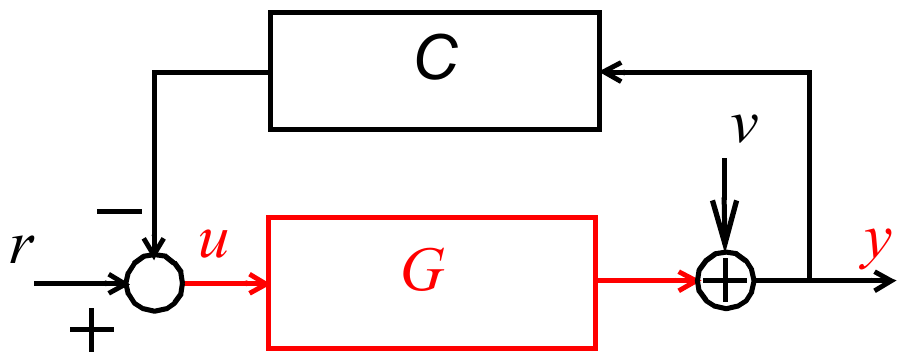}\caption{Measuring under feedback conditions. Observe that the input signal
$u(t)$ does not only depend on the reference signal $r(t)$, it is
also affected by the disturbance $v(t)$, so that the input is no
longer independent of the disturbances as it is the case under open-loop
measurement conditions.\label{fig:Feedback set-up}}
\end{figure}

\setcounter{subsection}{0}

\subsection*{Characterizing nonlinear distortions under closed-loop conditions
using a first-order correction\label{subsec:Characterizing-nonlinear-distortions under closed loop (DIrect method)}}

If the nonlinear distortions are not too large, and the SNR is high
(for example, more than 20 dB), it can be safely assumed that, at
the excited frequencies, the reference signal dominates the disturbances,
and hence, as explained in ``Measuring the FRF of a linear system
under closed-loop conditions'', $\tilde{G}\approx G.$ This measured
value is used to compensate for the presence of the feedback contributions
at those frequencies where the input was assumed to be zero. The direct
feed through of these disturbing terms on the output at the detection
lines $k_{det}$ can then be compensated for \cite{Van Hoenacker (2003) correction NL distortion analysis,Pintelon 2012 book,Schoukens 2012  Exercises book}

\[
Y_{corr}(k_{det})=Y(k_{det})-\check{G}(k_{det})U(k_{det}),
\]
 where $\check{G}(k_{det})$ is an interpolated value that is obtained
from the neighbouring excited frequencies. At those frequencies, by
using that $R(k_{det})=0,$ the corrected output equals

\[
Y_{corr}=Y-\check{G}U=\frac{1}{1+GC}V-\frac{-GC}{1+GC}V=V.
\]

Hence, the original value of the disturbance is retrieved. The sensitivity
function $1/(1+GC)$ of the closed loop is removed, which shows that
the correction results in an 'opening' of the closed loop.

\subsection*{Experimental illustration on an operational amplifier}

The results in this section are obtained from the work reported in
the publications \cite{Pintelon (2004) OPAMP1,Pintelon (2004) OPAMP 2};
a detailed description of the experimental setup is given in these
articles. 

\begin{figure}
\centering{}\includegraphics[scale=0.6]{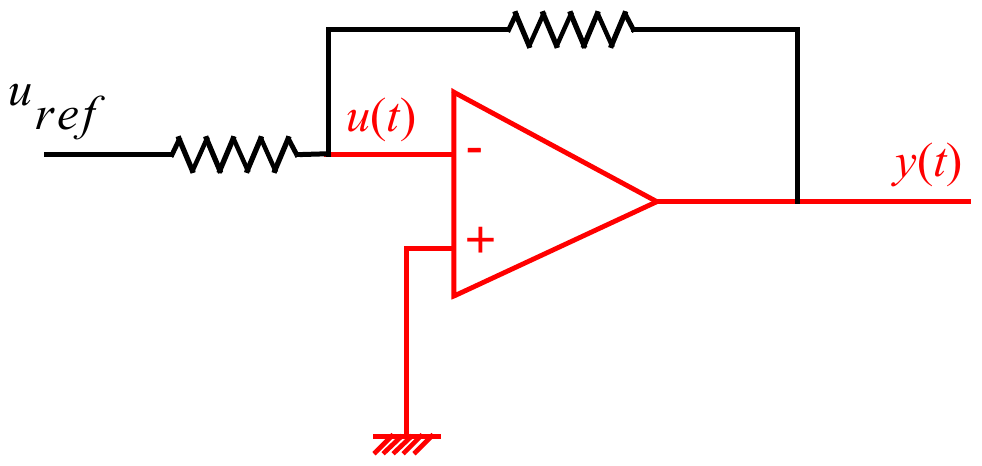}\caption{Operational amplifier, captured in a feedback loop. The signals $u,y$
are measured.\label{fig:Feedback OPAM setup}}
\end{figure}

The previously explained methods are applied to the characterization
of an operational amplifier (OPAMP). Such a device cannot operate
in open loop due to the very high gain at low frequencies (10 000
or more). For that reason the OPAMP under test is captured in a feedback
loop, as shown in Figure \ref{fig:Feedback OPAM setup}. 

\begin{figure}
\centering{}\includegraphics[scale=0.6]{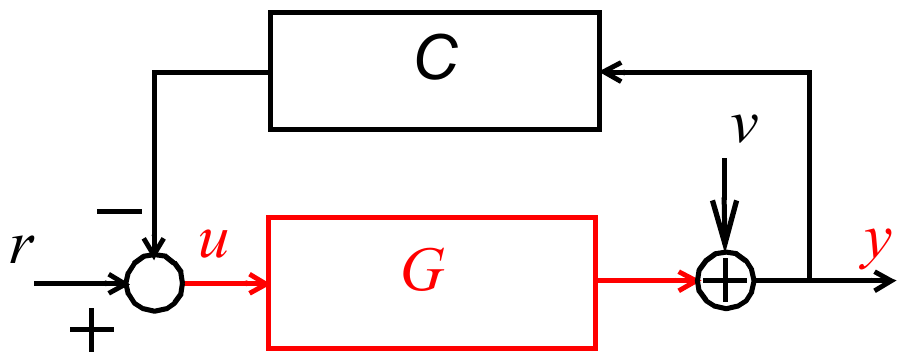}\caption{Linear equivalent representation of the operational amplifier setup
in Figure \ref{fig:Feedback OPAM setup}.\label{fig: Feedback OPAMP linear representation}}
\end{figure}

Replacing the nonlinear system by its BLA plus the nonlinear noise
term $y_{S}$ results in the equivalent representation for the OPAMP
setup in Figure \ref{fig: Feedback OPAMP linear representation}.

\begin{figure}
\centering{}\includegraphics[scale=0.35]{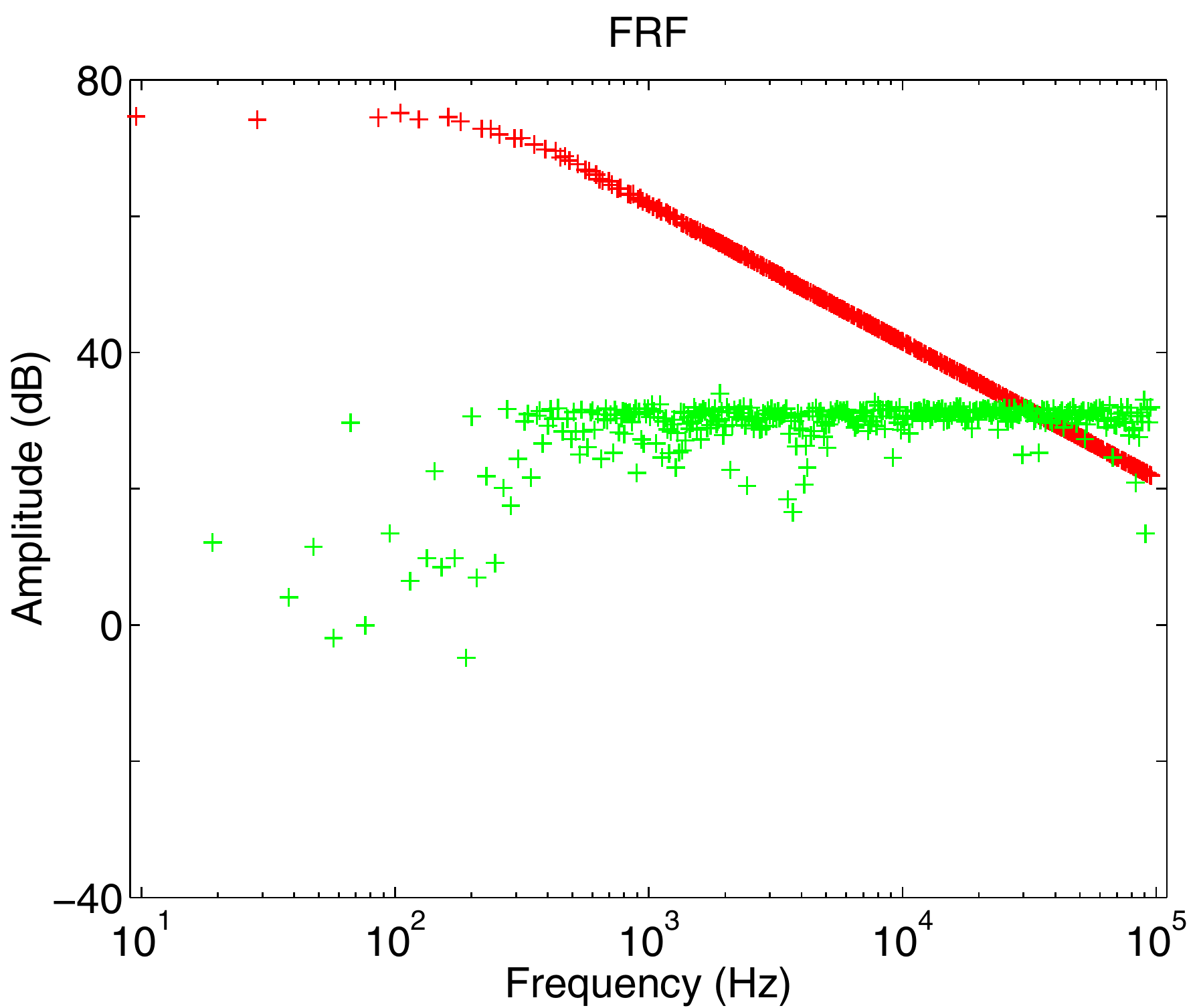}\caption{Nonlinear distortion analysis of an operational amplifier (OPAMP),
captured in a closed-loop setup. The signals $u(t),y(t)$ are measured.
The feedback effects are eliminated using a compensation algorithm.
A random odd excitation is used. At the excited frequencies, the open-loop
frequency response function FRF of the OPAMP is measured (red). At
the unexcited frequencies, the inverse of the feedback loop is measured
(green crosses). As expected, the OPAMP has a large gain at low frequencies,
but above the crossover frequency around 200 Hz, the amplitude rolls
off with 6 dB/octave. The FRF of the inverse transfer function of
the feedback $1/G_{FB}$ (green) remains constant over the frequency.
At low frequencies the measurements are strongly scattered because
the nonlinear distortions to noise ratio is very low at those frequencies.\label{fig:Feedback: OPAMP FRF measurement}}
\end{figure}

\begin{figure}
\begin{centering}
\includegraphics[scale=0.35]{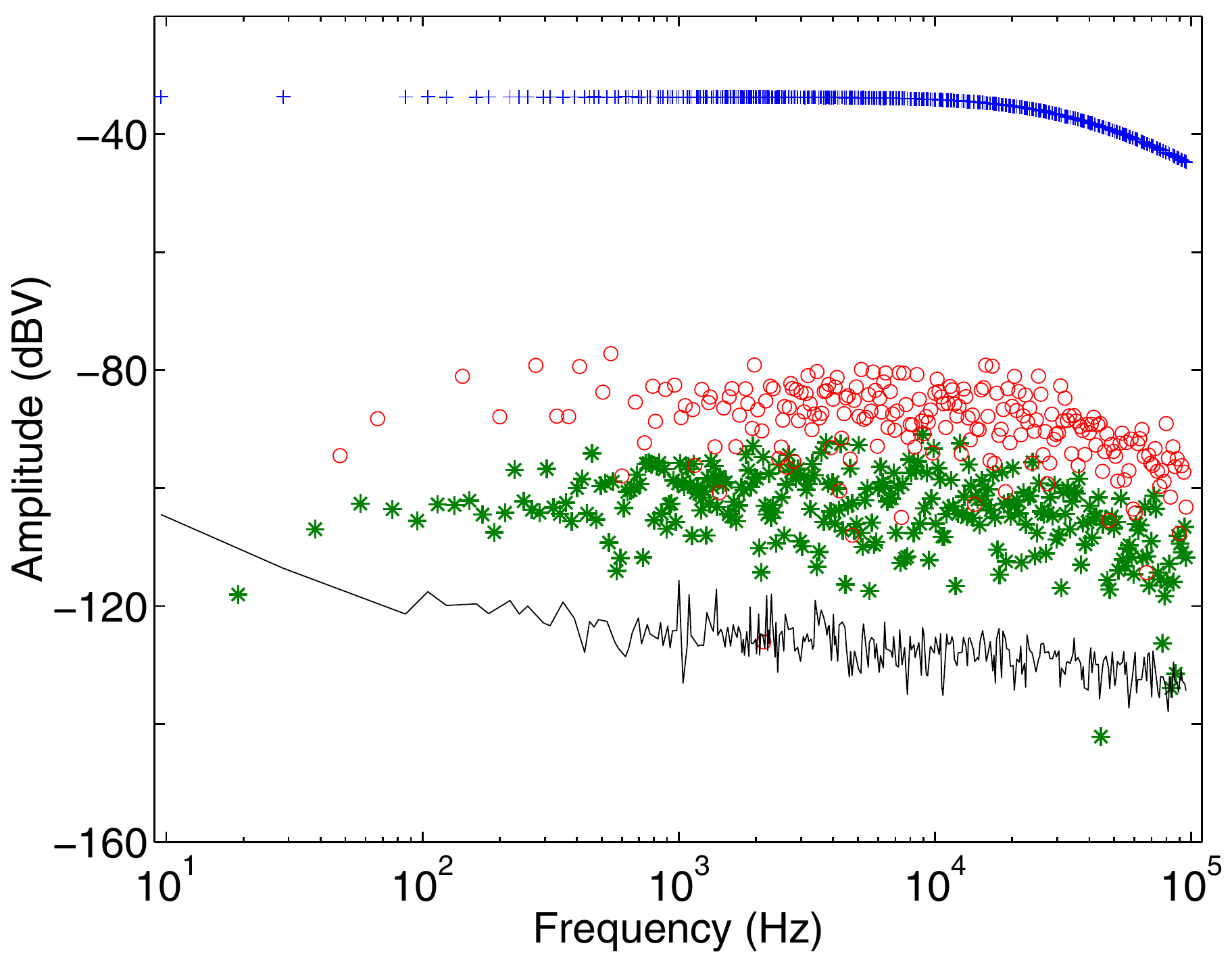}
\par\end{centering}
\begin{centering}
(a)
\par\end{centering}
\begin{centering}
\includegraphics[scale=0.35]{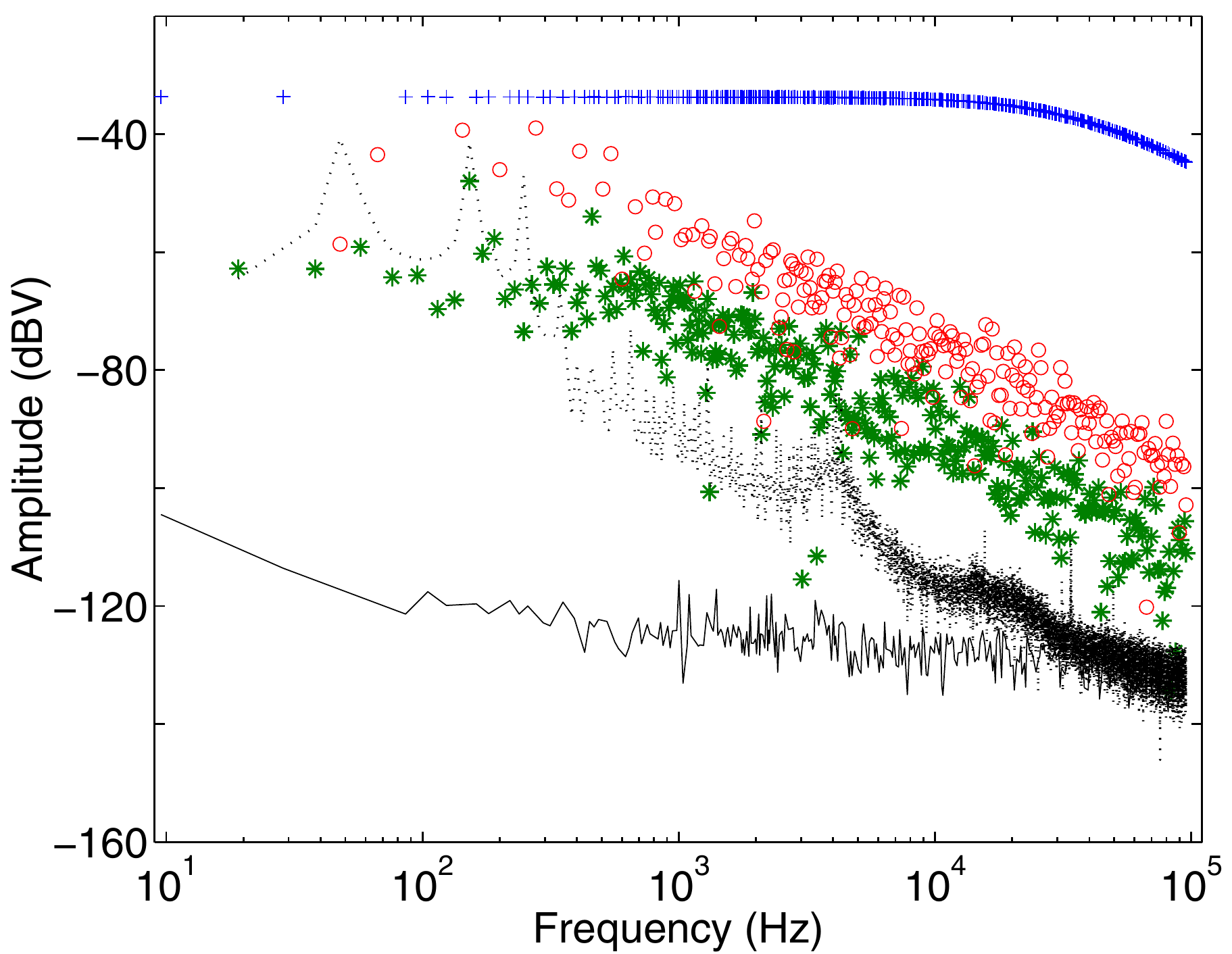}
\par\end{centering}
\begin{centering}
(b)
\par\end{centering}
\centering{}\caption{Nonlinear distortion analysis at the output of the OPAMP before (a)
and after (b) the software elimination of the feedback effects (see
Figure \ref{fig:Feedback: OPAMP FRF measurement}). The figure shows
the output measured at the excited frequencies (blue), the odd (red)
and even (green) nonlinear distortions, and the disturbing noise level
(black). The broken black line gives the disturbing noise level after
compensation. The nonlinear distortions in (a) are smaller than in
(b). The feedback is suppressing the nonlinear distortions. In (b),
the odd nonlinear distortions become as large as the output at the
excited frequencies. This shows that an OPAMP is a heavily nonlinear
component that is linearized by the feedback loop at a cost of the
gain of the amplifier. \label{fig:Feedback OPAMP nonlinear Distortions Results}}
\end{figure}

The reference signal $r(t)$ that excites the feedback circuit is
again designed as explained earlier in this article. At the excited
frequencies, the reference signal $r$ dominates, and the open-loop
characteristic of the OPAMP will be measured. At the detection lines
the disturbances $y_{S}$ dominate, and hence the inverse controller
characteristic will be obtained. The results are shown in Figure \ref{fig:Feedback: OPAMP FRF measurement}.
Using the color red for the feedforward and green for the feedback,
the different FRFs become  visible. As expected, the OPAMP has a
large gain at low frequencies, but above the cross-over frequency
around 200 Hz, the amplitude rolls off with 6 dB/octave. This is well
in agreement with the results from the textbooks. The FRF of the inversed
feedback (green) remains constant over the frequency, which is again
in agreement with the resistive feedback network in Figure \ref{fig:Feedback OPAM setup}.
At low frequencies, the measurements are strongly scattered; it will
be shown below that this is because the nonlinear distortions to noise
ratio is very low at those frequencies (Figure \ref{fig:Feedback OPAMP nonlinear Distortions Results}).

The results of the nonparametric nonlinear distortion analysis are
shown in Figure \ref{fig:Feedback OPAMP nonlinear Distortions Results},
before applying the compensation in Figure(a), while the compensated
results are given in Figure (b). The most obvious difference is the
strong increase of the nonlinear distortion level. The high-gain feedback
loop results in a very strong disturbance suppression. The high gain
is exchanged for an improved linear behavior. Without this high disturbance
rejection of the feedback loop, the nonlinearities at low frequencies
would become as large as the actual output of the OPAMP. So it can
be concluded that the nonlinearity level will set the maximum gain
that can be obtained with an OPAMP circuit.

\subsection*{Characterizing nonlinear distortions under closed-loop conditions:
an extended framework\label{subsec:Characterizing-nonlinear-distortions in Closed Loop (indirect method)}}

As explained before, two problems are faced to deal with measurements
under closed-loop conditions: i) the FRF measurement is biased; ii)
the actual excitation signal $u(t)$ in Figure \ref{fig:Feedback set-up}
is disturbed by the nonlinear distortions that are fed back to the
input so that the detection lines are also excited, which is in conflict
with their definition. In the previous sections, a simple linear correction
method was proposed to reduce the effects on the distortion analysis.
Here an extended framework is shortly presented that eliminates the
bias on the FRF measurements, and generalizes the concepts of BLA
and the stochastic nonlinear contributions to closed-loop systems
(the actuator, the plant, and the controller can be nonlinear). Here
an intuitive explanation is given; see \cite{Pintelon (2013) I=000026M NL dist under NL feedback}
for a detailed and formal discussion. The basic idea is to use not
only the measured input and output $u,y$, but to make also explicit
use of the availability of the reference signal $r$ (see Figure \ref{fig:Feedback set-up}). 

\subsubsection*{The indirect FRF-measurement method}

When a direct measurement of the BLA $G_{BLA}$ is made \eqref{eq:Gbla SYU/SUU},
the nonlinear distortions $Y_{S}$ in $Y=G_{BLA}U+Y_{s}$ will create
a bias

\[
\hat{G}_{BLA}=\frac{S_{YU}}{S_{UU}}=G_{BLA}+\frac{S_{Y_{S}U}}{S_{UU}}.
\]
(the frequency index $k$ is dropped to simplify the notations).
In general, $S_{Y_{S}U}\neq0$ because, through the feedback path,
the input $U$ depends on the nonlinear distortions $Y_{S}$. The
bias term $S_{Y_{S}U}/S_{UU}$ can be eliminated by making an indirect
measurement of the BLA. The FRF $G_{BLA}$ is estimated as the division
of the BLA from the reference signal to the input $G_{ur}$ and from
the reference to the output $G_{yr}$
\[
G_{BLA,r}=\frac{G_{yr}}{G_{ur}}=\frac{S_{YR}}{S_{UR}}.
\]

\subsubsection*{Nonlinear distortion analysis using the indirect method}

Define the stochastic nonlinear contributions $\tilde{U}_{S},\tilde{Y}_{S}$
with respect to the reference signal $r$ as
\[
\begin{array}{c}
Y=G_{yr}R+\tilde{Y}_{S},\\
U=G_{ur}R+\tilde{U}_{S}.
\end{array}
\]
A generalized definition for the stochastic nonlinearities $Y_{S}$
of the plant, captured in the closed loop is then
\[
Y_{S}=\tilde{Y}_{S}-G_{BLA,r}\tilde{U}_{S}.
\]
The following properties of $G_{BLA,r}$ and the generalized nonlinear
distortions are shown to hold \cite{Pintelon (2013) I=000026M NL dist under NL feedback}
\begin{itemize}
\item \emph{Open Loop, Nonlinear System, and Linear Actuator}: the extended
concepts of the BLA and the nonlinear distortions become identical
to the previously defined open-loop concepts.
\item \emph{Closed Loop, Linear System, Nonlinear Actuator, and Nonlinear
Feedback}: In this case, $G_{BLA}$ equals the FRF of the linear system
transfer function. The generalized stochastic nonlinearities $Y_{S}$
are equal to zero. $\tilde{U}_{S},\tilde{Y}_{S}$ will be different
from zero, pointing to the global nonlinear behavior of the loop.
However, it will be detected that the plant is linear. This allows the nonlinear contributions in the loop to be assigned to the controller..
\item \emph{Closed Loop, Nonlinear System, Nonlinear Actuator, and Nonlinear
Feedback}: The level of the nonlinear behavior of the plant is detected.
Some precautions should be taken when interpreting the presence of
even and odd nonlinear contributions. Precise conditions, that can
be easily verified in practice, are given in \cite{Pintelon (2013) I=000026M NL dist under NL feedback},
to check if the results are reliable.
\end{itemize}
These results confirm that the simplified procedure that was explained
earlier in this article can be used safely if the SNR is more than
10 dB (bias below 10\%) or 20 dB (bias below 1\%).

\section*{Publicly available software}

All the results in this article can be reproduced using publicly available
Matlab toolboxes. The motivational example was produced using the
System Identification toolbox of Matlab (Mathworks). Alternatively,
the freely available frequency-domain identification toolbox FDIDENT
could be used to obtain similar results (http://home.mit.bme.hu/\textasciitilde{}kollar/fdident/).
This toolbox also includes the tools to design the random-phase multisines
and to perform the nonparametric nonlinear analysis. In \cite{Schoukens 2012  Exercises book},
all the procedures that are presented in this article are discussed
in full detail, and the related Matlab software can be freely downloaded
at (booksupport.wiley.com). 

\section*{Conclusions}

This article studied the problem of how to deal with nonlinear distortions
in the linear system identification framework. In a first step, nonparametric
tools were discussed to detect the presence and the level of the
nonlinear distortions, and to analyze their nature (even or odd).
Next, the concept of the BLA was introduced. The dependency of the BLA
on the user choices was studied (choice of the excitation signal,
and choice of the approximation criterion). Optimal measurement strategies
to measure the FRF of the BLA were presented, and eventually the
impact of the nonlinear distortions on the linear parametric identification
approach were discussed. All these methods were illustrated on real-life
examples.

\section*{Acknowledgement}

This work was supported in part by the Fund for Scientific Research
(FWO-Vlaanderen), by the Flemish Government (Methusalem), the Belgian
Government through the Inter-university Poles of Attraction (IAP VII)
Program, and by the ERC advanced grant SNLSID, under contract 320378.
The authors thank Chris Criens, Technical University Eindhoven, The
Netherlands, for the contribution on the Section ``Characterization
of the air path of a diesel engine'', including Figures \ref{fig:Diesel Engine}
and \ref{fig:Diesel Engine results}; P.W.M.J. Nuij, Technical University
Eindhoven, The Netherlands, for the contribution to the Section ``Higher-order
sinusoidal input describing functions'', including Figures \ref{fig:Nuij Setup StickSlip}
and \ref{fig:Nuij HOSIDF 1 and 3}; E. Wernholt, Linkoping University,
Sweden, for the contribution to the Section ``Multiple-input multiple-output
FRF measurements on an industrial robot'', including the Figure \ref{fig:Robot-ABB};
G. Kerschen and J.P. Noel, Univerist� de Li�ge, Belgium, for the Sections
``Swept sine test'', including Figures \ref{fig:Ground-vibration-test Swept Sine}
and \ref{fig:Ground-vibration-test Time-Freq analysis}, and ``Ground
vibration test on an air fighter'', including Figures \ref{fig:F16-fighter NL dist}
and \ref{fig:F16-fighter-FRF}; B. Peeters, LMS International, part
of Siemens Product Lifecycle Management, Belgium, for the contributions
to the ``Ground vibration test on an air fighter'', including Figures
\ref{fig:F16-fighter NL dist} and \ref{fig:F16-fighter-FRF}.

\section*{Sidebar 1: A mathematical framework for nonlinear systems\label{subsec:Approximation NLS: A-mathematical-framework} }

In this section, some basic results of the Volterra theory \cite{Schetzen 2006}
are given, without proof, to provide the reader with an intuitive
insight into the behavior of nonlinear systems. The emphasis will
be to show how a nonlinear system is shifting the input power from
one frequency to the other. Three intermediate steps will be made:
\begin{itemize}
\item From a one-dimensional impulse response (linear theory) to a multidimensional
impulse response (nonlinear system).
\item multidimensional frequency description of nonlinear systems: a tool
for intuitive insight in nonlinear behavior \cite{Chua and Ng (1979),Chua (1979) NLS freq. domain Paper 2}.
\item Return to the physical world: collapsing the multidimensional frequency
description to a single dimension.
\end{itemize}

\subsection*{Using Volterra kernels as multidimensional impulse responses}

A linear system is characterized by its impulse response $g(t)$,
and the input-output relation is given by the convolution integral
\[
y(t)=\int_{0}^{\infty}g(\tau)u(t-\tau)d\tau.
\]
In the Volterra approach, the output of the nonlinear system is given
by the sum of the contributions of increasing nonlinear degree $\alpha$
\begin{equation}
y(t)=\sum_{\alpha=1}^{\infty}y^{\alpha}(t),\label{eq:Volterra sum yalfa}
\end{equation}
with $y^{\alpha}(t)=\int_{0}^{\infty}\ldots\int_{0}^{\infty}g_{\alpha}(\tau_{1},\ldots,\tau_{\alpha})u(t-\tau_{1})\ldots u(t-\tau_{\alpha})d\tau_{1}\ldots d\tau_{\alpha}.$
The kernel $g_{\alpha}(\tau_{1},\ldots,\tau_{\alpha})$ is the multidimensional
impulse response of degree $\alpha$.

\subsection*{Multidimensional frequency response functions}

The signal $y^{\alpha}(t)$ in \eqref{eq:Volterra sum yalfa} can
be generalized to a multidimensional time signal $y^{\alpha}(t_{1},\ldots,t_{\alpha})=\int_{0}^{\infty}\ldots\int_{0}^{\infty}g_{\alpha}(\tau_{1},\ldots,\tau_{\alpha})u(t_{1}-\tau_{1})\ldots u(t_{\alpha}-\tau_{\alpha})d\tau_{1}\ldots d\tau_{\alpha}.$
The original signal is retrieved by putting $t_{1}=\ldots=t_{\alpha}=t$.
The multidimensional representation in the frequency domain becomes
\begin{equation}
Y^{\alpha}(\omega_{1},\ldots,\omega_{\alpha})=G^{\alpha}(\omega_{1},\ldots,\omega_{\alpha})U(\omega_{1})\ldots U(\omega_{\alpha}),\label{eq:Volterra multi freq dom Y}
\end{equation}
with $G^{\alpha}(\omega_{1},\ldots,\omega_{\alpha})$ the multidimensional
Fourier transform of $g_{\alpha}(\tau_{1},\ldots,\tau_{\alpha})$.
This multidimensional representation in the frequency domain is very similar to the result for the
linear case: the output spectrum is obtained as the (multidimensional)
product of the transfer function with the input.

\subsection*{From a multidimensional to a one-dimensional frequency variable}

The one-dimensional spectrum $Y^{\alpha}(\omega)$ is retrieved by
looking for all frequency combinations $\omega_{1},\ldots,\omega_{\alpha}$,
such that $\omega_{1}+\ldots+\omega_{\alpha}=\omega$. These are retrieved
by $Y^{\alpha}(\omega)=\int_{-\infty}^{\infty}\ldots\int_{-\infty}^{\infty}Y^{\alpha}(\omega_{1},\ldots,\omega_{\alpha-1},\omega-\omega_{1}+\ldots+\omega_{\alpha-1})d\omega_{1}\ldots d\omega_{\alpha-1}.$
Observe that this is a generalization of the response of a nonlinear system to a sinusoid.

\subsection*{Measuring the Volterra kernels}

Although the Volterra representation is an attractive nonparametric
description of a general class of nonlinear systems, there are only a few
methods described in the literature to measure the Volterra kernels,
most of them focussing on systems with short memories. The major
reason for this lack of interest is the exploding number of parameters
to be identified due to the multidimensional nature of the kernels.
A first possibility is to use higher-order correlation methods \cite{Schetzen 2006},
often combined with an orthogonal representation of the Volterra series,
for example, using a Wiener representation. Methods were presented to avoid the long correlation times that are needed by the use of well-designed excitation signals \cite{Boyd (1983) measuring volterra kernels}.
The basic idea is to generate a multisine where the active frequencies
are selected such that the harmonic interference of the kernels of
different degree is eliminated up to a given degree (for example,
degree 4). The design of such 'no interharmonic distortion' signals
(NID-signals) is discussed in detail in \cite{Evans (1994) NID signals,Evans (1996) NID signals design}.
Methods to measure the multivariate FRF that make use of such signals
are discussed, for example, in \cite{Boyd (1983) measuring volterra kernels,Tan and Godfrey (2002) LIFRED}.

\section*{Sidebar 2: The best linear approximation of a Volterra system }

In this section, an explicit expression is given for the BLA for
a Volterra kernel of degree $\alpha$. The multidimensional output
is given in \eqref{eq:Volterra multi freq dom Y}. For a multisine
excitation, the contributions at a given frequency $k$ are retrieved
by looking for all frequency combinations such that $\sum_{i=1}^{\alpha}k_{i}=k$,
with $k_{i}$ an excited frequency, see also \eqref{eq:NLS output frequencies}.
These multivariate output contributions are given by 
\begin{equation}
Y(k_{1},k_{2},\ldots,k_{\alpha})=G^{\alpha}(k_{1},k_{2},\ldots,k_{\alpha})U(k_{1})U(k_{2})\ldots U(k_{\alpha}),\label{eq:BLA Volterra Y(w,w,w)}
\end{equation}
 for a kernel of degree $\alpha$. Among these, only the contributions
for which 
\begin{equation}
U(k_{1})U(k_{2})\ldots U(k_{\alpha})U(-k)=U(k_{1})U(k_{2})\ldots U(k_{\alpha})\bar{U}(k),\label{eq:BLA  YBLA contribution condition}
\end{equation}
 (where the overbar denotes the complex conjugate) do not depend on
the input phases $\angle U$ will contribute to the BLA. If this
product would still depend on the random-phases of the input, its
expected value over multiple realizations would be zero because, by
definition, for a random-phase multisine $E\{e^{j\varphi}\}=0$, and
the phases of a random-phase multisine are independent over the frequency.
So it should be possible to write \eqref{eq:BLA  YBLA contribution condition}
as a real number, which can only be done if all the components $U$
in this product can be combined in pairs of complex-conjugated inputs,
for example $U(k_{i})U(-k_{i})=|U(k_{i})|^{2}$, such that the input
phases are cancelled.

From this result, it can be seen that only kernels with an odd degree
$\alpha$ can contribute. For the odd kernels, the contribution of
degree $\alpha=2\beta+1$ to $Y_{BLA}(k)$ is
\[
Y_{BLA}(k)=\sum_{k_{1}}\ldots\sum_{k_{\beta}}C_{k,k_{1},-k_{1},\ldots,k_{\beta},-k_{\beta}}G_{\alpha}(k,k_{1},-k_{1},\ldots,k_{\beta},-k_{\beta})|U(k_{1})|^{2}\ldots|U(k_{\beta})|^{2}U(k),
\]
where the sum runs over all frequencies that are excited. The constant
$C_{k,k_{1},-k_{1},\ldots,k_{\beta},-k_{\beta}}$ accounts for the
number of all possible frequency combinations that are obtained by
changing the position of the frequencies in \foreignlanguage{english}{$G_{\alpha}(k,k_{1},-k_{1},\ldots,k_{\beta},-k_{\beta})$.}

\emph{Example: a Wiener-Hammerstein system}

\begin{figure}
\centering{}\includegraphics[scale=0.5]{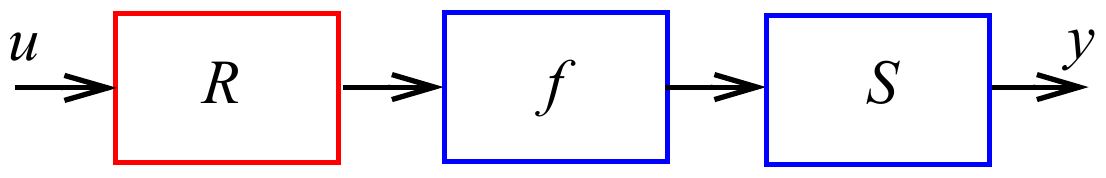}\caption{Wiener-Hammerstein system. A static nonlinear system $f$ is sandwiched
between the linear dynamic systems $R,S$.\label{fig:BLA: Wiener-Hammerstein-system.}}
\end{figure}

Consider the Wiener-Hammerstein system in Figure \ref{fig:BLA: Wiener-Hammerstein-system.}
excited by a random-phase multisine with $F$ excited frequencies.
For a static nonlinearity $f=x^{3}$, the multivariate output \eqref{eq:BLA Volterra Y(w,w,w)}
becomes
\[
Y(k_{1},k_{2},k_{3})=G^{\alpha}(k_{1},k_{2},k_{3})U(k_{1})U(k_{2})U(k_{3}).
\]

For the Wiener-Hammerstein system, this reduces further to
\[
Y(k_{1},k_{2},k_{3})=S(k_{1}+k_{2}+k_{3})R(k_{1})R(k_{2})R(k_{3})U(k_{1})U(k_{2})\ldots U(k_{3}).
\]

The contributions $Y_{BLA}(k)$ are then given by looking for all
combinations $k=k_{1}+k_{2}+k_{3}$ that depend only on the phase
of $U(k)$ 
\[
Y_{BLA}(k)=U(k)\times6S(k)R(k)\sum_{l=1}^{F}|S(l)|^{2}|U(l)|^{2}-E_{Y},
\]
with $E_{Y}=3S(k)R(k)|S(k)|^{2}|U(k)|^{2}U(k)$ a correction term
because for $l=k$, only 3 permutations are possible instead of 6
\cite{Evans and Rees (2000) O(N-1) effect in Gbla}. 

The stochastic nonlinear contributions $Y_{S}(k)$ are given by all
other terms where condition \eqref{eq:BLA  YBLA contribution condition}
does not hold, for example, the contribution $S(k)R(k-1)R(-k-2)R(k+3)U(k-1)U(-k-2)U(k+3)$.

\section*{Sidebar 3: Measuring the FRF of a linear system under closed-loop
conditions }

Measuring the FRF of a system under closed-loop conditions requires special precautions. Depending on the SNR of the measurements, either the FRF of the feedforward branch, the inverse feedback branch, or a combination of both results.
The most simple approach to measure the FRF of a system is to measure
the input and output signals $u(kT_{s}),y(kT_{s}),k=1,\ldots,N$,
with $T_{s}=1/f_{s}$ the sampling period, calculate the discrete
Fourier transforms $U(k),Y(k)$ of these signals, and divide the resulting
spectra to obtain an estimate $\hat{G}(k)=Y(k)/U(k)$ at the frequency
$kf_{s}/N$. The raw data need to be averaged to reduce the noise
and leakage errors. This should be done before dividing the spectra,
because large errors will occur at those frequencies where $U(k)$
becomes very small. For that reason, it is better to estimate first
the cross- and auto-spectrum $\hat{S}_{YU},\hat{S}_{UU}$, and the
FRF estimate at frequency $k$ is 
\begin{equation}
\hat{G}(k)=\frac{\hat{S}_{YU}(k)}{\hat{S}_{UU}(k)}.\label{eq:Gbla SYU/SUU}
\end{equation}
These methods became popular in the 1960s, especially in the combination
with pseudo random-binary excitation signals to generate multifrequency
excitations \cite{Bendat and Piersol (2010) book}-\cite{Wellstead (1981) Non-Parametric SI}.
To do so, the record is split in $P$ subrecords, and for each of
these the discrete Fourier transform $U^{[l]}(k),Y^{[l]}(k),l=1,\ldots,P$
is calculated. The cross- and auto-power spectrum estimate is then
obtained \cite{Schoukens 2012  Exercises book}

\[
\hat{S}_{YU}(k)=\frac{1}{P}\sum_{l=1}^{P}Y^{[l]}(k)\bar{U}^{[l]}(k),\hat{S}_{UU}(k)=\frac{1}{P}\sum_{l=1}^{P}|U^{[l]}(k)|^{2},
\]
where $\bar{U}$ is the complex conjugate of $U.$ When the measurement
is made under feedback conditions (see Figure \ref{fig:Feedback set-up}),
the output $y(t)$ depends on both the measured input $u(t)$ and
the disturbance source $v(t)$. Due to the presence of the feedback
loop, the signal $u$ depends also on the disturbance $v$. As a result,
the FRF measurement at frequency $k$ converges to \cite{Wellstead (1977) FRF in feedback}
\[
\tilde{G}=\frac{GS_{RR}-\bar{C}S_{VV}}{S_{RR}+|C|^{2}S_{VV}}.
\]

This expression reduces to $\tilde{G}=G,$ if $S_{VV}=0$ ($r$ dominates
over $v$), and $\tilde{G}=\frac{-1}{C},$ if $S_{RR}=0$ ($v$ dominates
over $r$). For mixed SNR, the estimate becomes a mixture of the feedforward
and feedback characteristics. 

Biographies

\emph{Johan Schoukens} received both the Master's degree in electrical
engineering in 1980, and the PhD degree in engineering sciences in
1985 from the Vrije Universiteit Brussel (VUB), Brussels, Belgium.
In 1991 he received the degree of Geaggregeerde voor het Hoger Onderwijs
from the VUB, and in 2014 the degree of Doctor of Science from The
University of Warwick. From 1981 to 2000, he was a researcher of the
Belgian National Fund for Scientific Research (FWO-Vlaanderen) at
the Electrical Engineering Department of the VUB where he is currently
a full-time professor in electrical engineering. Since 2009, he is
visiting professor at the department of Computer Sciences of the Katholieke
Universiteit Leuven. His main research interests include system identification,
signal processing, and measurement techniques. He has been a Fellow
of IEEE since 1997. He was the recipient of the 2002 Andrew R. Chi
Best Paper Award of the IEEE Transactions on Instrumentation and Measurement,
the 2002 Society Distinguished Service Award from the IEEE Instrumentation
and Measurement Society, and the 2007 Belgian Francqui Chair at the
Universit� Libre de Bruxelles (Belgium). Since 2010, he is a member
of Royal Flemish Academy of Belgium for Sciences and the Arts. In
2011 he received a Doctor Honoris Causa degree from the Budapest University
of Technology and Economics (Hungary). Since 2013, he is an honorary
professor of the University of Warwick. 

\emph{Mark Vaes} graduated as an Industrial Engineer in Electromechanics
at the Erasmushogeschool Brussel. In February 2013 he joined the ELEC
department as a PhD student. His main interests are in the field of
system identification of linear and nonlinear systems.

\emph{Rik Pintelon} received a Master's degree in electrical engineering
in 1982, a doctorate (PhD) in engineering in 1988, and the qualification
to teach at university level (geaggregeerde voor het hoger onderwijs)
in 1994, all from the Vrije Universiteit Brussel (VUB), Brussels,
Belgium. In 2014 he received the degree of Doctor of Science (DSc)
from the University of Warwick (UK) for publications with the collective
title \textquotedblleft Frequency Domain System Identification: A
Mature Modeling Tool\textquotedblright . From 1982 to 1984 and 1986
to 2000, he was a researcher with the Belgian National Fund for Scientific
Research (FWO-Vlaanderen) at the Electrical Engineering (ELEC) Department
of the VUB. From 1984 to 1986 he did his military service overseas
in Tunesia at the Institut National Agronomique de Tunis. From 1991
to 2000 he was a part-time lecturer at the VUB, and since 2000 he
is a full-time professor in electrical engineering at the same department.
Since 2009, he is visiting professor at the department of Computer
Sciences of the Katholieke Universiteit Leuven, and since 2013 he
is an honorary professor in the School of Engineering of the University
of Warwick. His main research interests include system identification,
signal processing, and measurement techniques. He is the coauthor
of 4 books on system identification and the coauthor of more than
200 articles in refereed international journals. He has been a Fellow
of IEEE since 1998 and was the recipient of the 2012 IEEE Joseph
F. Keithley Award in Instrumentation and Measurement (IEEE Technical
Field Award).

\begin{thebibliography}{References}
\bibitem{Ljung boek 1999}L. Ljung. \emph{System Identification:}
\emph{Theory for the User (second edition)}. Prentice Hall, Upper
Saddle River, New Jersey, 1999.

\bibitem{Soderstrom boek 1989}T. S�derstr�m, and P. Stoica. \emph{System
Identification. }Prentice-Hall, Englewood Cliffs, 1989.

\bibitem{Pintelon 2012 book}R. Pintelon, and J. Schoukens. \emph{System
Identification. A Frequency Domain Approach}, 2nd edition. Wiley-IEEE-press,
Piscataway, 2012.

\bibitem{Schoukens 2012  Exercises book}J. Schoukens, R. Pintelon,
Y. Rolain. \emph{Mastering System Identification in 100 Exercises}.
Wiley-IEEE-press, Piscataway, 2012.

\bibitem{Ewins Modal Testing 2000}D.J. Ewins. \emph{Modal Testing:
Theory, Practice and Application}. Research Studies Press LTD, Hertfordshire,
2000.

\bibitem{Oomen (2014) Waferstage}T. Oomen, R. van Herpen, S. Quist,
M. van de Wal, O. Bosgra, M. Steinbuch, ``Connecting system identification
and robust control for next-generation motion control of a wafer stage,''\emph{
IEEE Trans. on Control Systems Technology}, vol. 22, pp. 102-118,
2014.

\bibitem{Verbeeck (1999) Identification Synchronous Machines}J. Verbeeck,
R. Pintelon, P. Lataire, ``Identification of synchronous machine
parameters using a multiple input multiple output approach,'' \emph{IEEE
Trans. on Energy Conversion}, vol. 14, pp. 909-917, 1999.

\bibitem{Verbeeck (2000) Saturation Synchronous Machines}J. Verbeeck,
R. Pintelon, P. Lataire, ``Infulence of saturation on estimated synchronous
machine parameters in standstill frequency response tests,'' \emph{IEEE
Trans. on Energy Conversion,} vol. 15, pp. 277-283, 2000.

\bibitem{Dedene Nele (2002) FRF MIMO synchronous machine NL distortions}N.
Dedene, R. Pintelon, P. Lataire, ``Measurement of multivariable frequency
response functions in the presence of nonlinear distortions - Some
practical aspects,'' IEEE Trans. on Instrumentation and Measurement,
vol. 31, pp. 577-582, 2002.

\bibitem{Dedene Nele (2003) Global Synchronous Machine model MIMO}N.
Dedene, R. Pintelon, P. Lataire, ``Estimation of a global synchronous
machine model using a multiple-input multiple-output estimator,''\emph{
IEEE Trans. on Energy Conversion}, vol. 18, pp. 11-16, 2003.

\bibitem{Rivera Daniel (2009) Constrained Multisine  plant-friendly identification}D.E.
Rivera, H. Lee, H.D. Mittelmann, M.W. Braun, ``Constrained multisine
input signals for plant-friendly identification of chemical process
systems,'' \emph{Journal of Process Control}, vol. 19, pp. 623-635,
2009.

\bibitem{Peeters Bart (2003)MSSP  Comparative modal analysis on bridge}B.
Peeters, C.E. Ventura, ``Comparative study of modal analysis techniques
for bridge dynamic characteristics,'' \emph{Mechanical Systems and
Signal Processing}, vol. 17, pp. 965-988, 2003.

\bibitem{Westwick boek 2003}D.T. Westwick, and R.E. Kearney. \textit{Identification
of Nonlinear Physiological Systems}. IEEE-Wiley, 2003.

\bibitem{Billings book 2013}S.A. Billings. \emph{Nonlinear system
identification : NARMAX methods in the time, frequency, and spatio-temporal
domains.} John Wiley \& Sons Ltd, 2013.

\bibitem{Billings Fakhouri 1982}S.A. Billings, and S.Y. Fakhouri,
``Identification of systems containing linear dynamic and static
nonlinear elements,'' \textit{Automatica}, 18, pp. 15-26, 1982.

\bibitem{Kerschen (2006) MSSP groot NL paper}G. Kerschen, K. Worden,
A.F. Vakakis, J. Golinval, ``Past, present and future of nonlinear
systems identification in structural dynamics,'' \emph{Mechanical
Systems and Signal Processing}, vol. 20, pp. 505-592, 2006.

\bibitem{Giri and Bai book 2010}F. Giri, and E.W. Bai (Eds.). \textit{Block-oriented
Nonlinear System Identification}. Springer, 2010.

\bibitem{Schetzen 2006}M. Schetzen. \emph{The Volterra and Wiener
Theories of Nonlinear Systems}. Wiley and Sons, New York, 2006.

\bibitem{Nelles Book 2001}O. Nelles. Nonlinear System Identification.
Springer, 2001.

\bibitem{Haber and Kevicky}R. Haber and L. Kevicky. \emph{Nonlinear
System Identification - Input-Output Modeling Approach}. Volume 1\&2.
Kluwer Academic Publishers, 1999.

\bibitem{Paduart NLSS 2010}J. Paduart, L. Lauwers, J. Swevers, K.
Smolders, J. Schoukens, and R. Pintelon, ``Identification of nonlinear
systems using polynomial nonlinear state space models,'' \emph{Automatica},
46., pp. 647-656, 2010.

\bibitem{Schoukens AMC2014}J. Schoukens, ``System identification
in a real world,'' \emph{The 13th International workshop on Advanced
Motion Control AMC2014}, Yokohama, Japan, March 14-16, 2014.

\bibitem{Thompson and Stewart book Nonlinear Dynamics and Chaos}J.M.T.
Thompson and H.B. Stewart.\emph{ Nonlinear Dynamics and Chaos.} Wiley,
New York (1986).

\bibitem{Ueda (1991) Forced Duffing oscillator} Y. Ueda, ``Survey
of Regular and Chaotic Phenomena in the forced Duffing Oscillator,''
\emph{Chaos, Solutions \& Fractals}, vol. 1, pp. 199-231, 1991.

\bibitem{Forssell and Ljung (2000) Projection method} U. Forssell
and L. Ljung, ``A projection method for closed-loop identification\emph{,''
IEEE Trans. on Automatic Control}, vol. 45, pp. 2101-2106, 2000.

\bibitem{Chua and Ng (1979)}L.O. Chua and C.-Y. Ng, ``Frequence
domain analysis of nonlinear systems: General theory,'' \emph{IEE
Electron. Circuits Syst.}, 3, pp. 165-185, 1979.

\bibitem{Boyd Fading memory paper 1985} S. Boyd and L.O. Chua, ``Fading
Memory and the Problem of Approximating Nonlinear Operators with Volterra
Series,'' \emph{IEEE Trans. on Circuits and Systems}, 32, pp. 1150-1161,
1985.

\bibitem{Schoukens RiemanEquivalence}J. Schoukens, J. Lataire, R.
Pintelon, and G. Vandersteen, ``Robustness issues of the equivalent
linear representation of a nonlinear system,'' \emph{IEEE Trans.
Instrum. Meas.}, vol. 58, pp. 1737-1745, 2009.

\bibitem{Schoukens (2002) Survey excitations signals}J. Schoukens,
R. Pintelon, E. van der Ouderaa, and J. Renneboog, ``Survey of excitation
signals for FFT based signal analyzers,'' \emph{IEEE Trans. on Instrum.
and Meas.}, vol. 37, pp. 343-352, 1988.

\bibitem{Guillaume (2002) Crest-factor minimization}P. Guillaume,
J. Schoukens, R. Pintelon, and I. Kollar, ``Crest-factor minimization
using nonlinear Chebyshev approximation methods,''\emph{ IEEE Trans.
on Instrum. and Meas.}, vol. 40, 982-989, 2002.

\bibitem{Tan and Godfrey (2002) Binary excitations overview}A.H.
Tan, and K.R. Godrey, ``The generation of binary and near-binary
pseudorandom signals: an overview,'' \emph{IEEE Trans. on Instrum.
and Meas.}, vol. 51, pp. 583-588, 2002.

\bibitem{Wong Hin Kwan (2012) IEEE I=000026M paper 1  distribution}H.K.
Wong, J. Schoukens, K.R. Godfrey, ``Analysis of Best Linear Approximation
of a Wiener-Hammerstein System for Arbitrary Amplitude Distributions,''
\emph{IEEE Trans. on Instrum. and Meas.}, vol. 61, pp. 645-654, 2012.

\bibitem{Wong Hin Kwan (2013) IEEE I=000026M paper 2  PRBS}H.K. Wong,
J. Schoukens, K.R. Godfrey, ``Design of Multilevel Signals for Identifying
the Best Linear Approximation of Nonlinear Systems,'' \emph{IEEE
Trans. on Instrum. and Meas.}, vol. 62, pp. 519-524, 2013.

\bibitem{Geerardyn Egon 2013 Multisine design}E. Geerardyn, Y. Rolain,
J. Schoukens, ``Design of quasi-logarithmic multisine excitations
for robust broad frequency band measurements,'' \emph{IEEE Trans.
Instrum. Meas.}, vol. 62, pp. 1364-1372, 2013. 

\bibitem{Pintelon (2011) MSSP SISO LPM en NL analysis}R. Pintelon,
K. Barb�, G. Vandersteen, J. Schoukens, ``Improved (non-)parametric
identification of dynamic systems excited by periodic signals,''
\emph{Mechanical Systems and Signal Processing}, vol. 25, pp. 2683-2704,
2011. 

\bibitem{Pintelon (2011) MSSP  MIMO LPM and NL analysis}R. Pintelon,
G. Vandersteen, J. Schoukens, Y. Rolain, ``Improved (non-)parametric
identification of dynamic systems excited by periodic signals\textemdash The
multivariate case,'' \emph{Mechanical Systems and Signal Processing},
vol. 25, pp. 2892-2922, 2011. 

\bibitem{Anna vgl NN SVM P}A. Marconato, M. Schoukens, Y. Rolain
and J. Schoukens, ``Study of the effective number of parameters in
nonlinear identification benchmarks,'' \emph{52nd IEEE Conference
on Decision and Control}, December 10-13, 2013, Florence, Italy. 

\bibitem{Criens (2014) PhD}C. Criens. \emph{Air-path control of clean
diesel engines : for disturbance rejection on NOx, PM and fuel efficiency}.
PhD-thesis, TUEindhoven, 2014.

\bibitem{Criens et al. (2015) IJPT journal InPress}C. Criens, T.
Van Keulen, F. Willems, M. Steinbuch. A Control Oriented Multivariable
Identification Procedure for Turbocharged Diesel Engines. \emph{International
Journal of Powertrains, }In Press.

\bibitem{Criens (2010) Diesel paper}C. H.A. Criens, F.P.T. Willems,
T.A.C. van Keulen, and M. Steinbuch, ``Disturbance rejection in diesel
engines for low emissions and high fuel efficiency\emph{,'' IEEE
Trans. on Control Systems Technology,} vol. 23, 662-669, 2015\emph{.}

\bibitem{Govers (2014) ISMA ground vibration tests}Y. Govers, M.
B�swald, P. Lubrina, S. Giclais, C. Stephan, N. Botargues, ``Airbus
A350XWB ground vibration testing: efficient techniques for customer
oriented on-site modal identification,'' \emph{International Conference
on Noise and Vibration Engineering ISMA2014}, Leuven, Belgium, pp.
2495-2507, 2014.

\bibitem{Bendat and Piersol (2010) book}J.S. Bendat and A.G. Piersol.
\emph{Random Data: Analysis and Measurement Procedures}, Wiley, 2010.

\bibitem{Hazlerigg and Noton (1965) old correlation analysis paper for FRF estimation}A.D.G.
Hazlerigg, and A.R.M. Noton, ``Application of crosscorrelating equipment
to linear-system identification,'' \emph{Proceedings IEE}, vol. 112,
2385-2400, 1965.

\bibitem{Godfrey (1965) Correlation methods + prbs}K.R. Godfrey,
``The Theory of the Correlation Method of Dynamic Analysis and its
Application to Industrial Processes and Nuclear Power Plant,'' \emph{Measurement
and Control}, vol. 2, pp. 65-72, 1969.

\bibitem{Wellstead (1981) Non-Parametric SI}P.E. Wellstead, ``Non-Parametric
Methods of System Identification,'' \emph{Automatica}, vol. 17, pp.
55-69, 1981.

\bibitem{Wellstead (1977) FRF in feedback}P.E. Wellstead, ``Reference
Signals for Closed-Loop Identification,'' \emph{International Journal
of Control}, vol. 26, pp. 945-962, 1977.

\bibitem{Van Hoenacker (2003) correction NL distortion analysis}K.
Vanhoenacker, and J. Schoukens, ``Detection of nonlinear distortions
with multisine excitations in the case of nonideal behavior of the
input signal,''\emph{ IEEE Trans. on Instrumentation and Measurement},
vol. 52, pp. 748-753, 2003.

\bibitem{Pintelon (2004) OPAMP1} R. Pintelon, G. Vandersteen, L.
De Locht, Y. Rolain, and J. Schoukens, ``Experimental characterization
of operational amplifiers: A system identification approach - Part
I: Theory and simulations,'' \emph{IEEE Trans. on Instrumentation
and Measurement}, vol. 53, pp. 854-862, 2004. 

\bibitem{Pintelon (2004) OPAMP 2} R. Pintelon, Y. Rolain, G. Vandersteen,
and J. Schoukens, ``Experimental characterization of operational
amplifiers: A system identification approach - Part II: Calibration
and measurements,'' \emph{IEEE Trans. on Instrumentation and Measurement},
vol. 53, pp. 863-876, 2004. 

\bibitem{Pintelon (2013) I=000026M NL dist under NL feedback}R. Pintelon,
and J. Schoukens, ``FRF measurement in nonlinear systems operating
in closed loop,'' \emph{IEEE Trans. on Instrum. and Meas.}, vol.
62, pp. 1334-1345, 2013.

\bibitem{Vanhoenacker (2002) ISMA nonlinear tests}K. Vanhoenacker,
J. Schoukens, J. Swevers, and D. Vaes, ``Summary and comparing overview
of techniques for the detection of non-linear distortions,'' \emph{International
Conference on Noise and Vibration Engineering ISMA20}02, Leuven, Belgium,
pp. 1241-1255, 2002.

\bibitem{Vanhoenacker (2003) PhD}K. Vanhoenacker. \emph{Frequency
Response Function Measurements in the Presence of Non-Linear Distortions}.
PhD thesis, Vrije Universiteit Brussel, Belgium, 2003. 

\bibitem{Worden and Tomlinson (2001) book}K. Worden, and G.R. Tomlinson.
\emph{Non-linearity in Structural Dynamics, Detection, Identification
and Modelling}. Institute of Physics Publisching Ltd, London, England,
2001.

\bibitem{Silva and Maia (1997) book}J.M.M. Silva, and N.M.M. Maia.
\emph{Theoretical and Experimental Modal Analysis}. Research Studies
Press LTD, Taunton, England, 1997.

\bibitem{Choi and Chang (1984). Bispectral NL analysis}D. Choi, J.H.
Chang, R.O. Stearman, and E.J. Powers, ``Bispectral identification
of nonlinear mode interactions,'' \emph{Proceedings of the 2nd International
Modal Analysis Conference}, Florida, U.S., pp 602-609, 1984. 

\bibitem{Chang and Stearman (1985) Bispectral NL analysis}J.H. Chang,
R.O. Stearman, D. Choi and E.J. Powers, ``Identification of aero
elastic phenomenon employing bispectral analysis techniques,'' \emph{Proceedings
of the 3rd International Modal Analysis Conference}, Orlando, Florida
(U.S.), pp. 956-964, 1985. 

\bibitem{Tomlinson (1987) Hilbert transform}G.R. Tomlinson, ``Developments
in the use of the Hilbert transform for detecting and qualifying nonlinearities
associated with frequency response function measurements,'' \emph{Mechanical
Systems and Signal Processing, }vol. 1, pp. 151-171, 1987. 

\bibitem{Enqvist (2007) correlation nonlinearity tests}M. Enqvist,
J. Schoukens, R.Pintelon, ``Detection of unmodeled nonlinearities
using correlation methods,'' \emph{IEEE Instrumentation and Measurement
Technology Conference, IMTC 2007}, Warsaw, Poland, 2007.

\bibitem{McCormack (1994) correlation NL periodic signals}A.S. McCormack,
K.R. Godfrey, and J.O. Flower, ``The detection of an compensation
for nonlinear effects using periodic input signals,'' \emph{Control
94}, Warwick (UK), pp. 297-302, 1994.

\bibitem{Gelb (1969)}A. Gelb, and W. E. Vander Velde. \emph{Multiple-Input
Describing Functions and Nonlinear System Design}. \emph{McGraw Hill},
1968

\bibitem{Nuij (2006) MSSP  HOSIDF}P.W.J.M. Nuij, O.H. Bosgra, and
M. Steinbuch, ``Higher-order sinusoidal input describing functions
for the analysis of non-linear systems with harmonic responses,''
\emph{Mechanical Systems and Signal Processing}, vol. 20, pp. 1883-1904,
2006.

\bibitem{Nuij (2006) CEP  HOSIDF feedback}P.W.J.M. Nuij, M. Steinbuch,
and O.H. Bosgra, ``Measuring the higher order sinusoidal input describing
functions of a nonlinear-plant operating in feedback,'' \emph{Control
Engineering Practice}, vol. 16, pp. 101-113, 2008.

\bibitem{Nuij (2008) HOSIDF Stick/sliding}P.W.J.M. Nuij, M. Steinbuch,
and O.H. Bosgra, ``Experimental characterization of the stick/sliding
transition in a precision mechanical system using the third order
sinusoidal input describing function,'' \emph{Mechatronics}, vol.
18, pp. 100-100, 2008.

\bibitem{Noel (2014) NL analysis using swept sine}J.P. No�l, L. Renson,
and G. Kerschen, ``Complex dynamics of a nonlinear aerospace structure:
Experimental identification and modal interactions,'' \emph{Journal
of Sound and Vibration}, vol. 333, pp. 2588-2607, 2014.

\bibitem{Esa Standard (2012)}European Cooperation for Space Standardization.
\emph{Space engineering testing ECSS-E-ST-10-03C}. ESA ESTEC, 2012.

\bibitem{Roy (2012) Sweep rate impact on modal analysis}N. Roy, and
A. Girard, ``Revisiting the effect of sine sweep rate on modal identification''.
\emph{12th European Conference on Space Structures, Materials \& Environmental
Testing}, Noordwijk, The Netherlands, 2012.

\bibitem{Gloth (2012) Sweep rate analytical study}G. Gloth, and M.
Sinapius (2004). ``Analysis of swept-sine runs during modal identification'',
\emph{Mechanical Systems and Signal Processing}, pp. 1421-1441, 2004.

\bibitem{Pintelon MSSP part I}R. Pintelon, J. Schoukens, G. Vandersteen,
and K. Barbe, ``Estimation of nonparametric noise and FRF models
for multivariable systems-Part I: Theory,'' \emph{Mechanical Systems
and Signal Processing}, 24, pp. 573-595, 2010. 

\bibitem{Dhammika excitation signals CEP} W.D. Widanage, J. Stoev,
A. Van Mulders, J. Schoukens, G. Pinte, ``Nonlinear system-identification
of the filling phase of a wet-clutch system,'' \emph{Control Engineering
Practice}, 19, pp.1506-1516, 2011.

\bibitem{Chua (1979) NLS freq. domain Paper 2}L.O. Chua, and C.Y.Ng,
``Frequency domain analysis of nonlinear systems: formulation of
transfer functions,'' \emph{IEE Electronic Circuits and Systems},
vol. 3, pp. 165-185, 1979.

\bibitem{Evans (1994) NID signals}C. Evans, D. Rees, and L. Jones,
``Nonlinear disturbance errors in sustem identification using multisine
test signals,'' \emph{IEEE Trans. on Instrum. and Meas.}, vol. 43,
pp. 238-244, 1944.

\bibitem{Evans (1996) NID signals design}C. Evans, D. Rees, L. Jones,
and M. Weis, ``Periodic signals for measuring nonlinear Volterra
kernels,'' \emph{IEEE Trans. on Instrum Meas.}, vol. 45, pp. 362-371,
1996.

\bibitem{Boyd (1983) measuring volterra kernels}S. Boyd, Y.S. Tang,
and L.O. Chua, ``Measuring Volterra Kernels,'' \emph{IEEE Trans.
on Circuits and Systems}, vol. 30, pp. 571-577, 1983.

\bibitem{Tan and Godfrey (2002) LIFRED}A.H. Tan, and K. Godfrey,
``Identification of Wiener-Hammerstein models using linear interpolation
in the frequency domain (LIFRED),'' \emph{IEEE Trans. on Instrum.
and Meas}., vol. 51, pp. 509-521, 2002.

\bibitem{Enqvist 2005 Thesis}M. Enqvist. \emph{Linear Models of Nonlinear
systems}. PhD Thesis No. 985, Institute of technology, Link�ping University,
Sweden, 2005.

\bibitem{Enqvist Ljung 2005}M. Enqvist, and L. Ljung, ``Linear approximations
of nonlinear FIR systems for separable input processes,'' \emph{Automatica},
41, vol. 7, 459-473, 2005.

\bibitem{Makila 2004a}P.M. M�kil�, and J.R. Partington, ``Least-squares
LTI approximation of nonlinear systems and quasistationarity analysis,''
\emph{Automatica} 40, vol. 7, pp. 1157-1169, 2004.

\bibitem{Bussgang}J.J. Bussgang. \emph{Cross-correlation functions
of amplitude-distorted Gaussian signals}. Technical Report, 216, MIT
Laboratory of Electronics, 1952.

\bibitem{schoukens dobrowiecki NL dist}J. Schoukens, T. Dobrowiecki,
and R. Pintelon, ``Parametric and nonparametric identification of
linear systems in the presence of nonlinear distortions. A frequency
domain approach,'' \emph{IEEE Trans. Autom. Contr.}, 43, vol. 2,
176-190, 1998.

\bibitem{Schoukens Automatica 2005 plenary}J. Schoukens, R. Pintelon,
T. Dobroviecki, and Y. Rolain, ``Identification of linear systems
with nonlinear distortions,'' \emph{Automatica}, 41, pp. 491-504,
2005.

\bibitem{Schoukens (2010) Ys --> Ybla} J. Schoukens, T. Dobrowiecki,
Y. Rolain, and R. Pintelon, ``Upper Bounding Variations of Best Linear
Approximations of Nonlinear Systems in Power Sweep Measurements,''
IEEE Trans. on Instrum. and Meas., vol. 59, pp. 1141-1148, 2010.

\bibitem{Schoukens 2006 leakage}J. Schoukens, Y. Rolain, R. Pintelon,
and T. Dobrowiecki, ``Analysis of windowing/leakage effects in frequency
response function measurements,'' \emph{Automatica} 42, vol. 1, 27-38,
2006.

\bibitem{Evans and Rees (2000) O(N-1) effect in Gbla} C. Evans, and
D. Rees, ``Nonlinear distortions and multisine signals - Part I:
Measuring the best linear approximation,'' \emph{IEEE Trans. on Instrum.
and Meas.}, vol. 49, pp. 602-609, 2000.

\bibitem{Schoukens variance FRF BLA 2012}J. Schoukens, K. Barb�,
L. Vanbeylen and R. Pintelon, ``Nonlinear Induced Variance of Frequency
Response Function Measurements,'' \emph{IEEE Trans. on Instrum. and
Meas.}, 59, pp. 2468-2474, 2010.

\bibitem{Dobrowiecki 2007 automatica}T. Dobrowiecki, and J. Schoukens,
``Measuring a linear approximation to weakly nonlinear MIMO systems,''
\emph{Automatica}, vol. 43 (10), 1737-1751, 2007.

\bibitem{Schoukens variability parametric bla 2010}J. Schoukens and
R. Pintelon, ``Study of the Variance of Parametric Estimates of the
Best Linear Approximation of Nonlinear Systems,'' \emph{IEEE Trans.
on Instrum. and Meas}., 59, pp. 3156-3167, 2010.

\bibitem{Dhaen Tom 2005}T. D'Haene, R. Pintelon, J. Schoukens, and
E. Van Gheem, ``Variance analysis of frequency response function
measurements using periodic excitations,''\emph{ IEEE Trans. Instrum.
Meas}., vol. 54, pp. 1452-1456, 2005.

\bibitem{Wernholt (2008) I=000026M Robots ABB}E. Wernholt and S.
Gunnarsson, ``Estimation of Nonlinear Effects in Frequency Domain
Identification of Industrial Robots,'' \emph{IEEE Trans. on Instrum.
and Meas.}, vol. 57, pp. 856-863, 2008.

\bibitem{Wernholt (2008) PhD} E. Wernholt. \emph{Multivariable Frequency-Domain
Identification of Industrial Robots}. Link�ping Studies in Science
and Technology. Dissertations. No. 1138, Link�ping University, 2007.
\end{thebibliography}
\end{document}